\documentclass[12pt]{article}
\pdfoutput=1
\usepackage[colorlinks,citecolor=blue,urlcolor=blue,linkcolor = blue]{hyperref}
\usepackage[comma,sort&compress]{natbib}
\usepackage{amssymb, amsmath,bm,graphicx}
\usepackage{mathtools}
\usepackage{setspace}
\usepackage{mathrsfs,amsthm}
\usepackage[normalem]{ulem}
\usepackage{subcaption}
\usepackage{comment}
\usepackage{times,bm}
\usepackage{multirow}
\usepackage{anyfontsize}
\usepackage{graphicx,psfrag,epsf}
\usepackage{float}
\usepackage[font=small,labelfont=bf]{caption}
\usepackage{tabularx}
\usepackage{url}
\usepackage{booktabs}
\usepackage{hyperref}
\usepackage{color}
\usepackage{xcolor}
\usepackage{caption}
\usepackage{bm}
\usepackage{bbm}
\usepackage[mathscr]{euscript}
\usepackage{bigstrut,enumerate}
\usepackage[utf8]{inputenc}
\usepackage[ruled,vlined]{algorithm2e}
\usepackage{multirow}

\newtheorem{theorem}{Theorem}
\newtheorem{lemma}{Lemma}
\newtheorem{example}{Example}
\newtheorem{definition}{Definition}
\newtheorem{corollary}{corollary}
\newtheorem{proposition}{Proposition}
\newtheorem{remark}{Remark}

\newcommand{\red}[1]{{\textcolor{black}{#1}}}	
\newcommand{\rred}[1]{{\textcolor{black}{#1}}}	
\newcommand{\tb}[1]{{\textbf{#1}}}

\DeclareMathOperator*{\argmin}{arg\,min}

\title{Large-Scale Multiple Testing of Composite Null Hypotheses Under Heteroskedasticity}
\author{Bowen Gang and Trambak Banerjee\\
	Fudan University and University of Kansas}
\date{}
\begin{document}
	\def\spacingset#1{\renewcommand{\baselinestretch}%
		{#1}\small\normalsize} \spacingset{1.2}	
	%\vspace{-20pt}
	%\hspace{130pt}{\centering Updated version available \href{https://drive.google.com/file/d/1nxJe7RI9TxIAs7UxJjmDuNqgMk2Pto1u/view}{here} }
	\maketitle
\begin{abstract}
Heteroskedasticity poses several methodological challenges in designing valid and powerful procedures for simultaneous testing of composite null hypotheses. In particular, the conventional practice of standardizing or re-scaling heteroskedastic test statistics in this setting may severely affect the power of the underlying multiple testing procedure. Additionally, when the inferential parameter of interest is correlated with the variance of the test statistic, methods that ignore this dependence may fail to control the type I error at the desired level. We propose a new Heteroskedasticity Adjusted Multiple Testing (HAMT) procedure that avoids data reduction by standardization, and directly incorporates the side information from the variances into the testing procedure. Our approach relies on an improved nonparametric empirical Bayes deconvolution estimator that offers a practical strategy for capturing the dependence between the inferential parameter of interest and the variance of the test statistic. We develop theory to show that HAMT is asymptotically valid and optimal for FDR control. Simulation results demonstrate that HAMT outperforms existing procedures with substantial power gain across many settings at the same FDR level. The method is illustrated on an application involving the detection of engaged users on a mobile game app.
\end{abstract}
\noindent%
\spacingset{1.25} 
\section{Introduction}
\label{sec:intro}
% Multiple testing frameworks have become an indispensable tool for modern scientific inquiry. With improved data collection practices, the use of these methods permeate disparate fields, such as astronomy \citep{miller2001controlling}, educational research \citep{thissen2002quick}, %sports biomehcanics \citep{knudson2009significant}, 
% epidemiology \citep{glickman2014false}, meteorology \citep{wilks2006field}, where identifying interesting cases among thousands or even millions of examples is germane to the ongoing empirical analysis. The popularity of multiple testing methods emanate from the fact that they allow the detection of scientifically interesting cases within a rigorous statistical framework that controls the possibility of false rejections below a pre-specified level. For instance, methods based on the classical family-wise error rate (FWER) control the probability of committing any type I error across the multiple comparisons while those based on the false discovery rate (FDR)\citep{benjamini1995controlling} regulate the expected proportion of false discoveries among all discoveries. For large-scale problems however, FWER controlling methods exhibit very low power and procedures that control the FDR instead, are usually preferred.
Suppose $X_i$, $i=1, \cdots, m$, are independent summary statistics arising from the following random mixture model:
\begin{eqnarray}
	\label{eq:model1}
	%X_i&=&\mu_i+\epsilon_i,~\epsilon_i\stackrel{ind.}{\sim}N(0,\sigma_i^2)\\
	X_{i}&=&\mu_i+\sigma_i\epsilon_{i},~\red{\epsilon_{i}\stackrel{i.i.d.}{\sim}\eta(\cdot)},\\
	\label{eq:model2}
	\mu_i\mid\sigma_i&\stackrel{ind.}{\sim}&g_\mu(\cdot\mid\sigma_i),~\sigma_i\stackrel{i.i.d.}{\sim}g_\sigma(\cdot),
\end{eqnarray}
\red{where $\epsilon_{i}$ are i.i.d. $0$ mean random variables with a known probability density function (PDF) $\eta(\cdot)$} and $g_\mu(\cdot\mid\sigma_i)$, $g_\sigma(\cdot)$ are, respectively, the PDFs of the unknown mixing distributions of $\mu$ given $\sigma_i$ and $\sigma_i$. \red{Model \eqref{eq:model1} incorporates, for instance, the setting where $\epsilon_{i}$ are central $t-$distributed random variables with $\nu>2$ degrees of freedom, as well as the more common case of $X_{i}\mid\mu_i,\sigma_i\stackrel{ind.}{\sim}N(\mu_i,\sigma_i^2)$, which finds substantial use in large-scale inference problems \citep{efron2004large,jin2007estimating,efron2007testing,efron2012large}.} %In these applications, the Gaussian distribution assumption in Equation \eqref{eq:model1} often provides a good approximation to the distribution of the summary statistics $X_i$. 
Following \cite{xie2012sure,weinstein2018group,sun2012multiple,fu2022heteroscedasticity}, we assume that $\sigma_i$ are known or can be well estimated from the data. 
Upon observing the pair $(X_i, \sigma_i)$, the goal is to simultaneously test the following $m$ hypotheses: 
\begin{equation}
	\label{eq:model3}
	H_{0, i}: \mu_i \in \mathcal{A} \quad \text{versus} \quad  H_{1, i}: \mu_i \notin \mathcal{A},~i=1,\ldots,m, 
\end{equation}
%for some known interval $\mathcal A\subset \mathbb{R}$. 
where $\mathcal A$ represents the indifference region such that the researcher is indifferent to the effects in $\mathcal A$ \citep{sun2012multiple}. Here $H_{0,i}$ represents a composite null hypothesis as opposed to a simple null hypothesis when $\mathcal A$ is singleton. 

Much of the focus of extant multiple testing methods is directed towards simultaneously testing simple null hypotheses against composite alternatives. A typical example arises in genome-wide association studies involving millions of single nucleotide polymorphisms (SNPs), where the primary goal is to discover SNPs that are statistically associated with a specific trait or disease of interest \citep{basu2018weighted,uffelmann2021genome}. The simultaneous inference problem in these applications require testing $m$ hypotheses of the form $H_{0,i}:\mu_i=0~vs~H_{1,i}:\mu_i\ne 0$ where $\mu_i$ is the unknown effect of SNP $i$ on the disease response, such as cholesterol level.   %of gene $i$ in detecting if the $i^{th}$ parameter of interest $\mu_i\ne \mu_0$ for some known scientifically important constant $\mu_0$. \red{Give gene example}. 
However, across numerous medical and social science applications it is important to detect if $\mu_i\notin\mathcal A$. %where $\mathcal A$ represents the indifference region where the researcher is indifferent to the effects in $\mathcal A$ \citep{sun2012multiple}. %that captures the scientifically unimportant effects. 
For instance, \cite{gu2018oracle,pop2013going} study the effect of attending a more selective school on the exam grade of high-school students in Romania. There the inferential objective is to identify schools with a positive effect on the average exam grade and it is desirable for the null hypothesis to include both zero and negative effects, i.e., to test a one-sided composite null hypothesis $H_{0,i}:\mu_i\in \mathcal A$ against the alternative $H_{1,i}:\mu_i\notin \mathcal A,~i=1,\ldots,m$, where $\mathcal A=(-\infty,0]$. 
% Another example arises in disease outbreak detection across $m$ spatial locations, where one of the primary goals is to identify regions that are experiencing disease incidence rates in excess of a baseline threshold $\mu_0$ \citep{caldas2006controlling}. Thus, for each location $i$ the aim is to test $H_{0,i}:\mu_i\in \mathcal A~vs~H_{1,i}:\mu_i\notin \mathcal A$, where $\mathcal A=[0,\mu_0]$.
In high-throughput gene sequencing studies, a fundamental task is to discover genes that exhibit differential expression levels that exceed a biologically relevant threshold $\mu_0$ \citep{love2014moderated}. So, for each gene $i$ a two-sided composite null hypothesis $H_{0,i}:\mu_i\in\mathcal A$ is tested against the alternative $H_{1,i}:\mu_i\notin \mathcal A$  where $\mathcal A=[-\mu_{0},\mu_{0}]$.

The standard practice for simultaneously testing a large number of hypotheses involves %estimating a summary statistic $X_i$ of $\mu_i$ and then developing 
%methods that rely on significance indices, such as $p$-values
constructing significance indices, such as $p-$values %\citep{roeder2009genome,li2019multiple,ignatiadis2019bias} 
or local false discovery rate (Lfdr) statistics \citep{sun2012multiple,basu2018weighted,efron2012large,sun2007oracle}, for ranking the hypotheses and then estimating a threshold along the ranking for type I error control. 
%However, for testing composite null hypotheses, procedures based on $p-$values are conservative as \red{the $p-$values fail to adapt to the asymmetry of the alternative about the null \citep{sun2007oracle,sun2012multiple} and the}. 
However, for testing composite null hypotheses, procedures based on $p-$values are not as powerful since the $p-$values may
fail to adapt to the potential asymmetry of the alternative about the null \citep{sun2007oracle,sun2012multiple} and tend to concentrate near $1$ under the null. %, thus, leading to a conservative testing procedure.
The Lfdr statistic, on the contrary, {adapts to such asymmetry by incorporating information about the null as well as the alternative distribution of the test statistic.} %Notwithstanding, modern gargantuan datasets often exhibit heterogeneity which pose several methodological challenges in constructing the Lfdr statistic. We discuss those challenges below.
%\subsection{Issues with heteroskedasticity}
Given a summary statistic $X_i$ of $\mu_i$, the Lfdr statistic represents the posterior probability of a case being null and relies on the mixture density of $X_i$ under the null and the alternative. %relies on the density of $X_i$ under the null, and its mixture density under the null and the alternative. 
When testing composite null hypotheses, this density is unknown in practical applications and must be estimated from the available data. %Under heteroskedasticity, $X_i$ is usually accompanied by its standard deviation $\sigma_i$ that varies across the units. 
The heteroskedasticity in the summary statistics raises two main challenges in estimating the mixture density. 

\vspace{5pt}
\noindent\textit{Effect of heteroskedasticity on the inferential parameter of interest -} In heteroskedastic settings, the parameter $\mu_i$ may be correlated with $\sigma_i$ \citep{weinstein2018group}. For instance, in a restaurant rating app it is often the case that extremely good and extremely bad restaurants tend to receive a large number of reviews. Thus, if the goal is to identify restaurants within a certain rating range then both the mean $(\mu_i)$ and standard deviation $(\sigma_i)$ of the ratings are related to the number of reviews. A key to constructing reliable estimate of the mixture density depends on a deconvolution step that learns the distribution of $\mu_i$ from the data and can effectively capture the dependence between $\mu_i$ and $\sigma_i$. However, existing approaches for empirical Bayes deconvolution, such as \cite{koenker2014convex,efron2016empirical}, assume independence between $\mu_i$ and $\sigma_i$, which is often violated in practice. In Section \ref{sec:dependence}, we demonstrate via a numerical example that procedures for testing composite null hypotheses may incur power loss and even fail to control the FDR when their underlying deconvolution estimator ignores this dependence. 

\vspace{5pt}
\noindent\textit{Power distortion due to standardization -} The conventional approach to mitigate the impact of heteroskedasticity is to re-scale each $X_i$ by $\sigma_i$ and construct $z-$values $Z_i= X_i/\sigma_i$ so that the Lfdr statistics can be estimated using the homoskedastic $Z_i$'s. However, for two-sided composite null hypotheses standardization distorts the underlying scientific question \citep{sun2012multiple} and, recently, \cite{fu2022heteroscedasticity} demonstrate that such a data reduction step may severely affect the power of multiple testing procedures even in the case of simple null hypotheses. In Section \ref{sec:standardization}, we present illustrative examples to demonstrate that standardization may lead to considerable power loss while testing one-sided composite null hypotheses as the power of testing procedures can vary substantially with $\sigma_i$. 

%\subsection{Proposed approach}
In this article, we propose a new heteroskedasticity-adjusted multiple testing (HAMT) procedure for composite null hypotheses (Equation \eqref{eq:model3}). HAMT represents an effective strategy for incorporating the side-information in the standard deviations for simultaneous testing of composite nulls and it operates in two steps: in step (1) HAMT constructs a significance index for ranking the hypotheses and then in step (2) it estimates a threshold along the ranking for identifying interesting hypotheses. The significance index is a new Lfdr statistic that addresses the methodological challenges discussed earlier in dealing with heteroskedasticity. First, our Lfdr statistic utilizes the full data, namely the summary statistic and its standard deviation, thus avoiding standardization to $z$-values and the potential power distortion due to data reduction. Second, the construction of the Lfdr statistic relies on an improved nonparametric empirical Bayes deconvolution estimator that provides a practical strategy for \red{estimating $g_\mu(\cdot|\sigma_i)$ (Equation \eqref{eq:model2}) while} incorporating \red{its dependence on $\sigma_i$. %the dependence between $\mu_i$ and $\sigma_i$.
Third, while conventional empirical Bayes deconvolution techniques, such as \cite{efron2016empirical}, involve maximizing the marginal likelihood of the data, HAMT relies on a density matching approach (Section \ref{sec:method}) to learn $g_\mu(\cdot|\sigma_i)$, which ultimately yields a consistent estimate of the mixture density in the heteroskedastic setting.} %, and yields a consistent estimate of the mixture density in the heteroskedastic setting. %\red{The deconvolution estimator developed here can be of independent interest for large-scale inference.} 
HAMT is designed for problems where the number of hypotheses being tested is large, which allows the deconvolution estimator to efficiently learn the latent structural relationship between $\mu_i$ and $\sigma_i$ in the data. Our theoretical results (Section \ref{sec:theory}) show that for such large-scale problems, HAMT is valid for FDR control and is as powerful as the oracle procedure that has full knowledge of the underlying data generating process under our hierarchical model (Equations \eqref{eq:model1}--\eqref{eq:model2}). In our numerical experiments (sections \ref{sec:sims} and  \ref{sec:more_sims}), we find that HAMT exhibits substantial power gains over existing methods across many settings while controlling FDR at the target level. 

%\subsection{Connections to existing works and main contributions}
%\red{adaptgmm \citep{chao2021adapt}, camt\citep{zhang2022covariate} , adapt \citep{lei2018adapt} and zap\citep{leung2021zap}} 
Our work is closely related to \cite{sun2012multiple},\cite{stephens2017false} and \cite{gu2018oracle}. \cite{sun2012multiple} develop an FDR controlling procedure based on Lfdr statistics for testing composite null hypotheses under heteroskedasticity. However, HAMT differs on two important aspects. First, and in contrast to \cite{sun2012multiple}, we allow $\mu_i$ and $\sigma_i$ to be dependent in our hierarchical model, which presents a challenging deconvolution problem for estimating the mixture density. Second, the kernel method developed in \cite{sun2012multiple} for estimating this density is highly unstable \citep{fu2022heteroscedasticity}. Here, we develop a nonparametric empirical Bayes deconvolution estimator which is scalable to large problems and provides a consistent estimate of the mixture density. In the terminology of \cite{efron2014two}, our deconvolution estimator is related to the $g-$modeling strategy for empirical Bayes estimation. While existing $g-$modeling approaches, such as \cite{koenker2014convex,efron2016empirical}, ignore the dependence between $\mu_i$ and $\sigma_i$, we develop a simple yet effective technique for modeling such dependence while estimating the distribution of $\mu_i$. 
\red{\cite{stephens2017false} develop a multiple testing procedure, {ASH}, for FDR control. Similar to {HAMT}, their approach relies on using the bivariate sequence $(X_i,\sigma_i)$ instead of summarizing them to $p-$ or $z-$ values. The hierarchical model underlying {ASH} assumes that conditional on $\sigma_i$ and for an unknown constant $c\ge 0$, the distribution of $\mu_i/\sigma_i^c$ is unimodal and posits it as a mixture of a point mass at $0$ and a scale-mixture of zero-mean Gaussian distributions. %The mixing proportions and $c$ are then estimated by maximizing the marginal likelihood of $X_i/\sigma_i^c$ given $\sigma_i$ using a fix grid for the Gaussian scales. 
Extensions to {ASH} allow mixtures of uniforms, half-uniforms and half-normals, among others, depending on the prior knowledge regarding the sign and symmetry of the distribution of $\mu_i$'s. We note that there are two key differences between the hierarchical models underlying {ASH} and {HAMT}. First, {ASH} assumes that conditional on $\sigma_i$ the distribution of $\mu_i/\sigma_i^c$ is independent of $\sigma_i$ for some $c\ge 0$. In contrast, {HAMT} does not make such an assumption and provides a practical approach for estimating the unknown mixing density $g_\mu(\cdot|\sigma_i)$ that accounts for the dependence of $\mu_i$ on $\sigma_i$. Second, unlike {ASH}, {HAMT} does not require a unimodal assumption for the distribution of $\mu_i$ given $\sigma_i$. In particular, our theoretical analyses in Section \ref{sec:theory} only requires that \rred{(i) $g_\mu(\cdot|\sigma)$ 
is supported on a compact interval, and (ii) has bounded first and second order derivatives with respect to $\sigma$, %to satisfy some weak regularity conditions 
for providing valid FDR control as the number of hypotheses $m\to\infty$.}%in the case of \texttt{HAMT}, Second the dependence structure considered in \cite{stephens2017false} is more rigid than ours. Specifically, \cite{stephens2017false} assumes conditional distribution of $X_i/\sigma^\alpha_i$ given $\sigma_i$ is independent of $\sigma_i$ for some $\alpha$, while our model does not require such condition. 
} Recently, \cite{gu2018oracle} propose a FDR controlling method for one-sided composite null hypotheses. Their approach is based on $z-$values and relies on the deconvolution estimate obtained from nonparametric maximum likelihood \citep{kiefer1956consistency,laird1978nonparametric} techniques to estimate the Lfdr. The illustrative examples in Section \ref{sec:standardization} show that such an approach based on standardization may lead to substantial power loss when $\mu_i$ and $\sigma_i$ are correlated.

Since variance can be viewed as a covariate in multiple testing problems, our work is also connected to the rapidly expanding literature on multiple testing with generic covariates. Here, proposals for heteroskedasticity adjustment of multiple testing methods %have recently received significant attention. Proposals 
vary from using $\sigma_i$ as a potential covariate for pre-ordering the hypotheses \citep{g2016sequential,lei2016power,li2017accumulation,cao2022optimal} to grouping methods based on the magnitudes of $\sigma_i$ \citep{efron2008microarrays,cai2009simultaneous,hu2010false,liu2016new}. {However, such a pre-ordering or grouping based on $\sigma_i$ may not always be informative since a larger $\sigma_i$ does not necessarily imply a relatively higher or lower likelihood of rejecting the null hypothesis.} More recently, several methods have been proposed that seek to directly use the covariate information along with the $p-$values to develop powerful testing procedures (see for example \cite{boca2018direct,lei2018adapt,zhang2019fast,li2019multiple,ignatiadis2021covariate,chao2021adapt,zhang2022covariate} and the references therein). %\red{While testing composite null hypotheses, the $p-$values fail to adapt to the potential asymmetry of the alternative about the null and the null $p-$values may concentrate near 1, both of which can cause the aforementioned testing procedures to be overly conservative.} 
While testing composite null hypotheses, the aforementioned testing procedures, however, can suffer from low power when 
{the null $p$-values are overly conservative}.
%some of the null $p-$values concentrate near $1$ and when the $p-$values fail to adapt to the potential asymmetry of the alternative about the null.
Methods that estimate the Lfdr statistic utilizing test statistic $X_i$ and additional covariates have also been developed (see for instance \cite{scott2015false,tansey2018black,chao2021adapt, leung2021zap}). In particular, \cite{scott2015false,tansey2018black} use the covariate information to estimate the null proportion in an empirical Bayes two-groups model while \cite{chao2021adapt} posit a Gaussian mixture model with $K$ classes to model the conditional distribution of $\mu_i$ given the covariates, where only the class probabilities depend on the covariates. In contrast to these works, HAMT does not rely on any pre-ordering or grouping of the hypotheses based on the magnitude of $\sigma_i$. Instead, HAMT is based on a Lfdr statistic that directly characterizes the impact of heteroskedasticity on the mixture density of the test statistic. For estimating the Lfdr statistics, our approach utilizes an empirical Bayes deconvolution estimator that does not depend on any parametric representation of the distribution of $\mu_i$ conditional on $\sigma_i$. 

In the following sections, we formally describe the multiple testing problem involving composite null hypotheses, present the oracle procedure, and then introduce the HAMT procedure and its asymptotic properties. %and Section \ref{sec:theory} is dedicated to the theoretical analysis of HAMT.% for estimating the Lfdr statistic. %HAMT relies on a Lfdr statistic that . %\red{also say that HAMT has $\sigma_i$ in both the prior and the marginal density. Others don't. Also the first two don't have theory.}
\section{Multiple testing of composite null hypotheses}
\label{sec:problem}
\subsection{Problem formulation}
\label{sec:prob_formulation}
% We first introduce the multiple testing problem involving composite null hypotheses and then describe the oracle testing procedure in Section \ref{sec:oracle_procedure}. Thereafter, two illustrative examples are presented in Section \ref{sec:standardization} to discuss the effects of standardization on the power of multiple testing procedures.
Let $\theta_i=I(\mu_i\notin\mathcal{A})$ be an indicator function that gives the true state of the $i$th testing problem in Equation \eqref{eq:model3}. For instance, if $\theta_i=1$ then the alternative hypothesis $H_{1, i}$ is true. Let $\delta_i\in\{0, 1\}$ be the decision we make about hypothesis test $i$, with $\delta_i = 1$ being a decision to reject $H_{0,i}$. Denote the vector of all $m$ decisions $\bm\delta=(\delta_1, \cdots, \delta_m)\in\{0,1\}^m$. 
A selection error, or false positive, occurs if we assert that $\mu_i$ is not in $\mathcal{A}$ when it actually is. %By convention this incorrect choice is called a false positive decision. 
In large-scale multiple testing problems, false positive decisions are inevitable if we wish to discover interesting effects with a reasonable power. Instead of aiming to avoid any false positives, a practical goal is to keep the false discovery rate (FDR) \citep{benjamini1995controlling} small, which is the expected proportion of false positives among all selections,
\begin{equation*}
	\label{eq:fdr}
	\text{FDR}(\bm \delta) = E\left[\dfrac{\sum_{i=1}^{m}(1-\theta_i)\delta_i}{\max\{\sum_{i=1}^{m}\delta_i, 1\}}\right].
\end{equation*}
The power of a testing procedure is measured by the expected number of true positives (ETP) where,
\begin{equation*}
	\label{eq:ETP}
	\mbox{ETP}(\bm \delta)=E\left(\sum_{i=1}^{m} \theta_i \delta_i\right)=E\left(\sum_{i=1}^{m} I(\mu_i\notin \mathcal{A}) \delta_i\right).
\end{equation*}
Hence, the multiple testing problem in Equation \eqref{eq:model3} can be formulated as
$$
\text{maximize}_{\bm \delta}~\text{ETP}(\bm \delta)~\text{subject to }\mbox{FDR}(\bm \delta)\leq\alpha,
$$
where $\alpha\in(0,1)$ is a user-defined cap on the maximum acceptable FDR. A quantity that is closely related to the FDR is the marginal false discovery rate (mFDR) where,
\begin{equation*}
	\label{eq:mfdr}
	\text{mFDR}(\bm \delta) = \dfrac{E\{\sum_{i=1}^{m}(1-\theta_i)\delta_i\}}{E\{\sum_{i=1}^{m}\delta_i\}}.	
\end{equation*}
Under certain first and second-order conditions on the number of rejections, the mFDR and the FDR are asymptotically equivalent \citep{genovese2002operating,basu2018weighted}, and for theoretical convenience we will aim to control mFDR instead. Formally, we study the following problem for the rest of the article:
\begin{equation}
	\label{eq:problem}
	\text{maximize}_{\bm \delta}~\text{ETP}(\bm \delta)~\text{subject to }\mbox{mFDR}(\bm \delta)\leq\alpha.
\end{equation}
\subsection{Oracle procedure}
\label{sec:oracle_procedure}
In this section we assume that the mixing densities $g_\mu(\cdot\mid\sigma)$ and $g_\sigma(\cdot)$ in Model \eqref{eq:model2} are known by the oracle and present the oracle procedure that solves Problem \eqref{eq:problem}. There are two steps involved in the derivation of the oracle procedure: the first step constructs the
optimal ranking of hypotheses and the second step determines the best threshold along
the ranking that satisfies the mFDR constraint in Problem \eqref{eq:problem}.

To rank the $m$ hypotheses, consider the oracle conditional local FDR (Clfdr) statistic which is defined as,
\begin{equation}
	\label{lfdr_orc}
	T_i^{\sf OR}= T^{\sf OR}(x_i,\sigma_i)=P(\mu_i \in \mathcal{A}|x_i,\sigma_i)=\dfrac{f_{0}(x_i|\sigma_i)}{f(x_i|\sigma_i)},
\end{equation}
where 
\begin{equation}
	\label{eq:marg_dens}
	f_0(x|\sigma)=\int_{\mu \in \mathcal{A}}\eta_{\sigma}(x-\mu)g_\mu(\mu\mid\sigma)\mathrm{d}\mu~\text{and}~f(x|\sigma)=\int_{\mathbb R}\eta_{\sigma}(x-\mu)g_\mu(\mu\mid\sigma)\mathrm{d}\mu.
\end{equation}
%denote, respectively, the composite null density and the marginal density of $X$ given $\sigma$ under Model \eqref{eq:model1}--\eqref{eq:model2}. 
\red{In Equation \eqref{eq:marg_dens}, $f(x|\sigma)$ denotes the marginal density of $X$ given $\sigma$ under Model \eqref{eq:model1}--\eqref{eq:model2} and  $\eta_\sigma(x-\mu)=\sigma^{-1}\eta\{(x-\mu)/\sigma\}$ is the density of $X_i$ conditional on $(\mu_i,\sigma_i)$.} %a Gaussian random variable with mean $\mu$ and standard deviation $\sigma$. 
Next, to derive the best threshold, suppose $Q(t)$ denotes the mFDR level of the testing procedure $\bm \delta^{\sf OR}(t)=\{I(T_i^{\sf OR}\le t):1\le i\le m\}$ for some $t\in(0,1)$. We propose the following oracle procedure for Problem \eqref{eq:problem},
\begin{equation}
	\label{eq:oracle_procedure}
	\bm \delta^{\sf OR}(t^*)=\{I(T_i^{\sf OR}<t^*):1\le i\le m\},
\end{equation}
where $t^*=\sup\{t\in(0,1):Q(t)\le \alpha\}$.	
%Definition \ref{def:oracle_proc} presents the oracle testing procedure for Problem \eqref{eq:problem}.
%\begin{definition}
%\label{def:oracle_proc}
%Suppose for some $t\in(0,1)$, $Q(t)$ denotes the mFDR level of the testing procedure $\bm \delta^{\sf OR}(t)=\{I(T_i^{\sf OR}\le t):1\le i\le m\}$. The oracle procedure $\bm \delta^{\sf OR}(t^*)$ for Problem \eqref{eq:problem} is defined as follows:
%\begin{equation*}
%	\bm \delta^{\sf OR}(t^*)=\{I(T_i^{\sf OR}<t^*):1\le i\le m\},
%\end{equation*}
%where $t^*=\sup\{t\in(0,1):Q(t)\le \alpha\}$.	
%\end{definition}

Denote $\bm X=(X_1,\ldots,X_m)$ and $\bm \sigma=(\sigma_1,\ldots,\sigma_m)$. In Theorem \ref{thm_orcale} we show that $\bm \delta^{\sf OR}(t^*)$ has the highest power among all procedures based on $(\bm X,\bm \sigma)$ that control the mFDR at level $\alpha$.  
\begin{theorem}
	\label{thm_orcale}
	Consider Model \eqref{eq:model1}--\eqref{eq:model2}. The oracle procedure $\bm \delta^{\sf OR}(t^*)$ in Equation \eqref{eq:oracle_procedure} controls \text{mFDR} at level $\alpha$. Additionally if $\bm \delta$ is any other procedure based on $(\bm X,\bm \sigma)$ that controls \text{mFDR} at level $\alpha$ then we have $\text{ETP}\{\bm \delta^{\sf OR}(t^*)\}\ge \text{ETP}(\bm \delta)$.
\end{theorem}
Theorem \ref{thm_orcale} establishes that the oracle procedure $\bm\delta^{\sf OR}(t^*)$ is valid and optimal for mFDR control. However, $\bm\delta^{\sf OR}(t^*)$ is not implementable in practice since both $T_i^{\sf OR}$ and $t^*$ are unknown in practical applications. In Section \ref{sec:method} we describe the proposed HAMT procedure that relies on a nonparametric empirical Bayes deconvolution estimator of $g_\mu(\cdot|\sigma_i)$ to construct a data-driven estimate of $T_i^{\sf OR}$ and uses a step-wise procedure to estimate $t^*$.

\subsection{Power loss due to standardization: illustrative examples}
\label{sec:standardization}
While $\bm \delta^{\sf OR}(t^*)$ is the optimal solution to Problem \eqref{eq:problem} based on $(\bm X,\bm\sigma)$, a plausible approach for solving Problem \eqref{eq:problem} is to construct $z-$values $Z_i= X_i/\sigma_i$ and then reject the null hypothesis for suitably small values of $\mathcal{Z}^{\sf OR}_i$ where $\mathcal{Z}^{\sf OR}_i=P(\mu_i\in\mathcal A|z_i)$. In fact, \cite{sun2007oracle} show that this approach is the most powerful $z$-value method. The apparent advantage of this data reduction step is that it transforms the heteroskedastic multiple testing problem to a homoskedastic one, %since conditional on $(\mu_i,\sigma_i)$ $Z_i\stackrel{ind.}{\sim}N(\mu_i/\sigma_i,1)$.
and enables a like-for-like comparison of the $m$ study units under consideration. However, in the case of two-sided composite null hypothesis, such a standardization may distort the underlying scientific question \citep{sun2012multiple}. Moreover, \cite{fu2022heteroscedasticity} demonstrate that data reduction via standardization could lead to power loss for multiple testing procedures even in the case of simple null hypotheses. In this section we consider two illustrative examples to demonstrate that power loss due to standardization can be substantial while testing one-sided composite null hypotheses.

\begin{figure}[!t]
	\centering
	\begin{subfigure}{.45\textwidth}
		\centering
		\includegraphics[width=\linewidth]{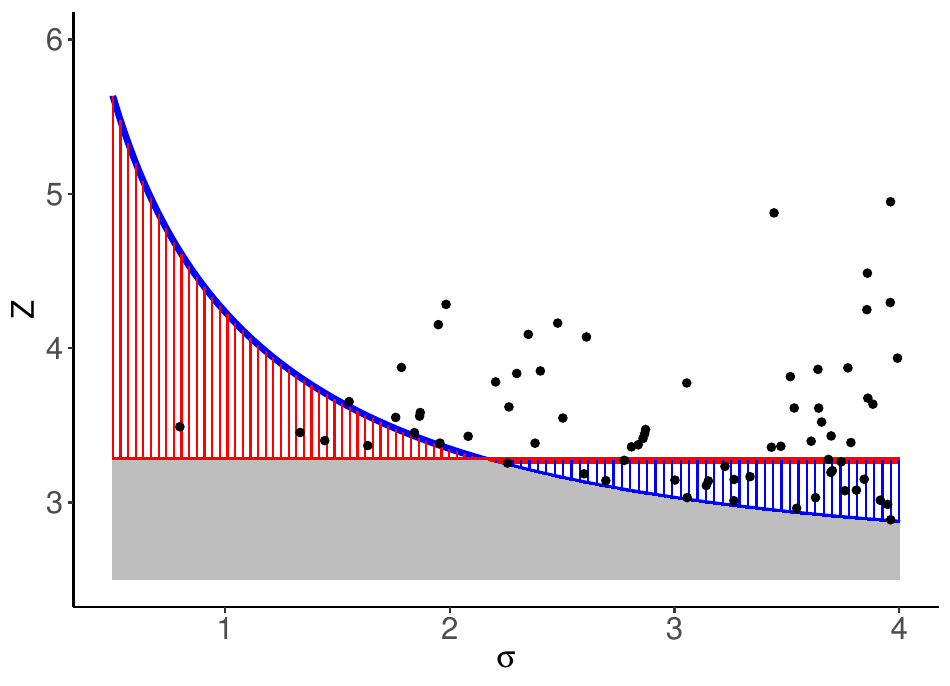}
		\caption{Rejection regions in Example 1.}
		\label{fig:sub1}
	\end{subfigure}%
	\begin{subfigure}{.45\textwidth}
		\centering
		\includegraphics[width=\linewidth]{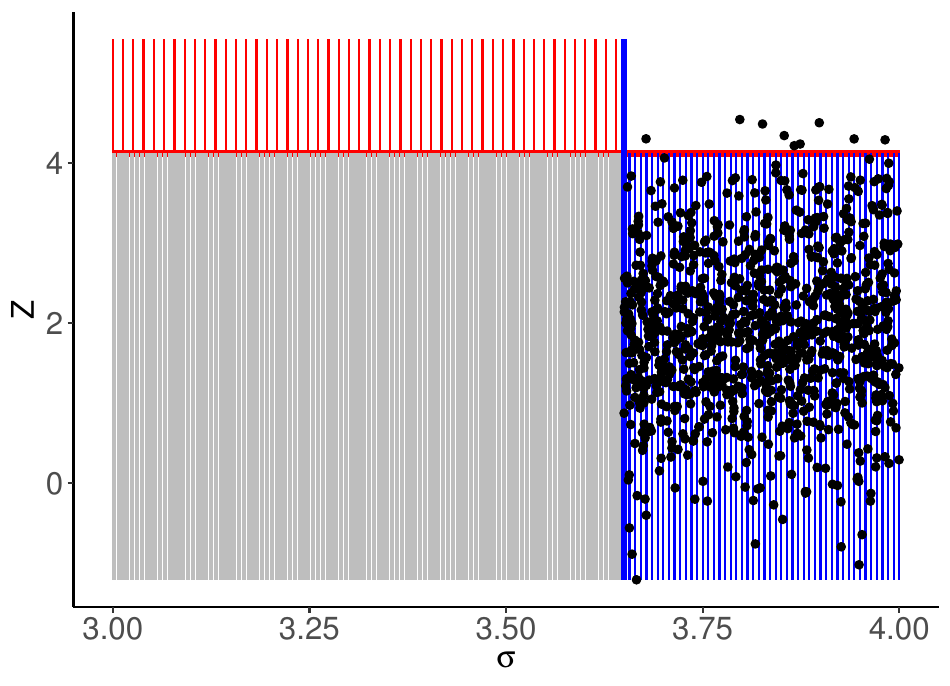}
		\caption{Rejection regions in Example 2.}
		\label{fig:sub2}
	\end{subfigure}
	\caption{In the red shaded region $\bm{\delta}^{\sf ZOR}(t_z)$ rejects while $\bm \delta^{\sf OR}(t^*)$ does not, in the blue region $\bm \delta^{\sf OR}(t^*)$ rejects while $\bm \delta^{\sf ZOR}(t_z)$ does not and both procedures reject in the white region. Finally, in the gray shaded region neither procedures reject. The black dots represent instances where the null hypothesis is false and fall within the three rejection regions.}
	\label{fig:rejection_region}
\end{figure}

\setcounter{example}{0}
\begin{example}
Suppose data are generated from Model \eqref{eq:model1}--\eqref{eq:model2} with $X_i\mid\mu_i,\sigma_i\stackrel{ind.}{\sim}N(\mu_i,\sigma_i^2)$, $\sigma_i\stackrel{i.i.d}{\sim }\text{Unif}(0.5,4)$ and $\mu_i\mid\sigma_i\stackrel{ind.}{\sim} 0.9\delta_0(\cdot)+0.1 \delta_{\sigma^{1.5}_i}(\cdot),$ where $\delta_a(\cdot)$ is a Dirac delta function indicating a point mass at $a$. In this example $\sigma_i$ controls the magnitude of the non-zero $\mu_i$ and we are interested in Problem \eqref{eq:model3} with $\mathcal A=(-\infty,0]$. 	We first consider the oracle procedure based on the $z-$values $\bm Z=(Z_1,\ldots,Z_m)$. In Section \ref{app:calculations} we show that this oracle procedure is a thresholding rule of the form $\bm \delta^{\sf ZOR}(t_z)=\{I(Z_i> t_z):1\le i\le m\}$ where $t_z=3.273$ at $\alpha=0.1$.
	Next, recall from Equation \eqref{eq:oracle_procedure} that the oracle procedure $\bm{\delta}^{\sf OR}(t^*)$ based on $(\bm X,\bm\sigma)$ is of the form $\{I(T_i^{\sf OR}<t^*):1\le i\le m\}$. This is equivalent to a thresholding rule $\{I(Z_i>\lambda_{\sigma_i}(t^*)):1\le i\le m\}$ (details provided in Section \ref{app:calculations}), where
	$$
	\lambda_\sigma(t)=\dfrac{1}{\sqrt{\sigma}}\Big[-\log\Bigl\{\dfrac{0.1t}{(1-t)0.9}\Bigr\} +0.5\sigma\Big],
	$$
	and $t^*=0.177$ at $\alpha=0.1$.
	
	While both $\bm \delta^{\sf ZOR}$ and $\bm \delta^{\sf OR}$ control the mFDR exactly at $\alpha$, their powers are substantially different in this example: power of $\bm{\delta}^{\sf ZOR}(t_z)$ is  $0.0432$ and that of $\bm{\delta}^{\sf OR}$ is $0.0611$. %, indicating that the oracle procedure based on $(X_i,\sigma_i)$ is more powerful than the procedure based on the standardized statistic $Z_i$. 
	To further examine the power gain of $\bm{\delta}^{\sf OR}(t^*)$, we consider the left panel of Figure \ref{fig:rejection_region} that plots the rejection regions of $\bm{\delta}^{\sf OR}(t^*)$ and $\bm{\delta}^{\sf ZOR}(t_z)$ as a function of $Z_i$ and $\sigma_i$. In the red shaded region $\bm{\delta}^{\sf ZOR}(t_z)$ rejects while $\bm \delta^{\sf OR}(t^*)$ does not, in the blue region $\bm \delta^{\sf OR}(t^*)$ rejects while $\bm \delta^{\sf ZOR}(t_z)$ does not and both procedures reject in the white region. Finally, in the gray shaded region neither procedures reject. The black dots represent instances where the null hypothesis is false and fall within the three rejection regions. While it is clear that a vast majority of the non-null cases appear in the white region, approximately $64\%$, the blue region captures relatively more non-null cases than the red region, $30\%$ versus $6\%$. Thus, $\bm \delta^{\sf OR}(t^*)$ rejects an overall higher percentage of the non-null cases than $\bm \delta^{\sf ZOR}(t_z)$, which explains the power gain of the former over the latter.
 \end{example}
\begin{example}
 Unlike the previous setting, in this example $\sigma_i$ controls the sparsity as well as the magnitude of the non-zero $\mu_i$. Data are generated from Model \eqref{eq:model1}--\eqref{eq:model2} with $X_i\mid\mu_i,\sigma_i\stackrel{ind.}{\sim}N(\mu_i,\sigma_i^2)$, $\sigma_i\stackrel{i.i.d}{\sim }\text{Unif}(0.5,4)$ and $\mu_i\mid\sigma_i\stackrel{ind.}{\sim} \delta_0(\cdot)I\{\sigma_i\le 3.65\}+\delta_{\sigma^{1.5}_i}(\cdot)I\{\sigma_i>3.65\},$ where $P(\sigma_i\le 3.65)=0.9$. We are interested in Problem \eqref{eq:model3} with $\mathcal A=(-\infty,0]$. The oracle procedure based on $\bm Z$ is of the form $\bm \delta^{\sf ZOR}(t_z)=\{I(Z_i> t_z):1\le i\le m\}$ where $t_z=4.124$ at $\alpha=0.1$ with power $0.0015$. In contrast, $\bm \delta^{\sf OR}(t^*)$ in this example simply observes if $\sigma_i>3.65$ to detect if $H_{0,i}$ is false and thus, provides a perfect classification rule with FDR equal to $0$ and power equal to $1$. The stark contrast in the power of these two procedures is further elucidated in the right panel of Figure \ref{fig:rejection_region}. Here, the rejection regions continue to have the same interpretation as in the left panel. However, the blue region now captures almost $99\%$ of all the non-null cases that fall within the three regions while the white region only accounts for the remaining $1\%$. Moreover, the red region does not capture any non-null case, thus explaining the substantially low power of $\bm \delta^{\sf ZOR}(t_z)$ in this setting.
\end{example}
The preceding examples illustrate that data reduction via standardization may lead to power loss even when testing one-sided composite null hypotheses. While standardization is a natural pre-processing step for testing heteroskedastic units, Examples 1 and 2 demonstrate that such a step suppresses the information contained in the standard deviations that can boast the power of these tests. Our numerical experiments in Section \ref{sec:sims_onesided} and Section \ref{sec:more_sims} of the supplement corroborate this observation where we find that $z-$value procedures are, in general, not as powerful as the proposed HAMT procedure which is based on $(\bm X,\bm \sigma)$. 
\section{Heteroskedasticty adjusted multiple testing procedure for composite null hypotheses}
\label{sec:method}
\subsection{Improved empirical Bayes deconvolution}
\label{sec:deconvolution}
This section develops a data-driven procedure to mimic the oracle. We discuss the estimation of $T_i^{\sf OR}$ and $t^*$, and present the HAMT procedure in Definition \ref{def:data_driven}.
Our approach for estimating $T_i^{\sf OR}$ involves constructing a nonparametric empirical Bayes deconvolution estimate of the unknown mixing density $g_\mu(\cdot\mid\sigma_i)$. While there are several popular approaches to estimating an unknown mixing density, we demonstrate in Section \ref{sec:dependence} that existing methods that fail to account for the dependence between $\mu_i$ and $\sigma_i$ can suffer from power loss and may not even provide FDR control. Here we present a practical approach for estimating $g_\mu(\cdot\mid\sigma_i)$ that effectively accounts for this dependence.

%Suppose $g_\mu(\cdot\mid\sigma_i)$ is continuous in $\sigma_i$ and the parameter space of $\mu_i$ is a finite discrete set $\mathcal T=\{u_1,\ldots,u_S\}$ of size $S$. The assumption on the discreteness of $\mathcal T$ is a convenience that aids with the practical implementation of our method. See for example \cite{efron2016empirical} for a similar assumption while defining their deconvolution estimator. Let $g_{j}(\sigma_i)= g_\mu(u_j\mid\sigma_i)$ denote the prior probability mass on $u_j$ conditional on $\sigma_i$ where $j=1,\ldots,S$. Since each $g_{j}(\sigma_i)$ depends on $\sigma_i$, we approximate $g_{j}(\sigma_i)$ as a linear combination of $K$ basis functions as follows:
%\begin{equation}
%	\label{eq:basis_rep}
%	g_j(\sigma_i)\approx\sum_{k=1}^{K}w_{jk}q_k(\sigma_i)=\bm w_j^T\bm q(\sigma_i).
%\end{equation}
{Suppose $g_\mu(\cdot\mid\sigma_i)$ is continuous in $\sigma_i$. We first approximate the parameter space of $\mu_i$ using a discrete \rred{equispaced grid} $\mathcal T=\{u_1,\ldots,u_S\}$ of size $S$. Then, the conditional prior density $g_\mu(\cdot|\sigma_i)$ can be approximated by a mixture of point masses as follows:
	$$
	g_\mu(\cdot|\sigma_i)\approx\sum_{j=1}^{S} g_j(\sigma_i)\delta_{u_j}(\cdot),
	$$
	where  $\delta_{u_j}(\cdot)$ is the point mass at $u_j$ and $g_{j}(\sigma_i)= g_\mu(u_j\mid\sigma_i)$ is the prior probability mass on $u_j$ conditional on $\sigma_i$. A formal statement justifying this approximation is presented in Lemma \ref{lem1} in the supplement. We view $g_j(\sigma_i)$ as a continuous function of $\sigma_i$ and approximate it as a linear combination of $K$ basis functions as follows:
	\begin{equation}
		\label{eq:basis_rep}
		g_j(\sigma_i)\approx\sum_{k=1}^{K}w_{jk}q_k(\sigma_i)=\bm w_j^T\bm q(\sigma_i).
\end{equation}}
In Equation \eqref{eq:basis_rep}, $\bm w_j$ is a $K-$dimensional vector of unknown weights and $\bm q(\sigma_i)$ is a known vector of basis functions that depend on $\sigma_i$. We discuss the choice of these basis functions in Section \ref{sec:implement}. In this discrete setting, and using Equation \eqref{eq:basis_rep}, the quantities in Equation \eqref{eq:marg_dens} have the following representation:
$$\tilde{f}_{0}(x\mid\sigma_i)=\sum_{j:u_j \in \mathcal{A}}\eta_{\sigma_i}(x-u_j)\bm w_j^T\bm q(\sigma_i),~\tilde{f}(x\mid\sigma_i)=\sum_{j=1}^{S}\eta_{\sigma_i}(x-u_j)\bm w_j^T\bm q(\sigma_i).
$$
%Denote $\mathcal S^{S}=\{\bm \eta\in \bm{R}^S:\bm 1^T\bm \eta=1,~\bm \eta\succeq \bm 0\}$ as the $S-$dimensional unit simplex. Our goal is to estimate the $KS-$dimensional vector $\mathcal W=(\bm w_1^T,\ldots,\bm w_S^T)^T$ such that $\bm g_i=\{\bm w_j^T\bm q(\sigma_i):1\le j\le S\}\in \mathcal S^S$ for $i=1,\ldots,m$. 
\red{Our goal is to estimate the $K-$dimensional vectors $\{\bm w_1,\ldots,\bm w_S\}$ such that (i) $\sum_{j=1}^{S}\bm w_j^T\bm q(\sigma_i)=1~\text{for}~i=1,\ldots,m$, and (ii) $\bm w_j^T\bm q(\sigma_i)\ge 0 ~\text{for}~j=1,\ldots,S,~i=1,\ldots,m$. A standard approach involves maximizing the marginal log-likelihood of the data with respect to the $\bm w_j$s under the above two constraints as follows:
\begin{equation}
	\label{eq:npmle}
	\begin{split}
	&\min_{\{\bm w_1,\ldots,\bm w_S\}\in\mathbb R^{K}}-\sum_{i=1}^{m}\log\sum_{j=1}^{S}\eta_{\sigma_i}(x_i-u_j)\bm w_j^T\bm q(\sigma_i)\\
	&\mbox{subject\ to:}~\sum_{j=1}^{S}\bm w_j^T\bm q(\sigma_i)=1~\text{for}~i=1,\ldots,m.\\
	&~~~~~~~~~~~~~~~~~~\bm w_j^T\bm q(\sigma_i)\ge 0 ~\text{for}~j=1,\ldots,S~\text{and}~i=1,\ldots,m.
\end{split}
\end{equation}
Equation \eqref{eq:npmle} is a convex optimization problem that can be solved, relatively efficiently, using solvers, such as MOSEK \citep{mosek}. We provide those details in Section \ref{sec:npmleb_details} of the supplement. In this article, however, we take a different approach and estimate $\bm w_j$s using a density matching technique, which involves minimizing the average squared error loss between $\tilde{f}(x_i|\sigma_i)$ and a pilot estimate of the true marginal density ${f}(x_i\mid\sigma_i)$ with respect to the $\bm w_j$s. We first describe our approach and then discuss its advantage over Problem \eqref{eq:npmle}.}

\red{If $f(x_i|\sigma_i)$ were known, the proposed density matching technique would involve the following minimization problem with respect to the $\bm w_j$s:
%\begin{equation}
%	\label{eq:opt_oracle}
%	\min_{\mathcal W\in\mathbb R^{KS}} \sum_{i=1}^{m}\Bigl\{{f}(x_i\mid\sigma_i)-\tilde{f}(x_i\mid\sigma_i)\Bigr\}^2 \quad \mbox{subject\ to}\quad  \bm g_i\in\mathcal S^S~\text{for}~i=1,\ldots,m.
%\end{equation}
\begin{equation}
	\label{eq:opt_oracle}
	\begin{split}
	&\min_{\{\bm w_1,\ldots,\bm w_S\}\in\mathbb R^{K}} \dfrac{1}{2m}\sum_{i=1}^{m}\Bigl\{{f}(x_i\mid\sigma_i)-\sum_{j=1}^{S}\eta_{\sigma_i}(x_i-u_j)\bm w_j^T\bm q(\sigma_i)\Bigr\}^2\\
	&\mbox{subject\ to:}~\sum_{j=1}^{S}\bm w_j^T\bm q(\sigma_i)=1~\text{for}~i=1,\ldots,m.\\
	&~~~~~~~~~~~~~~~~~~\bm w_j^T\bm q(\sigma_i)\ge 0 ~\text{for}~j=1,\ldots,S~\text{and}~i=1,\ldots,m.
	\end{split}
\end{equation}}
However, ${f}(x_i\mid\sigma_i)$s in Problem \eqref{eq:opt_oracle} are not known in practice and estimating them directly from the data is difficult as we only have one pair of observation $(X_i,\sigma_i)$ for estimating each density. Recently, \cite{fu2022heteroscedasticity} consider a heteroskedasticity adjusted bivariate kernel density estimator $\hat{\varphi}^m(x,\sigma_i)$ for ${f}(x\mid\sigma_i)$ where
\begin{equation}
	\label{eq:kde_hart}
	\hat{\varphi}^m(x,\sigma_i)=\sum_{j=1}^{m}\dfrac{\phi_{h_\sigma}(\sigma_i-\sigma_j)  }{\sum_{k=1}^{m}\phi_{h_\sigma}(\sigma_i-\sigma_k) }\phi_{h_{xj}}(x-x_j).
\end{equation}
In Equation \eqref{eq:kde_hart}, $\phi_\sigma(\cdot)$ is the density of a Gaussian random variable with mean $0$ and standard deviation $\sigma$, $h_{xj}=h_x\sigma_j$ and $\bm h=(h_x,h_\sigma)$ is a pair of bandwidths. The weights $\phi_{h_\sigma}(\sigma_i-\sigma_j)/\sum_{k=1}^{m}\phi_{h_\sigma}(\sigma_i-\sigma_k)$ are designed to borrow strength from observations with variability close to $\sigma_i$, while placing little
weight on points where $\sigma_i$ and $\sigma_j$ are far apart. The variable bandwidth $h_{xj}$ adjusts for the heteroskedasticity in the data by inducing flatter kernels for data points that are observed with a higher variance. Furthermore, \cite{fu2022heteroscedasticity} show that $\hat{\varphi}^m(x,\sigma_i)$ is a consistent estimator of ${f}(x\mid\sigma_i)$ in the sense that $E\int\{\hat{\varphi}^m(x,\sigma_i)-f(x\mid\sigma_i)\}^2\mathrm{d}x\to 0$ as $m\to\infty$ for all $\sigma_i>0$. {In our analysis, we solve Problem \eqref{eq:opt_oracle} with 
a jacknifed version $\hat{\varphi}^m_{(-i)}(x_i,\sigma_i)$ of $\hat{\varphi}^m(x_i,\sigma_i)$ as a pilot estimate of $f(x_i\mid\sigma_i)$, where
\begin{equation}
	\label{eq:kde_hart_jacknife}
\hat{\varphi}^m_{(-i)}(x_i,\sigma_i)=\sum_{j\neq i=1}^{m}\dfrac{\phi_{h_\sigma}(\sigma_i-\sigma_j)}{\sum_{k\neq i=1}^{m}\phi_{h_\sigma}(\sigma_i-\sigma_k)}\phi_{h_{xj}}(x_i-x_j).
\end{equation}
}
{Denote $\hat{\bm \varphi}^m=[\hat{\varphi}^m_{(-1)}(x_1,\sigma_1),\ldots,\hat{\varphi}^m_{(-m)}(x_m,\sigma_m)]$, $\bm a_{ij}=\eta_{\sigma_i}(x_i-u_j)\bm q(\sigma_i)$, $\bm a_i=(\bm a_{i1}^T,\ldots,\bm a_{iS}^T)^T$ and $A=(\bm a_1,\ldots,\bm a_m)^T$. Additionally, let $C=[\bm 1_S\otimes \bm q(\sigma_1)\ldots \bm 1_S\otimes \bm q(\sigma_m)]^T$ where $\otimes$ denotes the usual Kronecker product of two matrices, and \rred{denote $\bm c_r$ as an $r-$dimensional vector with all entries equal to the scalar $c\in\mathbb R$.}
%$\bm 1_r$ is an $r-$dimensional vector with all entries equal to $1$. 
Finally, for $s=1,\ldots,S$, let $B_s$ denote a $m\times KS$ matrix whose entries are $0$ except for the $\{K(s-1)+1\}$th column which equals $[\bm q(\sigma_1),\ldots \bm q(\sigma_m)]^T$ and let $B=[B_1^T\ldots B_S^T]^T$. Problem \eqref{eq:opt_oracle}, with $f(x_i|\sigma_i)$ replaced by $\hat{\varphi}_{(-i)}^m(x_i,\sigma_i)$, is then equivalent to the following convex optimization problem with respect to $\mathcal W=(\bm w_1^T,\ldots,\bm w_S^T)^T$:
\begin{equation}
	\label{eq:opt}
	\begin{split}
	\min_{\mathcal W\in\mathbb R^{KS}} \dfrac{1}{2m}\Big\|\hat{\bm \varphi}^m-A\mathcal W\Big\|^2_2 \quad \mbox{subject\ to}~B\mathcal W\succeq \bm 0_{Sm},~C\mathcal{W}=\bm 1_m.
	\end{split}
\end{equation}
The density matching approach (Problem \eqref{eq:opt}) to estimating $\mathcal W$ has a distinct advantage over the one that estimates $\mathcal W$ by maximizing the marginal log-likelihood of the data (Problem \eqref{eq:npmle}). \rred{When the distribution of $\mu$ depends on $\sigma_i$, Problem \eqref{eq:opt} is a relatively superior strategy for learning $g_\mu(\mu|\sigma_i)$ than maximizing the marginal log-likelihood since the extra information encoded in $\bm \sigma$ can be better exploited by the bivariate kernel density estimator $\hat{\varphi}_{(-i)}^m(x_i,\sigma_i)$ (Equation \eqref{eq:kde_hart_jacknife}) of $f(x_i\mid\sigma_i)$. This intuition is supported by the numerical studies in sections \ref{sec:sims_onesided} and \ref{sec:more_sims} where, across many settings, the proposed procedure HAMT, that relies on the solution $\hat{\mathcal W}_m=(\hat{\bm w}_{1,m},\ldots,\hat{\bm w}_{S,m})$ from Problem \eqref{eq:opt}, demonstrates substantially higher power at the same FDR level than the competing procedure NPMLE B, which uses Problem \eqref{eq:npmle} to learn $g_\mu(\mu|\sigma_i)$.}
Furthermore, in Proposition \ref{prop1} (Section \ref{sec:theory}) we show that $\hat{\mathcal W}_m$ yields a consistent estimator of $f(x_i|\sigma_i)$, which plays an important role in establishing the validity and optimality of the data-driven {HAMT} procedure. \rred{Parallel results establishing the validity of NPMLE B is, to the best of our knowledge, still unknown at the time of writing this article. A recent development in this direction is by \cite{chen2022empirical} who model the distribution of $\mu$ given $\sigma_i$ as a flexible location-scale family and show that with high probability the average squared Hellinger distance between the estimated and the true marginal densities is small. However, their methodological development does not cover the problem of multiple testing. %Infact, our simulation experiments in Section \ref{sec:sims_onesided} and Appendix \ref{sec:more_sims} reveal some settings where the multiple testing procedure relying on $\mathcal W$ recovered from Problem \eqref{eq:npmle}, denoted NPMLE B, may fail to control the FDR at the desired level. 
Much research is needed to fully comprehend the conditions under which NPMLE B can guarantee asymptotic FDR control and how the hyper-parameters $S$ and $K$ impact its power. In Section \ref{sec:insights} of the supplement we provide additional insights on when NPMLE B can be expected to outperform HAMT in power at the same FDR level.} 

In the next section, we present our data-driven HAMT procedure that relies on the solution $\hat{\mathcal W}_m$ to Problem \eqref{eq:opt}. Section \ref{sec:implement} \rred{provides the recommended choices for $\mathcal T$, $S$ and $K$ in the context of HAMT while Section \ref{sec:npmleb_details} in the supplement includes the implementation details for solving Problem \eqref{eq:opt}.} 
\subsection{Proposed HAMT procedure}
\label{sec:data-driven}
We first present the estimator of the oracle Clfdr statistic $T_i^{\sf OR}$ in Definition \ref{def:clfdr}.
\begin{definition}
	\label{def:clfdr}
	Let $\hat{\mathcal W}_m=(\hat{\bm w}_{1,m},\ldots,\hat{\bm w}_{S,m})$ be the solution to Problem \eqref{eq:opt}. % and denote $\hat{g}_{ij,m}=\hat{\bm w}_{j,m}^T\bm q(\sigma_i)$.
	The data-driven Clfdr statistic is given by
	$$\hat{T}_{i,m}=\dfrac{\hat{f}^m_{0}(x_i\mid\sigma_i)}{\hat{f}^m(x_i\mid\sigma_i)},\quad \text{where}$$
	$$\hat{f}^m_{0}(x\mid\sigma_i)=\sum_{j:u_j \in \mathcal{A}}\eta_{\sigma_i}(x-u_j)\hat{\bm w}_{j,m}^T\bm q(\sigma_i),~\hat{f}^m(x\mid\sigma_i)=\sum_{j=1}^{S}\eta_{\sigma_i}(x-u_j)\hat{\bm w}_{j,m}^T\bm q(\sigma_i).
	$$
\end{definition}
%In Section \ref{sec:theory} we show that uniformly in $i$, $\hat{T}_{i,m}$ is asymptotically close to $T_i^{\sf OR}$ for large $m$. 
Next, in Definition \ref{def:data_driven} we present the proposed HAMT procedure that relies on the estimate $\hat{T}_{i,m}$ and uses a step-wise procedure from \cite{sun2012multiple} to estimate $t^*$.
\begin{definition}(HAMT procedure)
	\label{def:data_driven}
	Denote $\hat{T}_{(1),m}\le\ldots\le \hat{T}_{(m),m}$ the sorted Clfdr statistics and $H_{(1)},\ldots,H_{(m)}$ the corresponding hypotheses. Suppose
	$$r=\max\Bigl\{j:\dfrac{1}{j}\sum_{i=1}^{j}\hat{T}_{(i),m}\le \alpha\Bigl\}.
	$$
	Then, the HAMT procedure rejects the ordered hypotheses $H_{(1)},\ldots,H_{(r)}$. Furthermore, in comparison to the oracle procedure $\bm\delta^{\sf OR}(t^*)$ in Equation \eqref{eq:oracle_procedure}, HAMT has the following form:
	$$\bm \delta^{\sf HAMT}(\hat{t}^{*}_m)=\{I(\hat{T}_{i,m}<\hat{t}^{*}_m):1\le i\le m\},~\text{where}~\hat{t}^{*}_m=\hat{T}_{(r),m}.
	$$
\end{definition}
In Definition \ref{def:data_driven}, the estimate $\hat{t}^{*}_m$ of $t^*$ is based on the intuition that when the first $j$ ordered hypotheses are rejected then a good estimate of the false discovery proportion is given by the moving average $(1/j)\sum_{i=1}^{j}\hat{T}_{(i),m}$ and the condition $(1/j)\sum_{i=1}^{j}\hat{T}_{(i),m}\le \alpha$ then helps fulfill the FDR constraint. In Section \ref{sec:theory} we show that for large $m$, $\hat{T}_{i,m}$ is asymptotically close to $T_i^{\sf OR}$ uniformly in $i$, and the HAMT procedure in Definition \ref{def:data_driven} is a good approximation to the oracle procedure $\bm\delta^{\sf OR}(t^*)$.

\subsection{Effect of ignoring the dependence between $\mu_i$ and $\sigma_i$}
\label{sec:dependence}
%The deconvolution estimator of $g_\mu(\cdot\mid\sigma)$ plays a vital role in constructing data-driven procedures for testing composite null hypotheses. 
Here, we consider a numerical example to illustrate the effect on the power and validity of multiple testing procedures if the underlying deconvolution estimator for constructing the Clfdr statistics ignores the dependence between $\mu_i$ and $\sigma_i$ .

We fix $m=10^4$ and, for $i=1,\ldots,m$, sample $X_i$ independently from $N(\mu_i,\sigma_i^2)$ with $\mu_i=3\sigma_i$ and $\sigma_i\stackrel{i.i.d.}{\sim}\text{Unif}(0.5,2)$. The goal is to test $H_{0,i}:\mu_i\in \mathcal A ~vs~H_{1,i}:\mu_i\notin \mathcal A$ where $\mathcal A=(-\infty,4]$ and $\alpha=0.1$. The following three testing procedures are evaluated in this example: the procedure that relies on the deconvolution estimate obtained from nonparametric maximum likelihood \citep{kiefer1956consistency,laird1978nonparametric,koenker2017rebayes} (NPMLE) techniques to estimate the Clfdr statistic, the procedure that uses the deconvolution estimate from \cite{efron2016empirical} (DECONV) to estimate $T_i^{\sf OR}$ and the HAMT procedure from Definition \ref{def:data_driven}. While these procedures employ different methods for estimating $T_i^{\sf OR}$, they all rely on Definition \ref{def:data_driven} to estimate the threshold $t^*$.

The first row of Figure \ref{fig:motfig_test} highlights in red the hypotheses that were rejected by the three procedures. Here the dotted horizontal line is $\sigma=4/3$ and represents the oracle decision rule which rejects any hypothesis above that line. The rightmost panel presents the hypotheses that were rejected by HAMT and appears to correctly discover a substantially larger proportion of the non-null cases than NPMLE and DECONV while safeguarding, at the same time, the number of false discoveries. For instance, across $200$ repetitions of this multiple testing problem the average false discovery proportions for NPMLE, DECONV and HAMT are, respectively, $0.157$, $0.186$ and $0.010$ while their average proportion of true discoveries are $0.142$, $0.231$ and $0.845$.
\begin{figure}[!h]
	\centering
	\includegraphics[width=0.85\linewidth]{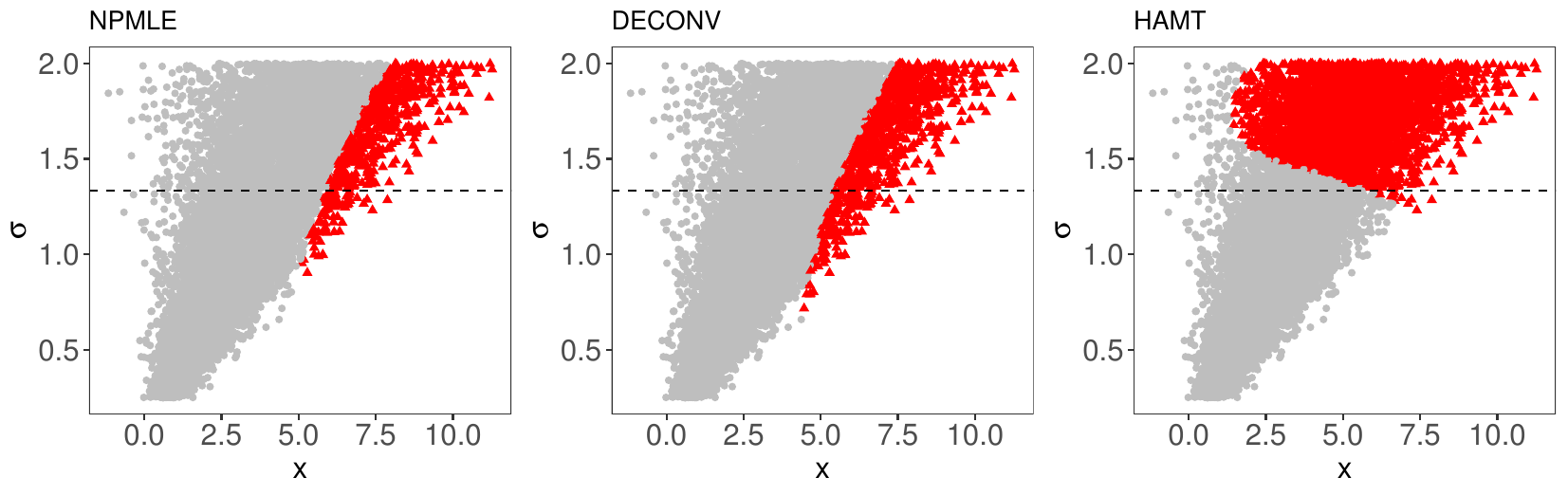}
	\caption{We test $H_{0,i}:\mu_i\le 4~vs~H_{1,i}:\mu_i>4,~i=1,\ldots,m$, where $X_i\stackrel{ind}{\sim}N(\mu_i,\sigma_i^2)$, $\mu_i=3\sigma_i,~\sigma_i\stackrel{i.i.d.}{\sim}\text{Unif}(0.5,2)$ and $m=10,000$. Across the three panels, in red are the hypotheses that were rejected by the testing procedures at $\alpha=0.1$. The dotted horizontal line is the oracle decision rule which rejects any hypothesis above that line. The left and center panels depict testing procedures that rely, respectively, on NPMLE's and DECONV's deconvolution estimates. The rightmost panel presents the HAMT procedure.}
	\label{fig:motfig_test}
	%		\includegraphics[width=\linewidth]{./images/motfig_density}
	%		\caption{estimated marginal density $f_i(x)$ of $X$ conditional on $\sigma_i\in\{1,1.5,2\}$.}
	%		\label{fig:motfig_density}
\end{figure}
\begin{figure}[!h]
	\centering
	\includegraphics[width=0.85\linewidth]{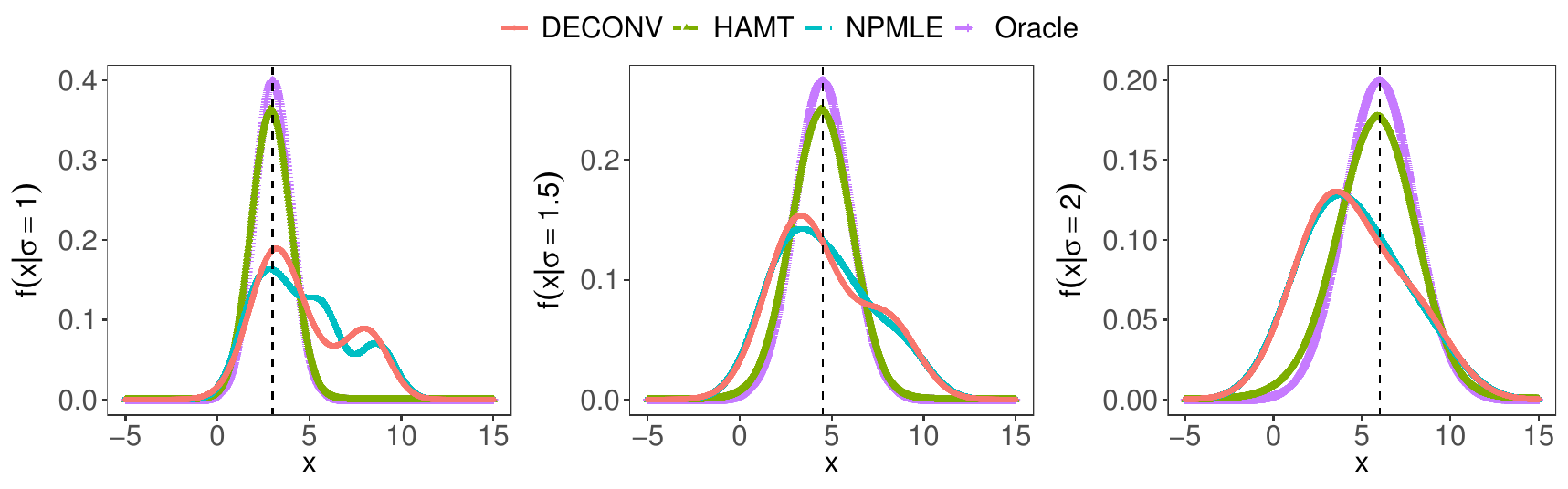}
	\caption{The oracle marginal density $f(x\mid\sigma)$ in green and the estimated marginal densities from the deconvolution estimates of NPMLE, DECONV and HAMT for $\sigma\in\{1,1.5,2\}$. The dotted vertical line represents the mean of the distribution of $X$ given $\sigma$.}
	\label{fig:motfig_density}
\end{figure}
%The estimates of the composite null and marginal densities play a vital role in testing composite null hypotheses and 
The relatively poorer performance of NPMLE and DECONV in this example is related to the fact that the underlying decovolution estimator for both these procedures ignore the dependence between $\mu_i$ and $\sigma_i$. To see that, we present the estimate of $f(\cdot\mid\sigma)$ for $\sigma\in\{1,1.5,2\}$ in Figure \ref{fig:motfig_density}. Across the three panels, the deconvolution estimates from NPMLE and DECONV result in marginal density estimates that are substantially different from the ground truth. The deconvolution estimator underlying the HAMT procedure, on the other hand, seems to generate marginal density estimates that are relatively closer to $f(\cdot\mid\sigma)$. In Section \ref{sec:theory} we present formal theories supporting this intuition and establish that $\hat{f}^m_{0}(\cdot\mid\sigma_i)$ and $\hat{f}^m(\cdot\mid\sigma_i)$ in Definition \ref{def:clfdr} are, in fact, consistent estimators of ${f}_0(\cdot\mid\sigma_i)$ and ${f}(\cdot\mid\sigma_i)$, respectively, as $m\to\infty$.

\section{Theory}
\label{sec:theory}
In this section we study the asymptotic properties of HAMT under the setting where the grid size $S= S(m)$ and the number of bases $K= K(m)$ vary with $m$. \red{For two real sequences $a_m,b_m$, we will use $a_m\asymp b_m$ to mean that there exists constants $c_2\ge c_1>0$ such that $c_1b_m\leq a_m\leq c_2b_m$ for large $m$.}
%In this section we present our main theoretical result, Theorem \ref{thm:dd}, which shows that the data driven procedure $\bm\delta^{\sf DD}$ asymptotically achieves the performance of the oracle procedure $\bm \delta^{\sf OR}$ (Equation \eqref{eq:oracle_proc}) as $m\to\infty$. 
The following regularity conditions are needed in our technical analysis. %For two sequences $a_m$ and $b_m$, we will denote $a_m\asymp b_m$ to mean $d_1a_m\le b_m\le d_2 a_m$ for large $m$ and some constants $d_2\ge d_1>0$.
\begin{description}	
	\item\textbf{(A1)} The density $g_\mu(\cdot|\sigma)$ is supported on a compact interval $[-M,M]$ for some $M<\infty$ and has bounded first and second derivatives with respect to $\sigma$ for all $\mu\in[-M,M]$.  
	\item\textbf{(A2)} The density $g_\sigma(\cdot)$ satisfies $0<C_1<g_\sigma(\sigma)<C_2<\infty$ and $|g'_\sigma(\sigma)|<C_3$ for some fixed constants $C_1$, $C_2$ and $C_3$ on $\text{supp}(g_\sigma)$ where 
 $\text{supp}(g_\sigma)\subset[M_1,M_2]$, for some $M_2<\infty$ and $M_1>0$.
	\item\textbf{(A3)} The bandwidths $(h_x,h_{\sigma})$ satisfy $h_x\asymp m^{-\nu_x}$, $h_\sigma\asymp  m^{-\nu_\sigma}$ where $\nu_x$ and $\nu_\sigma$ are small positive constants such that $0<\nu_\sigma+\nu_x<1$.
	%	\item\textbf{(A4)}  $\theta_j(\cdot)$'s are sufficiently smooth, that is $\theta_j(\cdot)=\sum_{i=1}^{\infty}w_{ji}B_i(\cdot)$ where $B_i(\cdot)$ is the $i$th cosine basis function and $\{w_{ji}\}_{i=1}^{\infty}$ belongs to the Sobolev ellipsoid $\Theta(\gamma,c)$. \red{Discuss}.
	%	\item\textbf{(A3)}	$\lim\limits_{\epsilon\rightarrow0}P(|E(\mu_i|x_i)-\mu_0|<\epsilon)=0$.
	\item\textbf{(A4)} The density $\eta(\cdot)$ is smooth.
 
 %\prp{$\eta_{\sigma}\{(\cdot)\sigma^{-1}\}$ is continuous in $\sigma$ and $\text{Var}[\eta_{\sigma_i}\{(\cdot)\sigma_i^{-1}\}]<\infty$ for all $i$. BOWEN: please check notations and clarify with respect to what distribution the variance is being taken.}
\end{description}
%\red{BOWEN: please discuss the assumptions A1-A2 here. Assumption (A2) on the compactness of the supports of $g_\mu(\cdot|\sigma)$ and $h_\sigma(\cdot)$ has been made in \citep{dicker2016high}.}
Assumption (A1) on the first and second derivatives of $g_\mu(\cdot|\sigma)$ with respect to $\sigma$ is a necessary condition in our proofs for information pooling across the heteroskedastic units and controlling the bias of of pilot estimate. The compactness of the supports of $g_\mu(\cdot|\sigma)$ and $g_\sigma(\cdot)$ in Assumptions (A1) and (A2) are standard regularity conditions for empirical Bayes deconvolution problems (see for example \cite{dicker2016high}) and are satisfied in most practical scenarios where the true mean $\mu$ often represents a score. The boundedness of $g_\sigma(\cdot)$ in (A2) is needed for controlling the variance of the pilot estimate.
Assumption (A3) is satisfied by common choices of bandwidths in \cite{wand1994kernel,silverman1986density}. \red{Assumption (A4) holds for typical error distributions such as Gaussian, T, and Laplace distributions.} 

Proposition \ref{prop1} formally establishes the asymptotic consistency of $\hat{f}^m_{0}(\cdot|\sigma)$ and $\hat{f}^m(\cdot|\sigma)$ as $m\to\infty$.
\begin{proposition}
	\label{prop1}
	{Consider Model \eqref{eq:model1}--\eqref{eq:model2} and suppose assumptions (A1) -- (A4) hold. Then as $m,S(m),K(m)\to\infty$ and $\sigma\sim g_\sigma$, we have, 
		\begin{eqnarray}
			E \|\hat
			{f}^m(\cdot|\sigma)-f(\cdot|\sigma)\|^2&=&E\int\{\hat{f}^m(x|\sigma)-f(x|\sigma)\}^2\mathrm{d}x\to 0~\text{and}~\nonumber\\
			E \|\hat
			{f}^m_{0}(\cdot|\sigma)-f^m_{0}(\cdot|\sigma)\|^2&=&E\int\{\hat{f}^m_{0}(x|\sigma)-f^m_{0}(x|\sigma)\}^2\mathrm{d}x\to 0,\nonumber
		\end{eqnarray}
		where the expectation is taken over $(\bm X, \bm \sigma)$ and $\sigma$.}
\end{proposition}
While Proposition \ref{prop1} does not place any restrictions on the rates at which $K(m)$ and $S(m)\rightarrow\infty$, the sizes of $K(m)$ and $S(m)$ do affect the quality of $\hat{f}^m_0(\cdot|\sigma)$ and $\hat{f}^m(\cdot|\sigma)$ for large $m$. To see that, first note that since $\hat{f}^m(\cdot|\sigma)$ is approximating $\hat{\varphi}^m(\cdot,\sigma)$, the order of $E \|\hat
{f}^m(\cdot|\sigma)-f(\cdot|\sigma)\|^2$ is upper bounded by the order of $E\|\hat{\varphi}^m(\cdot,\sigma)-f(\cdot|\sigma)\|^2$. It was shown in \cite{wand1994kernel} that when $h_x\asymp m^{-1/6}$ and $h_\sigma\asymp m^{-1/6}$, the optimal rate at which $E\|\hat{\varphi}^m(\cdot,\sigma)-f(\cdot|\sigma)\|^2\rightarrow 0$ is $O(m^{-2/3})$.  
To achieve this rate in our context, suppose, for instance, that $\eta(\cdot)$ is the density of a standard Gaussian random variable. Then, it is sufficient for $S(m)\asymp m^{1/3}\sqrt{\log m}$. This is formally established in Remark \ref{gridsize} in the supplement. Moreover, on the appropriate choice of the number of basis functions $K(m)$ in this setting, Remark \ref{basis} (in Section \ref{app:prf_prop1} of the supplement) shows that if
$g_j(\cdot)=\sum_{k=1}^{\infty}w_{jk}q_k(\cdot)$ and $\bm w_j=\{w_{jk}:k=1,2,\ldots,\}$ belong to the Sobolev ellipsoid $\Theta(\gamma,c)$ with order $\gamma>0$ and radius $c<\infty$ for $j=1,\ldots,S(m)$, then $K(m)\asymp m^{1/(2\gamma)}(\log m)^{1/(4\gamma)}$. 
Thus, while larger $K(m),~S(m)$ will not improve the quality of $\hat{f}^m(\cdot|\sigma)$, smaller $K(m),~S(m)$ may result in a slower convergence rate of $E \|\hat{f}^m(\cdot|\sigma)-f(\cdot|\sigma)\|^2$ to $0$. %However, we emphasize that the validity of Proposition 1 does not depend on the rates at which $K(m)$ and $S(m)$ approach $\infty$ since those rates only affect the convergence rates of $E \|\hat{f}^m(\cdot|\sigma)-f(\cdot|\sigma)\|^2$ and $E \|\hat{f}^m_{0}(\cdot|\sigma)-f^m_{0}(\cdot|\sigma)\|^2$ to $0$.
In Section 5.1 we provide recommendations on the practical choices of $S(m)$ and $K(m)$ that work well in our numerical experiments and real data analyses.

A consequence of Proposition \ref{prop1} is Corollary \ref{coro1} which establishes that the data-driven Clfdr statistic $\hat{T}_{i,m}$ in Definition \ref{def:data_driven} converges in probability to its oracle counterpart as $m\to\infty$.
\begin{corollary}\label{coro1}
	{Under the conditions of Proposition \ref{prop1} and uniformly in $i$, $\hat{T}_{i,m}{\rightarrow}T_i^{\sf OR}$ in probability as $m\to\infty$.}	
\end{corollary}
Next, we state the main theorem of this section which is related to the asymptotic performance of HAMT as $m\to\infty$.
\begin{theorem} 
	\label{thm:dd} 
	Consider Model \eqref{eq:model1}--\eqref{eq:model2}. Under assumptions (A1) -- (A4) and as $m\to\infty$, we have (i) the mFDR and FDR of $\bm\delta^{\sf HAMT}(\hat{t}_m)$ are controlled at level $\alpha + o(1)$, and \newline (ii) $ETP\{\bm \delta^{\sf HAMT}(\hat{t}_m)\}/ETP\{\bm \delta^{\sf OR}(t^*)\}=1+o(1)$.
\end{theorem}
Together with Theorem \ref{thm_orcale}, Theorem \ref{thm:dd} establishes that the proposed HAMT procedure is asymptotically valid for FDR control and attains the performance of the oracle procedure as $m\to\infty$.
\section{Numerical experiments}
\label{sec:sims}
\subsection{Recommended choices of $\mathcal{T},S,K$ and $q(\sigma_i)$} 
\label{sec:implement}
%We first discuss Problem \eqref{eq:opt}. 
While Section \ref{sec:theory} provides guidance on the asymptotic choices of the grid size $S(m)$ and the number of basis functions $K(m)$, in our implementation we fix $S=50$ and $K=10$, which work well in all of our numerical and real data examples. For the grid support $\mathcal T$, HAMT uses $S$ equispaced points in $[X_{(1)},X_{(m)}]$ where $X_{(1)}=\min\{X_1,\ldots,X_m\}$ and $X_{(m)}=\max\{X_1,\ldots,X_m\}$. %Finally, the conic interior-point optimizer in MOSEK \citep{mosek} solves Problem \eqref{eq:opt}. %The most expensive step in this program is computing the derivative and hessian of the objective, which have complexity $O(mK)$ and $Q(K^2)$ respectively (CHECK). Faster, first order methods, such as the one pursued in \cite{kim2020fast,zhang2022efficient} can be developed however we do not pursue that direction in this article. 
Next, for the basis functions $\bm q(\sigma_i)=(q_{1,i},\ldots,q_{K,i})$ in Equation \eqref{eq:basis_rep} we use the cosine basis $q_{k,i}=0.5\{1+\cos(k\sigma_i)\}$. Since we assume the dependence of $g_\mu(\cdot|\sigma)$ on $\sigma$ is continuous, the number of cosine basis functions used in Equation \eqref{eq:basis_rep} can be interpreted as the user's belief about the smoothness of such dependence. 
Lastly, the pilot estimator $\hat{\varphi}_{(-i)}^m(x_i,\sigma_i)$ in Equation \eqref{eq:kde_hart_jacknife} is borrowed from \cite{fu2022heteroscedasticity} and depends on a pair of bandwidths $\bm h = (h_x,h_\sigma)$. We follow the author's recommendation in choosing these bandwidths which rely on Silverman's rule of thumb \citep{silverman1986density}. Additional details related to the implementation of HAMT are provided in Section \ref{sec:npmleb_details} of the supplement. The R code for reproducing all numerical results in the paper is available at \url{https://github.com/trambakbanerjee/HAMT_paper}.
\subsection{Experiments involving one-sided composite null hypotheses}
\label{sec:sims_onesided}
In this section we assess the numerical performance of HAMT for one-sided composite null hypotheses. Specifically, we test $m=10^4$ hypotheses of the form $H_{0i}:\mu_i\in\mathcal A~vs~H_{1i}:\mu_i\notin\mathcal A$ where $\mathcal A=(-\infty, \mu_0]$. We consider eleven competing testing procedures of which four are $p-$value based methods and seven rely on an Lfdr estimate for ranking the hypotheses. The four $p-$value procedures are: \textbf{BH} - the Benjamini-Hochberg procedure from \cite{benjamini2006adaptive} which is designed to overcome the conservativeness of the original \cite{benjamini1995controlling} procedure by including a correction in size, \textbf{AdaPTGMM} - the procedure from \cite{chao2021adapt} that uses $\bm \sigma$ as an additional covariate, \textbf{CAMT} - the method from \cite{zhang2022covariate} and \textbf{IHW} - the procedure from \cite{ignatiadis2021covariate}. Along with \textbf{AdaPTGMM}, both \tb{CAMT} and \tb{IHW} are designed to exploit the side information in $\bm \sigma$ for testing the $m$ hypotheses. The remaining seven methods are the following Lfdr based procedures: \textbf{OR} -  the oracle procedure from Equation \eqref{eq:oracle_procedure}, \textbf{DECONV} - this procedure ignores the dependence between $\mu_i$ and $\sigma_i$, uses the empirical Bayes deconvolution method from \cite{efron2016empirical} to estimate $T_i^{\sf OR}$ and then relies on Definition \ref{def:data_driven} to estimate the threshold $t^*$, \tb{NPMLE B} - this method is similar to HAMT where $g_j(\sigma_i)=\bm w_j^T\bm q(\sigma_i)$ but we estimate the $\bm w_j$'s by solving Problem \eqref{eq:npmle}, \textbf{GS 1} - the testing procedure from \cite{gu2018oracle} that is based on the standardized statistic $Z_i=(X_i-\mu_0)/\sigma_i$ and relies on the deconvolution estimate obtained from nonparametric maximum likelihood estimation to construct the Lfdr, \textbf{GS 2} - another procedure from \cite{gu2018oracle} that allows for the possibility that in some applications, there might be a non-trivial probability mass at $\mu_0$ which may lead to poor FDR control if not accounted for while estimating the marginal density of $Z_i$, \tb{ASH} -  the method from \cite{stephens2017false} with $c=0$ and \tb{ASH 1} - the same method from \cite{stephens2017false} but with $c=1$. \red{Section \ref{sec:npmleb_details} of the supplement includes more details related to the implementation of \tb{NPMLE B}, \tb{ASH} and \tb{ASH 1}.}

The aforementioned procedures are evaluated on five different simulation settings with $\alpha$ fixed at $0.1$. For each simulation setting, the data are generated from Model \eqref{eq:model1}-\eqref{eq:model2} with $\epsilon_{i}\stackrel{i.i.d.}{\sim}N(0,1)$, and the average false discovery proportion FDP$(\bm \delta)=\sum_{i=1}^{m}\{(1-\theta_i)\delta_i\}/\max(\sum_{i=1}^{m}\delta_i,1)$ and the average proportion of true positives discovered PTP$(\bm \delta)=\sum_{i=1}^{m}\theta_i\delta_i/\max(\sum_{i=1}^{m}\theta_i,1)$ across $200$ Monte-Carlo (MC) repetitions are reported.

\begin{figure}[!h]
	\centering
	\includegraphics[width=0.9\linewidth]{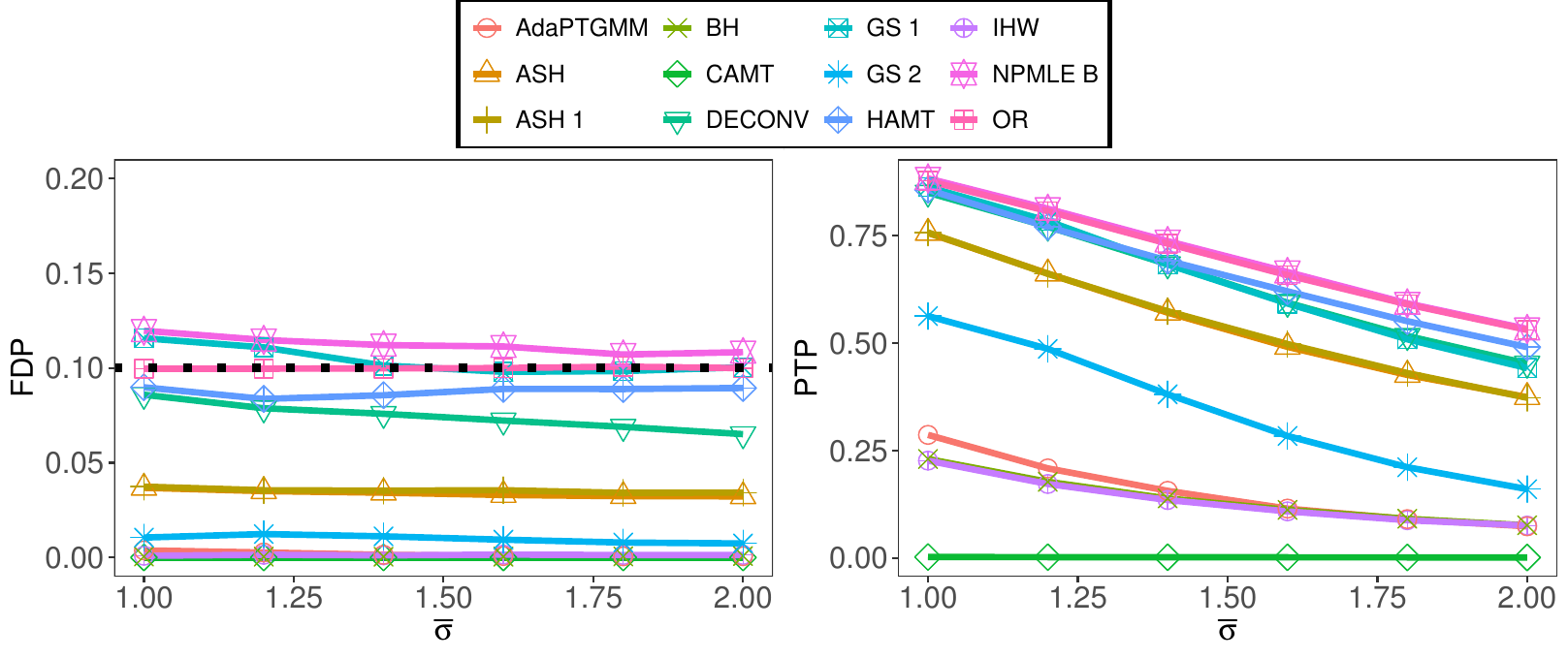}
	\caption{ Setting 1 - $\sigma_i\stackrel{i.i.d.}{\sim}\text{Unif}(0.5,\bar\sigma)$, $\mu_i=0$ with probability $0.9$ and $\mu_i\stackrel{i.i.d.}{\sim}N(3,1)$ with probability $0.1$. Here $\mathcal{A}=(-\infty,2]$. For ASH and ASH 1, \texttt{mixcompdist = "normal"}.% 880 by 380. 1035 by 435
	}
	\label{fig:sim_exp1_test}
\end{figure}
In the first setting  $(\mu_i,\sigma_i)$ are independent. We sample $\sigma_i\stackrel{i.i.d.}{\sim}\text{Unif}(0.5,\bar\sigma)$ and let $\mu_i=0$ with probability $0.9$ and $\mu_i\stackrel{i.i.d.}{\sim}N(3,1)$ with probability $0.1$. We vary $\bar\sigma\in\{1,1.2,1.4,1.6,1.8,2\}$ and take $\mu_0=2$. Figure \ref{fig:sim_exp1_test} reports the average FDP and PTP for the competing procedures in this setting. We observe that the four $p-$value methods are the most conservative with CAMT exhibiting the least power. The procedure GS 2 closely follows the $p-$value methods in FDR control but exhibits substantially better power. The remaining methods have an overall similar performance in this setting although GS1 and NPMLE B marginally fail to control the FDR level at $10\%$ for small values of $\bar\sigma$.
\begin{figure}[!h]
	\centering
	\includegraphics[width=0.9\linewidth]{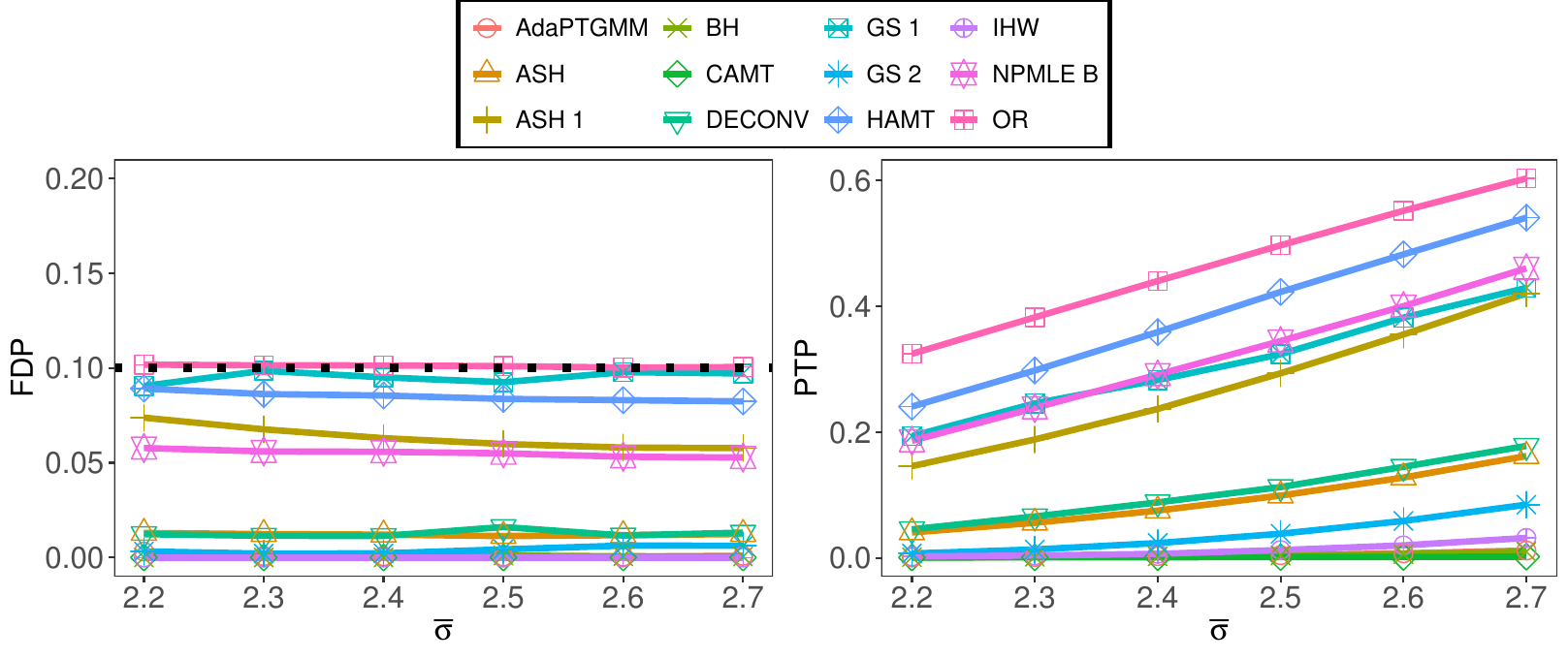}
	\caption{Setting 2 - $\sigma_i\stackrel{i.i.d.}{\sim}(1/3)I_{(0.5)}+(1/3)I_{(1)}+(1/3)I_{(2)}$ and conditional on $\sigma_i$, $\mu_i=0$ with probability $0.9$ or $\mu_i=\bar\sigma\sigma_i$ with probability $0.1$. Here $\mathcal{A}=(-\infty,2]$. For ASH and ASH 1, \texttt{mixcompdist = "+uniform"}.}
	\label{fig:sim_exp2_test}
\end{figure}

\red{The second setting represents a scenario where $\mu_i$ and $\sigma_i$ are correlated and have discrete distributions. Setting 2 is presented in Figure \ref{fig:sim_exp2_test} where $\sigma_i$ can take three values $\{0.5,1,2\}$ with equal probabilities. Conditional on $\sigma_i$, $\mu_i=0$ with probability $0.9$ or $\mu_i=\bar\sigma\sigma_i$ with probability $0.1$. We set $\mu_0=2$ and find that all methods control the FDR level in Figure \ref{fig:sim_exp2_test}. Among the data-driven procedures, HAMT has the highest power and is closely followed by NPMLE B, GS 1 and ASH 1. DECONV, which completely ignores the dependence between $\mu_i$ and $\sigma_i$, exhibits a substantially lower power than HAMT in this setting.}

\begin{figure}[!h]
	\centering
	\includegraphics[width=0.9\linewidth]{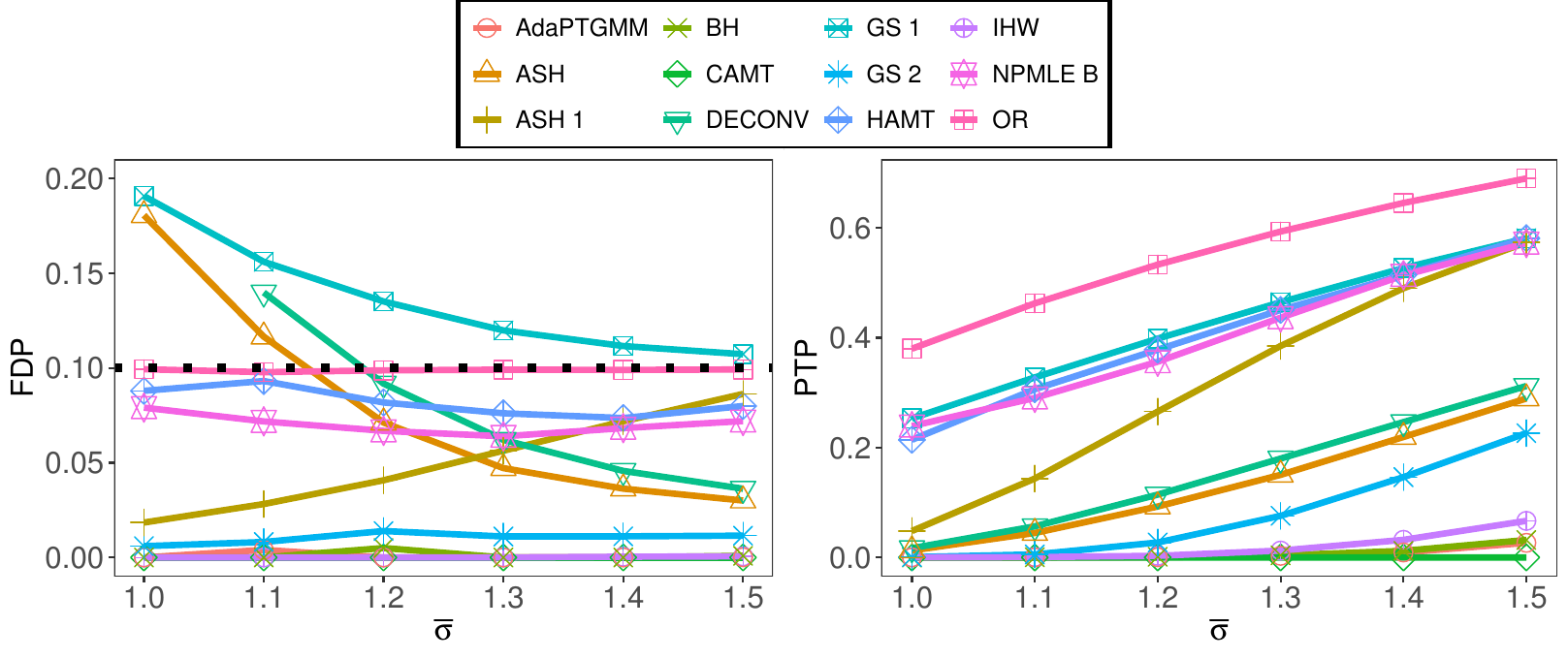}
	\caption{Setting 3: $\sigma_i\stackrel{i.i.d.}{\sim}\text{Unif}(0.5,\bar\sigma)$ and conditional on $\sigma_i$, $\mu_i\stackrel{ind.}{\sim}0.9N(-\sigma_i,0.5)+0.1\delta_{(2\sigma_i^2)}$, where $\delta_{(a)}$ represents a point mass at $a$. Here $\mathcal{A}=(-\infty,1]$. For ASH and ASH 1, \texttt{mixcompdist = "+uniform"}. At $\bar{\sigma}=1$, DECONV exhibits an FDR bigger than $0.2$.}
	\label{fig:sim_exp5_test}
\end{figure}
For Setting 3, $\sigma_i\stackrel{i.i.d.}{\sim}\text{Unif}(0.5,\bar\sigma)$ and conditional on $\sigma_i$, $\mu_i\stackrel{ind.}{\sim}0.9N(-\sigma_i,0.5)+0.1\delta_{(2\sigma_i^2)}$, where $\delta_{(a)}$ represents a point mass at $a$. Setting 3 is presented in Figure \ref{fig:sim_exp5_test} where $\mu_0=1$. We find that GS 1 fails to control the FDR level at $10\%$ while ASH and DECONV control the FDR at all values of $\bar{\sigma}$, except the first two where the latter exhibits an FDR value bigger than $0.2$ at $\bar{\sigma}=1$. HAMT effectively captures the dependence between $\mu_i$ and $\sigma_i$ and is, by far, the best testing procedure in this setting along with NPMLE B.

The remaining two settings present scenarios where HAMT provides a substantial improvement over all competing methods, both in terms of FDR control and power. In the fourth setting, $\sigma_i\stackrel{i.i.d}{\sim}0.9 \text{Unif}(0.5,1)+0.1 \text{Unif}(1,u)$, $\mu_i=0$, if $\sigma_i\le 1$ and $2/\sigma_i$, otherwise. Thus, in this setting $\sigma_i$ controls both the sparsity level of $\mu_i$ and the distribution of its non-zero effects. The performance of the competing methods is presented in Figure \ref{fig:sim_exp4_test} where $\mu_0=1$. The oracle procedure (OR) in this setting perfectly classifies each $\mu_i$ as satisfying $\mu_i\le \mu_0$ or $\mu_i>\mu_0$ simply by observing if $\sigma_i\le 1$ or $\sigma_i>1$ and $2/\sigma_i>\mu_0$. Thus in Figure \ref{fig:sim_exp4_test}, OR has FDP equal to 0 and PTP equal to 1 for all $\bar{\sigma}$. While, all other methods control the FDR at $10\%$, HAMT exhibits a substantially higher power in this setting for all values of $\bar{\sigma}$. 

\begin{figure}[!h]
	\centering
	\includegraphics[width=0.9\linewidth]{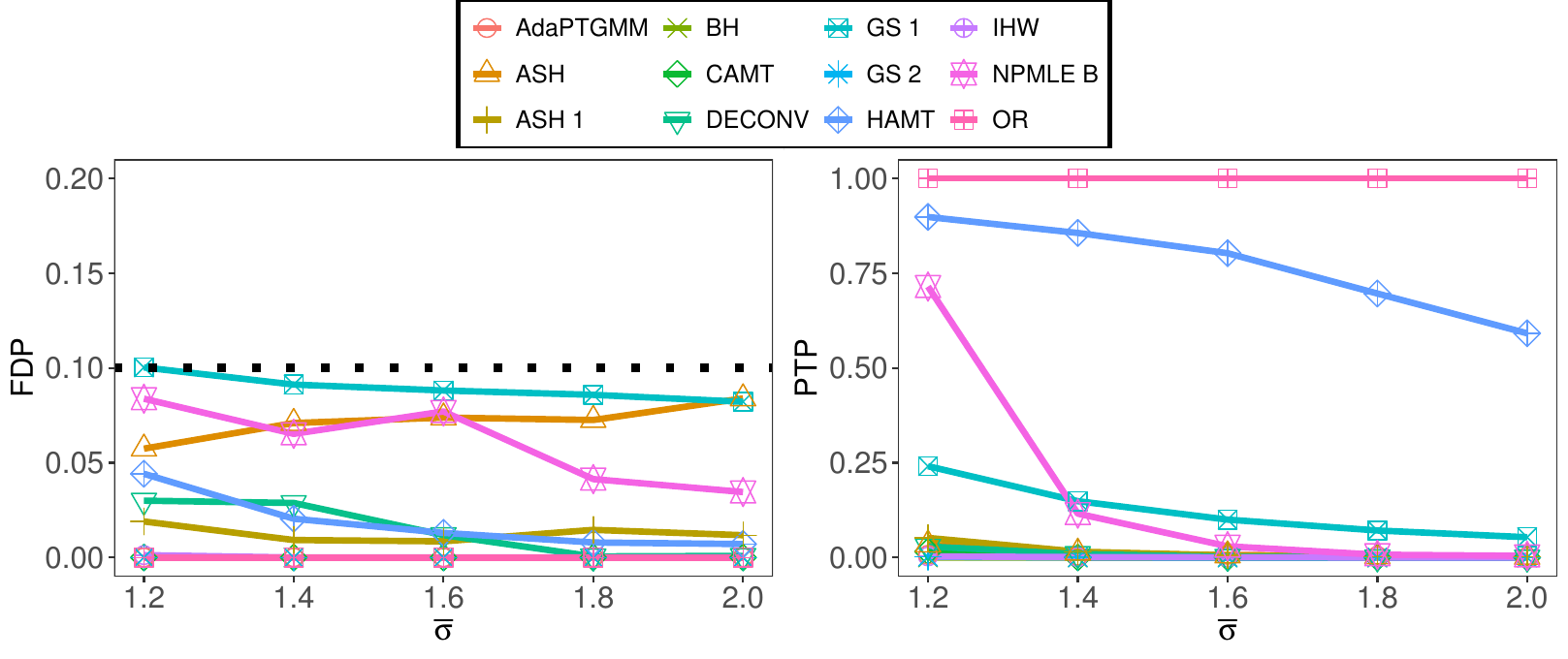}
	\caption{Setting 4 - $\sigma_i\stackrel{i.i.d}{\sim}0.9 \text{Unif}(0.5,1)+0.1 \text{Unif}(1,\bar\sigma)$ and $\mu_i=0$, if $\sigma_i\le 1$ and $2/\sigma_i$, otherwise. Here $\mathcal{A}=(-\infty,1]$. For ASH and ASH 1, \texttt{mixcompdist = "halfuniform"}.}
	\label{fig:sim_exp4_test}
%\end{figure}
%\begin{figure}[!h]
	\centering
	\includegraphics[width=0.9\linewidth]{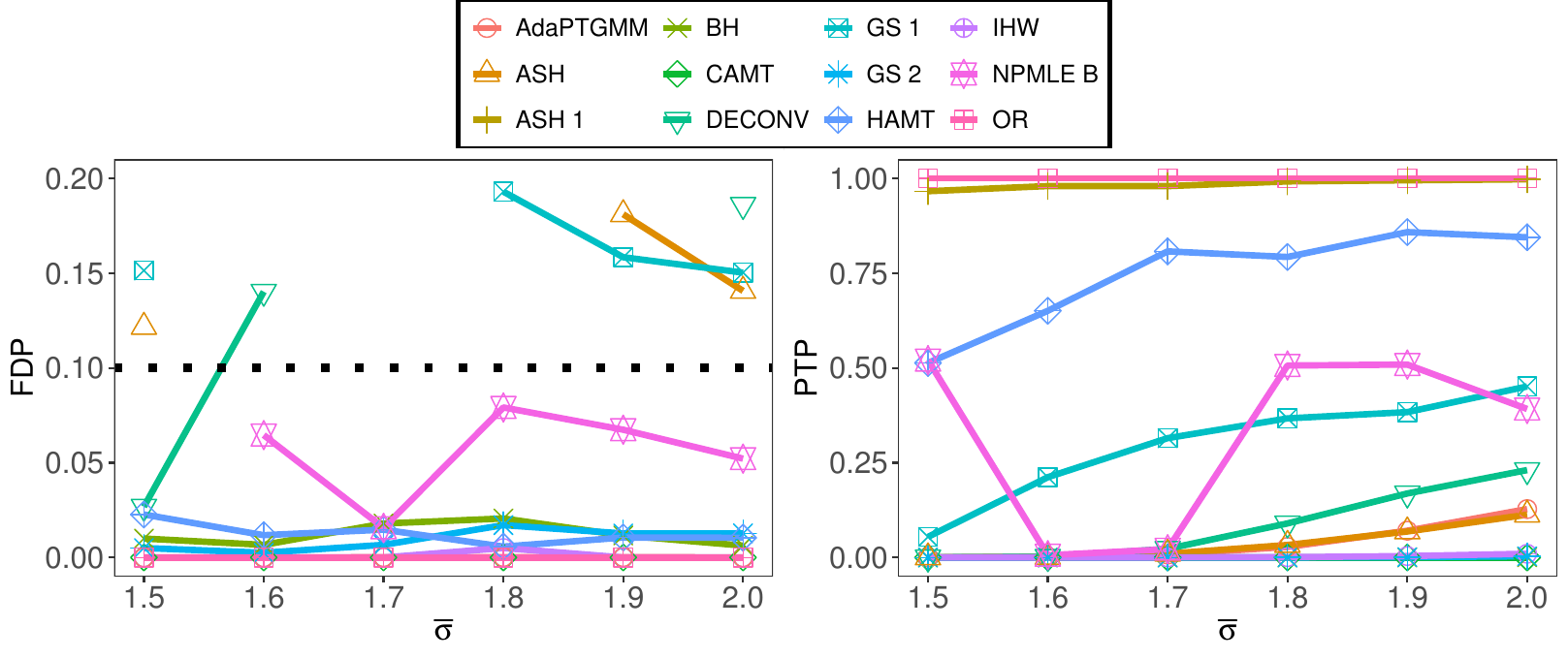}
	\caption{ Setting 5 - $\sigma_i\stackrel{i.i.d.}{\sim}\text{Unif}(0.25,\bar\sigma)$ and conditional on $\sigma_i$, $\mu_i=3\sigma_i$. Here $\mathcal{A}=(-\infty,4]$. For ASH and ASH 1, \texttt{mixcompdist = "+uniform"}.}
	\label{fig:sim_exp6_test}
\end{figure}
In the fifth setting, we allow $\mu_i$ and $\sigma_i$ to be perfectly correlated. Specifically, $\sigma_i\stackrel{i.i.d.}{\sim}\text{Unif}(0.25,u)$, $\mu_i=3\sigma_i$ and $\mu_0=4$. In Figure \ref{fig:sim_exp6_test}, \red{several methods} fail to control the FDR at $10\%$ and for some values of $\bar{\sigma}$, they exhibit FDP values bigger than $0.2$. The left panel of Figure \ref{fig:sim_exp6_test} excludes those values of $\bar{\sigma}$ for such methods. \rred{NPMLE B, in particular, exhibits a relatively high MC error in its FDP distribution for the first three values of $\bar\sigma$ and so for each $\bar\sigma$ we report its median FDP and PTP across the 200 MC repetitions.} The oracle procedure in this setting simply observes if $3\sigma_i>\mu_0$ for rejecting the null hypothesis and its data-driven counterpart, HAMT, has the highest power among all other testing procedures considered here. %\rred{While {NPMLE B} and {HAMT} are similar in the manner in which they model the unknown prior $g(\cdot|\sigma_i)$, the inability of the former to control the FDR is not entirely surprising. This is because there is no guarantee that the Clfdr statistic underlying {NPMLE B} is a consistent estimator of the oracle counterpart as $m\to\infty$. The Clfdr statistic underlying {HAMT}, in contrast, is consistent (Proposition \ref{prop1} and Corollary \ref{coro1}) which ultimately ensures that {HAMT} guarantees asymptotic FDR control and attains the performance of the oracle procedure (Theorem \ref{thm:dd}). In this context, the heteroskedastic bivariate kernel density estimator $\hat{\varphi}^m(\cdot|\sigma_i)$ (Equation \eqref{eq:kde_hart}) and the density matching approach to estimating $\mathcal W$ (Problem \eqref{eq:opt}) play a crucial role in establishing the validity and optimality of the {HAMT} procedure.}

\red{Overall, the aforementioned simulation experiments reveal that the $p-$value procedures considered here are considerably more conservative than their Lfdr counterparts, while the Lfdr methods that ignore the dependence between $\mu_i$ and $\sigma_i$ may even fail to control the FDR at the desired level. In contrast, HAMT, which relies on an improved deconvolution estimator for constructing the Clfdr statistic, provides an alternative multiple testing procedure that is often more powerful than competing methods, such as ASH, ASH 1, at the same FDR level. In Section \ref{sec:more_sims} of the supplement we present more simulation studies to assess the numerical performance of HAMT in settings involving (i) two-sided composite nulls, (ii) non-Gaussian likelihoods and (iii) unknown $\bm \sigma$ and Section \ref{sec:more_discuss} includes an additional discussion on the empirical performances of HAMT, NPMLE B and ASH reported here. A real data application is presented in Section \ref{sec:realdata}.}
\section{Discussion}
\label{sec:discuss}
Heteroskedasticity presents a challenging setting for designing valid and powerful multiple testing procedures. For testing composite null hypotheses, we show that the conventional practice of standardizing heteroskedastic test statistics may severely affect the power of the underlying testing procedure. Additionally, when the inferential parameter of interest is correlated with the variance of the test statistic, existing methods that ignore this dependence may fail to control the type I error at the desired level. In this article, we propose HAMT which is a general framework for simultaneously testing composite null hypotheses under heteroskedasticity. HAMT avoids data reduction by standardization and directly incorporates the side information from the variances into the testing procedure. It ranks the hypotheses using Clfdr statistics that rely on a carefully designed deconvolution estimator that captures the dependence between $\mu_i$ and $\sigma_i$. Our asymptotic analysis establishes that HAMT is valid and optimal for FDR control. In the numerical experiments, HAMT demonstrates substantial power gain against competing methods, particularly in the settings where $\mu_i$ and $\sigma_i$ are correlated.

\red{We conclude this article with a brief discussion on the limitations of HAMT and potential areas for future research.  
\textit{First}, it is of tremendous interest to develop powerful and valid multiple testing procedures that can pool side information from several covariate sequences (see for example \cite{chao2021adapt,zhang2022covariate} and the references therein). %Some very recent works in this direction include  among others. 
In the context of testing composite null hypotheses, HAMT can handle just one such sequence given by the $\sigma_i$'s and it is desirable to develop methods that can incorporate other side information, such as a grouping structure, in addition to heteroskedasticity. Given a $p-$dimensional side information vector $\bm Y_i\in\mathbb R^p$ for hypothesis $i$, the hierarchical Model \eqref{eq:model1}--\eqref{eq:model2} may be modified as follows: 
$$X_i=\mu_i+\sigma_i\epsilon_i,~\epsilon_i\stackrel{i.i.d.}{\sim}\eta(\cdot),~	\mu_i\mid(\sigma_i,\bm y_i)\stackrel{ind.}{\sim}g_\mu(\cdot|\sigma_i,\bm y_i),~(\sigma_i,\bm Y_i)\stackrel{i.i.d}{\sim}g_{\sigma,\bm y}(\cdot),$$
where $g_\mu(\cdot\mid\sigma_i,\bm y_i)$ and $g_{\sigma,\bm y}(\cdot)$ are, respectively, the probability density functions of the unknown mixing distributions of $\mu$ given $(\sigma_i,\bm y_i)$ and $(\sigma_i,\bm Y_i)$. A major methodological challenge towards extending HAMT in this direction will be to develop a reliable deconvolution estimator of $g_\mu(\cdot\mid\sigma_i,\bm y_i)$ for constructing the Clfdr statistic. This is especially important because for $p\ge 3$, the bivariate kernel density estimator of \cite{fu2022heteroscedasticity} may no longer provide a stable pilot estimate of the marginal density $f(x_i|\sigma_i,\bm y_i)$. \textit{Second}, our testing framework assumes that $\sigma_i$ are known and while numerical experiments in Section \ref{sec:sims_unknownvar} of the supplement show that using sample variances HAMT still controls the FDR level, it would be of great interest to further study the impact of estimating $\sigma_i$ on the power and validity of multiple testing procedures. In particular, the bandwidths $(h_{xj},h_\sigma)$ used in the pilot density estimator $\hat{\varphi}^m(\cdot,\sigma_i)$ depend on $\sigma_i$ and play an important role in establishing the consistency of $\hat{f}^m(\cdot|\sigma_i)$. When $\sigma_i$ are unknown, a natural alternative is to simply use the sample variance to determine the bandwidths but the corresponding density estimator may be highly unstable at the tails under such a choice, especially when the number of replicates available for each hypothesis is relatively small. Another strategy is to use variance estimates from empirical Bayes methods, such as those proposed in \cite{lu2019empirical,banerjee2020nonparametric}, however much research is needed to understand the theoretical properties of the resulting pilot density estimator. \textit{Finally}, while HAMT is guaranteed to provide asymptotic FDR control, it will be of interest to modify HAMT so that it can provably control FDR in finite samples, particularly when additional covariate sequences $\bm Y_i$ are available. Promising ideas in this direction include the construction of knockoffs or mirror sequences as done in \cite{barber2015controlling, leung2021zap}, or the use of conformal techniques as pursued in \cite{Batetal21, guan2022prediction}. } 
\section*{Acknowledgement}
%The authors thank the Editor, the Associate Editor and two anonymous referees whose comments and suggestions have improved the quality of this article. %for introducing them to this topic and for several interesting discussions around multiple testing. The authors also thank Yongyao Shi for helping with coding.
%The authors thank Professor Wenguang Sun for introducing them to this topic and for several interesting discussions around multiple testing. The authors also thank Yongyao Shi for helping with coding. 
B. Gang's research was supported by National Natural Science Foundation of China grant 12201123. T. Banerjee was partially supported by the University of Kansas General Research Fund allocation \#2302216. 
\newpage
\appendix
\begin{center}
	\Large{Supplement to ``Large-Scale Multiple Testing of Composite Null Hypotheses Under Heteroskedasticity''}
\end{center}
This supplement is organized as follows: the calculations for Examples 1 and 2 in Section \ref{sec:standardization} are presented in Section \ref{app:calculations}. The proofs of all other theoretical results in the paper are presented in Section \ref{app:proofs}. {Further details related to the implementation of HAMT, NPMLE B, ASH and ASH 1 are available in Section \ref{sec:npmleb_details}} and additional numerical experiments are provided in Section \ref{sec:more_sims}. {Section \ref{sec:more_discuss} includes more discussion on the empirical performances of the various methods considered in this article.} A real data application is discussed in Section \ref{sec:realdata}. 
%\appendixone
\section{Calculations for Section \ref{sec:standardization}}
\label{app:calculations}
\noindent\textbf{Example 1 - } we first consider the oracle rule based on the standardized statistic $Z_i=X_i/\sigma_i$.  The marginal density function of $Z_i$ under the alternative is 
$$
f_a(z)=\int_{0.5}^{4}\frac{1}{3.5\sqrt{2\pi}}\exp\left\{  -\frac{(z-\sqrt{\sigma})^2}{2}\right\}d\sigma,
$$
and the distribution function of $Z_i$ under the alternative is 
$$
F_a(t)=P(Z<t)=\int_{-\infty}^{t}f_a(z)dz=\int_{0.5}^{4}\frac{1}{3.5}\Phi(t-\sqrt{\sigma})d\sigma,
$$
where $\Phi$ is the distribution function of $N(0,1)$.
Then, using the definition of mFDR, it is not hard to see that the oracle procedure based on $\bm Z=(Z_1,\ldots,Z_m)$ is of the form $\bm{\delta}^{\sf ZOR}(t_z)=\{I(Z_i>t_z):1\le i\le m\}$ where,
$$
t_z=\inf\left\{ t>0: \dfrac{0.9\{1-\Phi(t)\}}{0.9\{1-\Phi(t)\}+0.1\{1-F_a(t)\}}\leq \alpha\right\}.
$$
When $\alpha=0.1$, the above display can be solved numerically for $t$ to get $t_z=3.273$ and the power of $\bm{\delta}^{\sf ZOR}(t_z)$ is $1-F_a(t_z)=0.0432$.

Next, consider the oracle rule $\bm{\delta}^{\sf OR}(t^*)$. Recall that $\bm{\delta}^{\sf OR}(t^*)$ is of the form $\{I(T_i^{\sf OR}<t^*):1\le i\le m\}$. Using the definition of Clfdr in Equation \eqref{lfdr_orc}, it is straightforward to show that this rule is equivalent to $\{I(Z_i>\lambda_\sigma(t^*)):1\le i\le m\}$, where
$$
\lambda_\sigma(t)=\dfrac{-\log(\frac{0.1t}{0.9(1-t)}) +\frac{1}{2}\sigma}{\sqrt{\sigma}},
$$
and 
$$
t^*=\sup \left[ t\in [0,1]:\dfrac{0.9\int( 1-\Phi \{\lambda_\sigma(t)\})\mathrm d\sigma}{0.9\int (1-\Phi \{\lambda_\sigma(t)\})\mathrm d\sigma +0.1\int( 1-\Phi\{\lambda_\sigma(t)-\sqrt{\sigma} \})\mathrm d\sigma}  \leq \alpha\right].
$$
When $\alpha=0.1$, the above display can be solved numerically to get  $t^*=0.177$ and the power of $\bm{\delta}^{\sf OR}(t^*)$ is given by
$(1/3.5)\int(1-\Phi\{\lambda_\sigma(t^*)-\sqrt{\sigma} \})\mathrm d\sigma=0.0611$.\hfill$\blacksquare$
\\~\\
\textbf{Example 2 - }for the oracle rule based on $Z_i$, the calculations from Example 1 give $t_z=4.124$ at $\alpha=0.1$ and the power of $\bm \delta^{\sf ZOR}(t_z)=0.0015$. Now, consider the oracle rule based on $T_i^{\sf OR}$. Note that $T_i^{\sf OR} =1$ if $\sigma_i\le 3.65$ and $0$ otherwise. So, $T_i^{\sf OR}$ perfectly classifies each case as being null or non-null based on $(X_i,\sigma_i)$. Consequently, the power of this procedure is $1$ while the FDR is $0$.
%\appendixtwo
\section{Proofs}
\label{app:proofs}
\subsection{Proof of Theorem \ref{thm_orcale}}
We divide the proof into two parts. In Part (a), we establish two properties of the testing rule $\pmb\delta^{\sf OR}(t)=\{I (T^{\sf OR}_i<t): 1 \leq i \leq m\}$ for an arbitrary $0<t<1$. In Part (b) we show that the oracle rule $\pmb\delta^{\sf OR}(t^*)$ attains the mFDR level exactly and is optimal amongst all mFDR procedures at level $\alpha$.

\medskip

\noindent\textbf{Part (a).} Denote $\alpha(t)$ the mFDR level of $\bm\delta^{\sf OR}(t)$. We shall show that (i) $\alpha(t) < t$ for all $0<t<1$ and that (ii) $\alpha(t)$ is nondecreasing in $t$. 

First, note that $E\left\{\sum_{i=1}^{m}(1-\theta_i)\delta^{\sf OR}_i(t)\right\}= E_{\bm{X, \sigma}}\{\sum_{i=1}^{m}T^{\sf OR}_i\delta^{\sf OR}_i(t)\}$. Then, according to the definition of $\alpha(t)$, we have
\begin{equation}\label{eq:rewritemFdr}
	E_{\bm{X, \sigma}}\left\{\sum_{i=1}^{m}\left\{T^{\sf OR}_i - \alpha(t)\right\}I(T^{\sf OR}_i \leq t)\right\} = 0.
\end{equation}
We claim that $\alpha(t)< t$. Otherwise if $\alpha(t)\geq t$, then we must have  $T^{\sf OR}_i < t \leq \alpha(t)$. It follows that the LHS must be negative, contradicting \eqref{eq:rewritemFdr}.

%To see this, consider that all non--zero terms with $T_i \leq t$. Consider three possible situations: (i) $\alpha \leq  T_i  < t$, (ii)  $T_i \leq \alpha < t$, and (iii) $T_i < t \leq \alpha$. Notice (i) produces zero or positive terms on the LHS of \eqref{eq:rewritemFdr}, (ii) produces zero or negative terms, and (iii) produces negative terms. If $\alpha_t \geq t$, then only (iii) is possible, which contradicts the RHS.

Next we show (ii), i.e, $\alpha(t)$ is nondecreasing in $t$. Let $\alpha(t_j) = \alpha_j$. We claim that if $t_1 < t_2$, then we must have $\alpha_{1} \leq \alpha_{2}$. We argue by contradiction. Suppose that $t_1 < t_2$ but $\alpha_1 > \alpha_2$. Then 	
\begin{eqnarray*}
	(T^{\sf OR}_i - \alpha_2) I(T^{\sf OR}_i < t_2) & = & (T^{\sf OR}_i - \alpha_1) I(T^{\sf OR}_i < t_1)   +  (\alpha_1 - \alpha_2)  I(T^{\sf OR}_i < t_1)\\
	&&+ (T^{\sf OR}_i - \alpha_2)  I(t_1 \leq T^{\sf OR}_i < t_2) \\
	& \geq & (T^{\sf OR}_i - \alpha_1)  I(T^{\sf OR}_i < t_1) + (\alpha_1 - \alpha_2)  I(T^{\sf OR}_i < t_1)\\
	&&+ (T^{\sf OR}_i - \alpha_1)  I(t_1 \leq T^{\sf OR}_i < t_2).
\end{eqnarray*}
It follows that $E \left\{\sum_{i=1}^{m}(T^{\sf OR}_i - \alpha_2) I(T^{\sf OR}_i < t_2) \right\} > 0$ since $E \left\{\sum_{i=1}^{m}(T^{\sf OR}_i - \alpha_1) I(T^{\sf OR}_i < t_1) \right\}=0$ according to \eqref{eq:rewritemFdr}, $\alpha_1 >\alpha_2$ and $T^{\sf OR}_i\geq t_1>\alpha_1$. However, this contradicts Equation \eqref{eq:rewritemFdr} and so we must have $\alpha_1 < \alpha_2.$

\medskip

\noindent\textbf{Part (b).} Let $\bar{\alpha} = \alpha(1)$. In Part (a), we showed that $\alpha(t)$ is non--decreasing in $t$. It follows that for all $\alpha < \bar{\alpha}$, there exists a $t^*$ such that $t^*=\sup\{t: \alpha(t^*) = \alpha\}$. By definition, $t^*$ is the oracle threshold. Consider an arbitrary decision rule $\bm \delta= (\delta_{1}, \ldots, \delta_{m})\in \{0, 1\}^m$ such that $\mbox{mFDR}(\bm \delta) \leq \alpha$. We have $\mathbb {E} \left\{\sum_{i=1}^{m}(T^{\sf OR}_i - \alpha) \delta^{\sf OR}_i(t^*) \right\} = 0 \text{ and } E \left\{\sum_{i=1}^{m}(T^{\sf OR}_i - \alpha) \delta_i \right\} \leq 0.$ Hence
\begin{equation}\label{eq:or.power1}
	E \Big[\sum_{i=1}^{m}\{\delta^{\sf OR}_i(t^*) - \delta_i\}(T^{\sf OR}_i - \alpha)  \Big] \geq 0.
\end{equation}
Consider the transformation $h(x) = (x-\alpha)/(1-x)$. Note that since $h(x)$ is monotone, we can rewrite  $\delta^{\sf OR}_i(t^*) = I\left[\left\{ (T^{\sf OR}_i - \alpha)/(1-T^{\sf OR}_i)\right\} < \lambda\right]$, where $\lambda =(t^* - \alpha)/(1-t^*)$. In Part (a) we have shown that $\alpha < t^* < 1$, which implies that $\lambda > 0$. Hence
\begin{equation}\label{eq:or.power2}
	E\left[\sum_{i=1}^{m}\left\{\delta^{\sf OR}_i(t^*)-\delta_i\right\}\left\{(T^{\sf OR}_i-\alpha) - \lambda(1-T^{\sf OR}_i)\right\}\right] \leq 0.
\end{equation}
To see this, consider the terms where $\delta^{\sf OR}_i(t^*)-\delta_i \neq 0$. Then we have two cases: (i) $\delta^{\sf OR}_i(t^*) > \delta_i$ or (ii) $\delta^{\sf OR}_i < \delta_i$. In case (i), $\delta^{\sf OR}_i(t^*) = 1$, implying that $\left\{(T^{\sf OR}_i - \alpha)/(1-T^{\sf OR}_i)\right\} < \lambda$. In case (ii), $\delta^{\sf OR}_i(t^*) = 0$, implying that $\left\{(T^{\sf OR}_i - \alpha)/(1-T^{\sf OR}_i)\right\} \geq \lambda$. Therefore, we always have $\{\delta^{\sf OR}_i(t^*)-\delta_i\}\{(T^{\sf OR}_i-\alpha) - \lambda(1-T^{\sf OR}_i)\} \leq 0$. Summing over the $m$ terms and taking the expectation yield \eqref{eq:or.power2}. 

Now, combining \eqref{eq:or.power1} and \eqref{eq:or.power2}, we obtain
$$
0 \leq E \left[\sum_{i=1}^{m}\{\delta^{\sf OR}_i(t^*) - \delta_i\}(T^{\sf OR}_i - \alpha)  \right]
\leq \lambda E \left[\sum_{i=1}^{m}\{\delta^{\sf OR}_i(t^*) - \delta_i\}(T^{\sf OR}_i - \alpha)  \right].
$$
Since $\lambda > 0$, it follows that $E \left[\sum_{i=1}^{m}\{\delta^{\sf OR}_i(t^*) - \delta_i\}(T^{\sf OR}_i - \alpha)  \right] > 0$. Finally, we apply the definition of ETP to conclude that $\mbox{ETP}\{\bm{\delta}^{\sf OR}(t^*)\} \geq \mbox{ETP}(\bm \delta)$ for all $\bm \delta\in\{0,1\}^m$ such that $\mbox{mFDR}(\bm \delta) \leq \alpha$.\hfill$\blacksquare$

\subsection{Proof of Proposition \ref{prop1}}
\label{app:prf_prop1}
We first state two useful lemmata where $\delta_{u}(\cdot)$ denotes a point mass at $u$
\begin{lemma}
	\label{lem1}
	Let $\eta_\sigma(\cdot)$ be the density function of $\sigma \gamma$ where $\gamma\sim \eta(\cdot)$ and suppose assumption (A4) holds. For any $g$ with support $\text{supp}(g)\subset[-M,M]$, and any $\epsilon>0$, $\tau>0$, with $S$ large enough (depending on $M, \epsilon,\tau$ only), there exists $g'\in \{\sum_{j=1}^{S}\theta_j\delta_{u_j}(\cdot)|\sum_{j=1}^{S}\theta_j=1,\ \theta_j\geq 0 \ \forall j  \}$ with $u_j=-M+2M(j-1)/(S-1)$ such that $|g*\eta_\tau(x)-g'*\eta_\tau(x)|^2<\epsilon$ for all $x$.
\end{lemma}
\begin{lemma}
	\label{lem2}
	Suppose $\hat{f}(x|\sigma)=\hat{g}*\eta_\sigma(x)$ and $f(x|\sigma)=g*\eta_\sigma(x)$, where $\eta_\sigma$ is as defined in Lemma \ref{lem1} and satisfies the assumptions therein. 
Then $E_{\pmb{x},\pmb{\sigma}}E_{x,\sigma}|\hat{f}(x|\sigma)-f(x|\sigma)|^2\rightarrow0 $ implies $E_\tau E_{\pmb{x},\pmb{\sigma}}\|\hat{g}*\eta_\tau-g*\eta_\tau\|^2_2\rightarrow 0$. Here $E_{\pmb{x},\pmb{\sigma}}$ is taken with respect to the data used to construct $\hat{f}$ and $\hat{g}$, $E_{x,\sigma}$ is taken with respect to the input for $\hat{f}$ and $f$, $E_\tau$ is taken with respect to $\tau\sim g_\sigma(\cdot)$.
\end{lemma}
Let $\tilde{f}(x_i|\sigma_i)$ be any bounded estimator of ${f}(x_i|\sigma_i)$.
	We begin by establishing 
 \begin{equation}\label{pf:1}
			\frac{1}{m}\sum_{i=1}^{m}\{\tilde{f}(x_i|\sigma_i)-\hat{\varphi}^m_{(-i)}(x_i,\sigma_i)\}^2 \overset{p}{\rightarrow} \frac{1}{m}\sum_{i=1}^{m}\{\tilde{f}(x_i|\sigma_i)-{f}(x_i|\sigma_i)\}^2 .
		\end{equation}
		Note that
		\begin{align*}
			&\frac{1}{m}\sum_{i=1}^{m}\{\tilde{f}(x_i|\sigma_i)-{f}(x_i|\sigma_i)\}^2\\
			=&\frac{1}{m}\sum_{i=1}^{m}\{\tilde{f}(x_i|\sigma_i)-\hat{\varphi}^m_{(-i)}(x_i,\sigma_i)+\hat{\varphi}^m_{(-i)}(x_i,\sigma_i)-{f}(x_i|\sigma_i)\}^2\\
			=&	\frac{1}{m}\sum_{i=1}^{m}\{\tilde{f}(x_i|\sigma_i)-\hat{\varphi}^m_{(-i)}(x_i,\sigma_i)\}^2 +\frac{2}{m}\sum_{i=1}^{m}\{\tilde{f}(x_i|\sigma_i)-\hat{\varphi}^m_{(-i)}(x_i,\sigma_i)\}\{\hat{\varphi}^m_{(-i)}(x_i,\sigma_i)-{f}(x_i|\sigma_i)\}\\
			+&\frac{1}{m}\sum_{i=1}^{m}\{\hat{\varphi}^m_{(-i)}(x_i,\sigma_i)-{f}(x_i|\sigma_i)\}^2
		\end{align*}
		Hence, to prove Equation \eqref{pf:1}, we only need to establish
		\begin{equation}\label{eq:2}
			\frac{2}{m}\sum_{i=1}^{m}\{\tilde{f}(x_i|\sigma_i)-\hat{\varphi}^m_{(-i)}(x_i,\sigma_i)\}\{\hat{\varphi}^m_{(-i)}(x_i,\sigma_i)-{f}(x_i|\sigma_i)\} \overset{p}{\rightarrow} 0,
		\end{equation}
		and
		\begin{equation}\label{eq:3}
			\frac{1}{m}\sum_{i=1}^{m}\{\hat{\varphi}^m_{(-i)}(x_i,\sigma_i)- {f}(x_i|\sigma_i)  \}^2\overset{p}{\rightarrow} 0.
		\end{equation}
			Since \rred{by Assumption (A2)} $\text{supp}(g_\sigma)\in [M_1,M_2]$ for some fixed $M_1>0, M_2<\infty$, $f(x_i|\sigma_i)=\int g_\mu (\mu)\phi_\sigma(x_i-\mu) d\mu $ is bounded by some fixed constant $C>0$. By capping $\tilde{f}$ and \rred{$\hat{\varphi}^m_{(-i)}$} at $C$ we assume, without loss of generality, that $|\tilde{f}(x_i|\sigma_i)-\rred{\hat{\varphi}^m_{(-i)}(x_i,\sigma_i)}|\leq C$. Hence, Equation \eqref{eq:2} is implied by
		\begin{equation}\label{eq:4}
			\frac{1}{m}\sum_{i=1}^{m}|\hat{\varphi}^m_{(-i)}(x_i,\sigma_i)- {f}(x_i|\sigma_i)  |\overset{p}{\rightarrow} 0.
		\end{equation}		
To prove equations \eqref{pf:1} and \eqref{eq:4} we borrow relevant results from the theory of kernel regression. Recall that for the following regression model
		\begin{equation*}\label{eq:nw}
			Y_i=m(\sigma_i)+\epsilon_i,
		\end{equation*}
the Nadaraya-Watson kernel estimator, upon observing $\{(\sigma_i,Y_i)\}_{i=1}^m$, is defined as
		\begin{equation}\label{eq:nw2}
			\hat{m}(\sigma)=\sum_{j=1 }^m\dfrac{\phi_{h_\sigma}(\sigma-\sigma_j)}{\sum_{k=1 }^m\phi_{h_\sigma}(\sigma-\sigma_k)}Y_j.
		\end{equation}
		It is well known that (Theorem 5.44 in \cite{wasserman2006all}) the estimator in Equation \eqref{eq:nw2} satisfies
		\begin{equation}\label{eq:b}
			E\{\hat{m}(\sigma)-m(\sigma)\}\asymp h_\sigma^2\left(m''(\sigma)+2\frac{m'(\sigma)g'_\sigma(\sigma)}{g_\sigma(\sigma)}\right),
		\end{equation}
		and 
		\begin{equation*}\label{eq:var}
			\text{Var}\{\hat{m}(\sigma)\}\asymp \frac{\gamma^2}{mh_\sigma}\{g_\sigma(\sigma)\}^{-1},
		\end{equation*}		
where $\gamma^2$ is an upper bound on the variance of $\epsilon_i$. To use this result, for a given $(x,\sigma)$ we simply take 
		\begin{equation*}\label{eq:bias}
			m(\sigma_j)=E\{\phi_{h_{x}\sigma_j}(x-x_j)\},\quad \text{and}\quad \epsilon_i=\phi_{h_{x}\sigma_j}(x-x_j)-E\{\phi_{h_{x}\sigma_j}(x-x_j)\},
		\end{equation*}
		where the expectation above is taken with respect to $x_j$ under the conditional density function $f(x|\sigma_j):=\int \eta_{\sigma_j}(x-\mu)g_\mu(\mu|\sigma_j)d\mu$ and $\eta_{\sigma_j}$ is the density function of $\sigma_j\epsilon_j$.
		Note that $m(\sigma_j)$ can be viewed as the density function of $\mu+\sigma_j\epsilon'_1+h_x\sigma_j\epsilon'_2$ evaluated at $x$, where $\mu\sim g_\mu(\cdot | \sigma_j)$, $\epsilon_1'\sim \eta(\cdot)$ and $\epsilon'_2\sim N(0,1)$ and $\epsilon_1', \epsilon_2'$ are independent.
	Since by Assumption (A1) $|\frac{\partial}{\partial \sigma}g_\mu(\mu|\sigma)|$ and $|\frac{\partial^2}{\partial \sigma^2}g_\mu(\mu|\sigma)|$ are bounded for all $\mu$, and $\eta(\cdot)$ is smooth by Assumption (A4), some elementary calculation shows $|m'(\sigma)|<C'$ and $|m''(\sigma)|<C''$ for some constants $C'$ and $C''$.
Furthermore, by Assumption (A2) we have
	 $|g'_\sigma(\sigma)|\leq C_3$ and $g_\sigma(\sigma)>C_1>0$ on $\text{supp}(g_\sigma)$. So the RHS of Equation \eqref{eq:b} has order $h_\sigma^2$. 
		
Next, observe that the density function of  $\sigma_j\epsilon'_1$ evaluated at $x$ is $\eta_{\sigma_j}(x)$ and the density function of $\sigma_j\epsilon'_1+h_x\sigma_j\epsilon'_2$ evaluated at $x$ is $\int \eta_{\sigma_j}(x-\epsilon)\phi_{h_x\sigma_j}(\epsilon)dx$. Some elementary calculation shows (page 20 of \cite{wand1994kernel}) 
		$$
		\Big|\int \eta_{\sigma_j}(x-\epsilon)\phi_{h_x\sigma_j}(\epsilon)dx-\eta(x)\Big|\asymp h^2_x.
		$$
		Hence, 	$\frac{1}{m}\sum_{i=1}^{m}|\hat{\varphi}^m_{(-i)}(x_i,\sigma_i)- {f}(x_i|\sigma_i)  |=O_p(h_x^2+h_\sigma^2).$ This establishes Equation \eqref{eq:4}. Using standard result from density estimation theory (page 21 of \cite{wand1994kernel}) we see that variance of $\epsilon_i$ is of order $h_x^{-1}$. Hence, we also have 
		$$
		\frac{1}{m}\sum_{i=1}^{m}\{\hat{\varphi}^m_{(-i)}(x_i,\sigma_i)- {f}(x_i|\sigma_i)  \}^2=O_p\left(h_x^4+h_\sigma^4+\frac{1}{mh_xh_\sigma}\right).
		$$
		By Assumption (A3) this establishes Equation \eqref{eq:3} and, hence, Equation \eqref{pf:1} follows.		
		
Next, for any $\epsilon>0$, since $\text{supp}\{g_\mu(\cdot|\sigma)\}\subset[-M,M]$ and $g_\mu(\cdot|\sigma)$ is continuous in $\sigma$, by Lemma \ref{lem1} there exists continuous functions $g_j,\ j=1,\ldots,S$ such that $g'_\mu(\cdot|\sigma)=\sum_{j=1}^{S}g_j(\sigma)\delta_{u_j}(\cdot)$ and $|g'_\mu(\cdot|\sigma)*\eta_\tau(x)-g_\mu(\cdot|\sigma)*\eta_\tau(x)|^2<\epsilon$ for all $x$. 
	
	Let $\{q_k\}_{k=1}^\infty$ be an orthonormal basis for $L^2[M_1,M_2]$. Since $g_j$'s are bounded and continuous they all belongs to $L^2[M_1, M_2]$, hence they can be written as $g_j(\sigma)=\sum_{k=1}^{\infty}w_{jk}q_k(\sigma)$. For each $g_j$ there exists $N_j$ such that we can find $\tilde{w}_{jk}$ with $ \|  g_j(\cdot)-\sum_{k=1}^{N_j}\tilde{w}_{jk}q_k(\cdot
	)\|^2_2<\epsilon/S.$ 
	Take $K=\max_j N_j$. Then, there exists $\tilde{w}_{jk}, j=1,\ldots,S,\ k=1,\ldots, K$, such that  $\|  g_j(\cdot)-\sum_{k=1}^{K}\tilde{w}_{jk}q_k(\cdot
	)\|^2_2<\epsilon/S$ for all $j$. Write $\tilde{g}_j(\cdot)= \sum_{k=1}^{K}\tilde{w}_{jk}q_k(\cdot
	)$. Let $\tilde{g}_\mu(\cdot|\sigma)=\sum_{j=1}^{K}\tilde{g}_j(\sigma)\delta_{u_j}(\cdot)$. Then for every $\sigma>0$ and any fixed $\tau>0$ we have 
	\begin{align*}
		\|g'_\mu(\cdot|\sigma)*\eta_\tau-\tilde{g}_\mu(\cdot|\sigma)*\eta_\tau \|^2_2&=\| \sum_{j=1}^{S} (\tilde{g}_j(\sigma)-g_j(\sigma))\eta_\tau(\cdot-u_j)\|^2_2=O(\epsilon).
	\end{align*}
	Hence, in the feasible region, it is possible to find $\tilde{f}$ such that
	$$
	\frac{1}{m}\sum_{i=1}^{m}\{\tilde{f}(x_i|\sigma_i)-{f}(x_i|\sigma_i)\}^2 \leq \epsilon.
	$$
	By Equation \eqref{pf:1} we also have with high probability $\frac{1}{m}\sum_{i=1}^{m}\{\tilde{f}(x_i|\sigma_i)-\rred{\hat{\varphi}^m_{(-i)}(x_i,\sigma_i)}\}^2 \overset{}{\rightarrow}0$. Hence, with high probability it is possible to find $\tilde{f}$ in the feasible region such that $\frac{1}{m}\sum_{i=1}^{m}\{\tilde{f}(x_i|\sigma_i)-\rred{\hat{\varphi}^m_{(-i)}(x_i,\sigma_i)}\}^2 \overset{}{\rightarrow}0$. By definition, 
	$$
	\hat{f}^m\in \argmin_{\tilde{f}} \sum_{i=1}^{m}\Bigl\{\rred{\hat{\varphi}^m_{(-i)}(x_i,\sigma_i)}-\tilde{f}(x_i\mid\sigma_i)\Bigr\}^2.
	$$
	Thus, we necessarily have $\frac{1}{m}\sum_{i=1}^{m}\{\hat{f}^m(x_i|\sigma_i)-\rred{\hat{\varphi}^m_{(-i)}(x_i,\sigma_i)}\}^2 \overset{p}{\rightarrow} 0$ and by Equation \eqref{pf:1} $\frac{1}{m}\sum_{i=1}^{m}\{\hat{f}^m(x_i|\sigma_i)-{f}(x_i|\sigma_i)\}^2\overset{p}{\rightarrow} 0$. The proposition then follows directly from Lemma \ref{lem2}.
%Note that the objective function in \eqref{cvx} converges to $E|\hat{f}_\sigma(x)-\hat{f}^m_\sigma(x)|^2$ (here $x$ and $\sigma$ are also regarded as random).

\hfill$\blacksquare$

\begin{remark} 
\label{gridsize}
		(Grid Size)
Note that $$
\frac{1}{m}\sum_{i=1}^{m}\{\hat{f}^m(x_i|\sigma_i)-{f}(x_i|\sigma_i)\}^2 =O(E\|\hat{f}^m-f\|_2^2)= O\{(mh_xh_\sigma)^{-1}+h^4_x+h^4_\sigma\}.
$$
The optimal rate of $(mh_xh_\sigma)^{-1}+h^4_x+h^4_\sigma$ is $m^{-2/3}$  and is achieved when $h_x\asymp h_\sigma \asymp m^{-1/6}$.
%In the proof of Lemma \ref{lem1} we used $\frac{1}{m}\sum_{j=1}^{m}\phi_\tau(x-u_{i(j)})$ to approximate $\hat{f}(x):=\frac{1}{m}\sum_{j=1}^{m}\phi_\tau(x-\mu_{j})$ where $\mu_j\overset{iid}{\sim } g$ for $j=1,\ldots,m$. We have also shown $E_{\pmb{\mu}}|f(x)-\hat{f}(x) |^2=O(1/m)$ where $f(x)=\int \phi_{\tau}(x-\mu)g(\mu)d\mu$. 
Hence, when choosing the grid size we only need 
$$
\big|\dfrac{1}{m}\sum_{j=1}^{m}\eta_\tau(x-\mu_j)-\dfrac{1}{m}\sum_{j=1}^{m}\eta_\tau(x-u_{i(j)})\big|^2=O(m^{-2/3}),
$$
where $u_{i(j)}\in \{u_1,\ldots, u_S\}$ is such that $|u_{i(j)}-\mu_{j}|=O(1/S)$.
Since $g_\mu(\cdot|
\sigma)$ has bounded support, such $u_{i(j)}$ can always be found. Let $\epsilon=|u_{i(j)}-\mu_{j}|$, then when $\eta_\tau(\cdot)$ is the density function of $N(0,\tau^2)$ we have
\begin{align}\label{rateeq}
	|\eta_\tau(x-\mu_j)-\eta_\tau(x-u_{i(j)})|^2&=\dfrac{1}{2\pi\tau^2}e^{-\frac{x^2}{\tau^2}}|1-e^{\frac{2x\epsilon-\epsilon^2}{2\tau^2}}  |^2.
\end{align} 
We want the above to be of order $O(m^{-2/3})$ uniformly for any $x$. If $x$ has order greater than $\sqrt{\log m}$ then the RHS of \eqref{rateeq} is $O(m^{-2/3})$. When $x$ has order less than $\sqrt{\log m}$, since $e^{-\frac{x^2}{\tau^2}}=O(1)$, we focus on $|1-e^{\frac{2x\epsilon-\epsilon^2}{2\tau^2}}  |^2.$ By Taylor expansion, 
\begin{equation*}\label{eq2}
	|1-e^{\frac{2x\epsilon-\epsilon^2}{2\tau^2}}  |^2=O\left\{ \left(\frac{2x\epsilon-\epsilon^2}{2\tau^2}\right)^2\right\}.
\end{equation*}
If $\epsilon=O(m^{-1/3}(\log m)^{-1/2})$ then the above is $O(m^{-2/3})$, and it follows that the grid size of $S(m)\asymp m^{1/3}(\log m)^{1/2}$ is sufficient.\hfill$\blacksquare$
\end{remark}
\begin{remark}
	\label{basis}
		In the proof of Proposition \ref{prop1} we used the fact that  for each $g_j$ there exists $N_j$ such that we can find $\tilde{w}_{jk}$ with $ \|  g_j(\cdot)-\sum_{k=1}^{N_j}\tilde{w}_{jk}q_k(\cdot
)\|^2_2<\epsilon/S.$ If we take $\tilde{w}_{ji}=w_{ji}$ then  
$$\|  g_j(\cdot)-\sum_{k=1}^{N_j}\tilde{w}_{jk}q_k(\cdot
)\|^2_2=\sum_{i=N_j}^{\infty} w^2_{ji}.
$$
Since $\{w_{ji}\}_{i=1}^\infty\in \Theta(\gamma,c)$ we have
$$
\sum_{i=N_j}^{\infty} w^2_{ji}=o\left( \int_{N_j}^{\infty} x^{-2\gamma-1} dx\right)=O(N^{-2\gamma}_j).
$$
Hence, for $\|g_j(\cdot)-\sum_{k=1}^{N_j}w_{jk}q_k(\cdot
)\|^2_2<\epsilon/S$ we only need $N_j^{2\gamma}>S/\epsilon$. Since we have argued that the order of $S$ does not have to be larger than $m^{1/3}(\log m)^{1/2}$, if we take $\epsilon$ to be of order $m^{-2/3}$  then the order of  $K(m)=\max_jN_j$ does not have to be larger than $m^{1/(2\gamma)}(\log m)^{1/(4\gamma)}$. \hfill$\blacksquare$
\end{remark}

\subsection{Proof of Corollary \ref{coro1}}
Note that $f(\cdot|\sigma)$ is continuous, then there exists $K_1=[-M,M]$ such that $P(x\in K^c_1)\rightarrow0$ as $M\rightarrow\infty$. Let $\inf_{x\in K_1}f(x|\sigma)=l_0$ and $A_{l_0}=\{x:|\hat{f}^m(x|\sigma)-f(x|\sigma)|\geq l_0/2\}$. Since $E\int| \hat{f}^m(x|\sigma)-f(x|\sigma) |^2dx\geq (l_0/2)^2P(A_{l_0})$; it follows that $P(A_{l_0})\rightarrow 0$. Thus $\hat{f}^m(\cdot|\sigma)$ and $f(\cdot|\sigma)$ are bounded below by a positive number for large $m,S,K$ except for an event that has a low probability. Similar arguments can be applied to the upper bound of $\hat{f}^m(\cdot|\sigma)$ and $f(\cdot|\sigma)$, as well as to the upper and lower bounds for $\hat{f}^m_{0}(\cdot|\sigma)$ and $f_{0}(\cdot|\sigma)$. Therefore, we conclude that $\hat{f}^m_{0}(\cdot|\sigma)$, $\hat{f}^m(\cdot|\sigma)$, $f_{0}(\cdot|\sigma)$ and $f(\cdot|\sigma)$. are all bounded in the interval $[l_a, l_b]$, $0<l_a<l_b<\infty$ for large $m,S,K$ except for an event, say $A_\epsilon$ that has low probability. 
Let $\hat{T}_m(x,\sigma)=\hat{f}^m_0(x|\sigma)/\hat{f}^m(x|\sigma)$ and $T^{\sf OR}(x,\sigma)=f_0(x|\sigma)/f(x|\sigma)$. We have $$\hat{T}_m(x,\sigma)-T^{\sf OR}(x,\sigma)=(\hat{f}^m_{0}(x|\sigma)f(x|\sigma)  -f_{0}(x|\sigma)\hat{f}^m(x|\sigma)   )/(\hat{f}^m(x|\sigma
)f(x|\sigma)).$$  It is easy to see that $( \hat{T}_m-T^{\sf OR} )^2$ is bounded by 1. Then
$$
E\{\hat{T}_m(x,\sigma)-T^{\sf OR}(x,\sigma)\}^2\leq P(A_{l_0})+c_1E\{ \hat{f}^m_{0}(x|\sigma)-f_0(x|\sigma)   \}^2+c_2E\{\hat{f}^m(x|\sigma)-f(x|\sigma) \}^2.
$$
Thus, $E\{\hat{T}_m(x,\sigma)-T^{\sf OR}(x,\sigma)\}^2\rightarrow 0.$ Let $B_\delta=\{x|\sigma: |\hat{T}_m(x,\sigma)- T^{\sf OR}(x,\sigma)|>\delta \}$. Then $\delta^2P(B_\delta)\leq E\{\hat{T}_m(x,\sigma)-T^{\sf OR}(x,\sigma)\}^2\rightarrow 0$, and the result follows.\hfill$\blacksquare$
\subsection{Proof of Theorem \ref{thm:dd}}
We begin with a summary of notation used throughout the proof:
%$\hat{T}_{i,m}{\rightarrow}T_i^{\sf OR}$,\newline  $ETP\{\bm \delta^{\sf HAMT}(\hat{t}_m)\}/ETP\{\bm \delta^{\sf OR}(t^*)\}=1+o(1)$.
\begin{itemize}
	\item $Q(t) = m^{-1}\sum_{i=1}^m (T_i^{\sf OR}-\alpha) I\{T_i^{\sf OR} < t\}$.
	\item $\hat{Q}(t) = m^{-1}\sum_{i=1}^m (\hat{T}_{i,m}-\alpha) I\{\hat{T}_{i,m} < t\}$.
	\item $Q_{\infty}(t) = E\{(T^{\sf OR}-\alpha)I\{T^{\sf OR}<t\}\}$.
	\item $t_{\infty} = \sup\{t \in (0,1): Q_{\infty}(t) \leq 0\}$: the ``ideal'' threshold.
\end{itemize}
For $\hat{T}_{(k),m} < t< \hat{T}_{(k+1),m}$, define a continuous version of $\hat{Q}(t)$ as
\[
\hat{Q}_{C}(t) = \frac{t-\hat{T}_{(k),m}}{\hat{T}_{(k+1),m} - \hat{T}_{(k),m}} \hat{Q}_{k} + \frac{\hat{T}_{(k+1),m}-t}{\hat{T}_{(k+1),m} - \hat{T}_{(k),m}} \hat{Q}_{k+1},
\]
where $\hat{Q}_{k} = \hat{Q}\left(\hat{T}_{(k),m}\right)$. Since $\hat{Q}_{C}(t)$ is continuous and monotone, its inverse $\hat{Q}_{C}^{ -1}$ is well--defined, continuous and monotone. Next we show the following two results in turn: (i) $\hat{Q}(t) \overset{p}\rightarrow Q_{\infty}(t)$ and (ii) $\hat{Q}_{C}^{-1}(0) \overset{p}\rightarrow t_{\infty}$.
To show (i), note that $Q(t) \overset{p}\rightarrow Q_{\infty}(t)$ 
by the WLLN, so that we only need to establish that  $\hat{Q}(t)-Q(t) \overset{p}\rightarrow 0$.

We need the following lemma, which is proven in Section \ref{prooflemma2}.
\begin{lemma}
	\label{lemma:2}
	Let $U_i = (T_i-\alpha) I(T_i < t)$ and $\hat{U}_i = (\hat T_i-\alpha)I\{\hat T_i < t\}$. Then $E\left(\hat{U}_i - U_i \right)^2 = o(1)$.
\end{lemma}
By Lemma \ref{lemma:2} and Cauchy-Schwartz inequality,  $E\left\{\left(\hat{U}_i-U_i\right)\left(\hat{U}_j-U_j\right)\right\} = o(1)$. Let 
$S_m = \sum_{i=1}^m \left(\hat{U}_i - U_i\right)$.

It follows that
$$
Var\left( m^{-1} S_m  \right) \leq m^{-2}\sum_{i=1}^{m} E\left\{ \left( \hat{U}_i - U_i\right)^2 \right\} +O\left(\frac{1}{m^2}\sum_{i,j:i\neq j} E\left\{\left(\hat{U}_i-U_i\right)\left(\hat{U}_j-U_j\right)\right\}\right)= o(1).
$$
By Corollary \ref{coro1}, $E(m^{-1}S_m)\rightarrow 0$, applying Chebyshev's inequality, we obtain $m^{-1}S_m = \hat{Q}(t) - Q(t) \overset{p}\rightarrow 0$. Hence (i) is proved. Notice that $Q_{\infty}(t)$ is continuous by construction, we also have $\hat{Q}(t)\overset{p}\rightarrow \hat{Q}_{C}(t)$.

Next we show (ii). Since  $\hat{Q}_{C}(t)$ is continuous, for any $\varepsilon>0$, we can find $\eta>0$ such that
$\left|\hat{Q}_{C}^{ -1}(0) - \hat{Q}_{C}^{ -1}\left\{\hat{Q}_{C}^{}\left(t_{\infty}^{}\right)\right\}\right| < \varepsilon$
if
$\left|\hat{Q}_{C}^{}\left(t_{\infty}^{}\right) \right|< \eta$. It follows that
\[
P\left\{ \left|\hat{Q}_{C}^{}\left(t_{\infty}^{}\right) \right|> \eta\right\} \geq P\left\{\left|\hat{Q}_{C}^{ -1}(0) - \hat{Q}_{C}^{ -1}\left\{\hat{Q}_{C}^{}\left(t_{\infty}^{}\right)\right\}\right| > \varepsilon\right\}.
\]
Corollary \ref{coro1} and the WLLN imply that $\hat{Q}_{C}^{}(t) \overset{p}\rightarrow Q_{\infty}^{}(t).$ Note that $Q_{\infty}^{}\left(t_{\infty}^{}\right) = 0$. Then $
P\left(\left|\hat{Q}_{C}^{}\left(t_{\infty}^{}\right) \right|>\eta\right) \rightarrow 0.$
Hence we have $
\hat{Q}_{C}^{-1}(0) \overset{p}\rightarrow \hat{Q}_{C}^{ -1} \left\{\hat{Q}_{C}\left(t_{\infty}\right)\right\} = t_{\infty}$, completing the proof of (ii).

To show FDR$(\bm \delta^{\sf HAMT}(\hat{t}_m)) =$ FDR$(\bm \delta^{\sf OR}(t^*)) + o(1) =\alpha + o(1)$, we only need to show mFDR$(\bm \delta^{\sf HAMT}(\hat{t}_m)) =$ mFDR$(\bm \delta^{\sf OR}(t^*)) + o(1)$. The result then follows from the asymptotic equivalence of FDR and mFDR, which was proven in \cite{basu2018weighted}.

Define the continuous version of $Q(t)$ as $Q_{C}(t)$ and the  corresponding threshold as $Q_{C}^{ -1}(0)$. Then by construction, we have
%$\hat{T}_{i,m}{\rightarrow}T_i^{\sf OR}$
\[
\bm \delta^{\sf HAMT}(\hat{t}_m)  = \left[I \left\{\hat T_{i,m} \leq \hat{Q}_{C}^{ -1}(0)\right\}: 1 \leq i \leq m\right] \quad \text{and} \quad \bm \delta^{\sf OR}(t^*)) = \left[I \left\{T_i\leq Q_{C}^{ -1}(0)\right\}: 1 \leq i \leq m\right].
\]
Following the previous arguments, we can show that $
Q_{C}^{ -1}(0) \overset{p}\rightarrow  t_{\infty}$. It follows that $\hat{Q}_{C}^{ -1}(0) = Q_{C}^{ -1}(0) + o_{p}(1).$ By construction $\mbox{mFDR}(\bm{\delta}^{\sf OR})=\alpha$. The mFDR level of $\bm{\delta}^{\sf HAMT}$ is
\[
\mbox{mFDR}(\bm{\delta}^{\sf HAMT}) = \frac{P_{H_0}\left\{\hat T_{i,m} \leq \hat{Q}_{C}^{ -1}(0)\right\}}{P\left\{\hat T_{i,m} \leq \hat{Q}_{C}^{ -1}(0)\right\}}.
\]
From Corollary \ref{coro1}, $\hat T_{i,m} \overset{p}\rightarrow T_i^{\sf OR}$. Using the continuous mapping theorem, $\mbox{mFDR}\left(\pmb{\delta}^{\sf HAMT}\right) = \mbox{mFDR}\left(\pmb{\delta}^{\sf OR}\right) + o(1)=\alpha+o(1)$. The desired result follows.

Finally, using the fact that  $\hat T_{i,m} \overset{p}\rightarrow T_i^{\sf OR}$ and $\hat{Q}_{C}^{ -1}(0)   \overset{p}\rightarrow Q_{C}^{ -1}(0)$, we can similarly show that $
\mbox{ETP}(\pmb{\delta}^{\sf HAMT})/\mbox{ETP}(\pmb{\delta}^{\sf OR}) =1 + o(1).
$\hfill$\blacksquare$

\subsection{Proof of Lemma \ref{lem1}}
Suppose $\mu_i\overset{iid}{\sim} g$, for $i=1,...,m$. Let $\hat{g}$ be the empirical density function $\sum_{i=1}^{m}\delta_{\mu_i}(\cdot)$. Let $f(x|\tau)=g*\eta_\tau(x)$ and $\hat{f}(x|\tau)=\hat{g}*\eta_\tau(x)$. Then 
$$
E\hat{f}(x|\tau)=E\sum_{i=1}^{m}\dfrac{1}{m}\eta_{\tau}(x-\mu_i)=E\eta_\tau(x-\mu)=\int_{-\infty}^{\infty}\eta_\tau(x-\mu)g(\mu)d\mu=f(x|\tau).
$$
Also since $\eta_\tau$ is smooth, thus  bounded, it follows that $\mbox{Var}\{\eta_\tau(x-\mu_i)\}<\infty$ (here we treat $\mu_i$ as random). Therefore 
\begin{align*}
	\mbox{Var}\hat{f}(x|\tau)&=\mbox{Var}\left\{\int_{-\infty}^{\infty}\eta_{\tau}(\mu-x)\hat{g}(\mu)d\mu\right\}\\
	&=\mbox{Var}\left\{ \dfrac{1}{m} \sum_{i=1}^{m}\eta_\tau(x-\mu_i)      \right\}\\
	&=\dfrac{1}{m}\mbox{Var}\{\eta_\tau(x-\mu_i)\}\rightarrow 0.
\end{align*}
It follows that $E_{\pmb{\mu}}|f(x|\tau)-\hat{f}(x|\tau) |^2\rightarrow 0$ as $n\rightarrow \infty$. 

The above implies it is possible to find a set $\{\mu_1,\ldots,\mu_m\}$ and $\hat{f}(x|\tau)=\frac{1}{m}\sum_{i=1}^{m}\eta_\tau(x-\mu_i)$ such that for all $x$, $|f(x|\tau)-\hat{f}(x|\tau) |^2\rightarrow 0$. 
Consider the set of functions $\{  \sum_{j=1}^{k}\theta_j\eta_{\tau}(x-u_j) | \sum_{j=1}^{k}\theta_j=1, \theta_j\geq 0 \ \ \forall j. \}$. {We can make the grid fine enough so that for any $\epsilon'>0$ and $j$, there exists $u_{i(j)}\in \{u_1,\ldots, u_k\}$ such that $|\mu_j-u_{i(j)}|<\epsilon'$. We can choose $\epsilon'$ small enough so that $| \eta_{\tau}(x-u_j)-\eta_{\tau}(x-\mu_{i(j)}) |^2<\epsilon$.} Hence,
\begin{align*}
	\big|\dfrac{1}{m}\sum_{j=1}^{m}\eta_\tau(x-\mu_j)-\dfrac{1}{m}\sum_{j=1}^{m}\eta_\tau(x-u_{i(j)})\big|^2&=\dfrac{1}{m^2}\big|\sum_{j=1}^{m}\eta_\tau(x-\mu_j)-\sum_{j=1}^{m}\eta_\tau(x-u_{i(j)})\big|^2\\
	&\leq \dfrac{1}{m}\sum_{j=1}^{m}| \eta_{\tau}(x-\mu_j)-\eta_{\tau}(x-u_{i(j)}) |^2\leq \epsilon.
\end{align*}
By the triangle inequality we have $|f(x|\tau)-\dfrac{1}{m}\sum_{j=1}^{m}\eta_\tau(x-u_{i(j)}) |^2\leq \epsilon$, we can let $g'(\cdot)=\frac{1}{m}\sum_{j=1}^{m}\delta_{u_{i(j)}}(\cdot)$.\hfill$\blacksquare$

\subsection{Proof of Lemma \ref{lem2}}
By Fubini's theorem, we have
\begin{align*}
	E|\hat{f}(x|\sigma)-f(x|\sigma)|^2&=E_\sigma E_{\pmb{x},\pmb{\sigma}} E_{x|\sigma}|\hat{f}(x|\sigma)-f(x|\sigma)|^2\\
	&=E_\sigma  E_{\pmb{x},\pmb{\sigma}}\int_{-\infty}^{\infty}|\hat{f}(x|\sigma)-f(x|\sigma)|^2f(x|\sigma)dx
\end{align*}
Hence, if $\tau\sim g_\sigma$ we must have 
$$
E\int_{-\infty}^{\infty}|\hat{g}*\eta_\tau(x)-g*\eta_\tau(x)|^2f(x|\tau)dx\rightarrow 0.
$$

$E\int_{-\infty}^{\infty}|\hat{g}*\eta_\tau(x)-g*\eta_\tau(x)|^2dx\nrightarrow 0$. Then there exists a sequence of sets $\mathcal{X}_m$ and $\epsilon_1>0$ such that $E|\hat{g}*\eta_\tau(x)-g*\eta_\tau(x)|>\epsilon_1$ on $\mathcal{X}_m$. If $\int_{\mathcal{X}_m} f(x|\tau)dx >0$ this would imply $E\int_{-\infty}^{\infty}|\hat{g}*\eta_\tau(x)-g*\eta_\tau(x)|^2f(x|\tau)dx\nrightarrow 0$ a contradiction. If $\int_{\mathcal{X}_m} f(x|\tau)dx \rightarrow0$, using the definition $f(x|\tau)=g*\eta_\tau(x)$ we have
$$E\int_{\mathcal{X}_m}|\hat{g}*\eta_\tau(x)-g*\eta_\tau(x)|^2dx\rightarrow E\int_{\mathcal{X}_m}|\hat{g}*\eta_\tau(x)|^2dx\nrightarrow 0$$
Since $\hat{g}*\eta_\tau(x)$ and ${g}*\eta_\tau(x)$ are both densities there must exists another sequence of sets $\mathcal{X'}_m$, $\epsilon>0$ and $\delta>0$ such that $g*\eta_\tau(x)>\hat{g}*\eta_\tau(x)+\epsilon$ on $\mathcal{X'}_m$ and $\int_{\mathcal{X}'_m}f(x|\tau)dx>\delta$, again contradicts the the fact that $
E\int_{-\infty}^{\infty}|\hat{g}*\eta_\tau(x)-g*\eta_\tau(x)|^2f(x|\tau)dx\rightarrow 0.
$
\hfill$\blacksquare$
\subsection{Proof of lemma \ref{lemma:2}}\label{prooflemma2}
Using the definitions of $\hat{U}_i$ and $U_i$, we can show that 
\begin{align*}
	\left(\hat{U}_i - U_i\right)^2 &=\left(\hat{T}_{i,m} - T_i^{\sf OR}\right)^2I \left(\hat{T}_{i,m} \leq t, T_i^{\sf OR} \leq t \right) + \left(\hat{T}_{i,m}-\alpha\right)^2I \left(\hat{T}_{i,m} \leq t, T_i^{\sf OR}> t \right)\\
	&+\left(T_i^{\sf OR}-\alpha\right)^2I \left(\hat{T}_{i,m} > t, T_i^{\sf OR} \leq t \right).
\end{align*}
Denote the three sums on the RHS as $I$, $II$, and $III$ respectively. By Corollary \ref{coro1}, $E(I) = o(1)$. Let $\varepsilon > 0$. Consider
\begin{align*}
	P\left(\hat{T}_{i,m} \leq t, T_i^{\sf OR}> t \right) &\leq P\left(\hat{T}_{i,m} \leq t, T_i^{\sf OR}\in \left(t, t+ \varepsilon \right) \right)+ P\left(\hat{T}_{i,m} \leq t, T_i^{\sf OR}\geq  t+ \varepsilon  \right) \\
	&\leq P\left\{T_i^{\sf OR} \in \left(t, t+ \varepsilon \right)\right\} + P\left(\left|\hat{T}_{i,m}- T_i^{\sf OR}\right| > \varepsilon \right)
\end{align*}
The first term on the right hand is vanishingly small as $\varepsilon \rightarrow 0$ because $T_i^{\sf OR}$ is a continuous random variable. The second term converges to $0$ by Corollary \ref{coro1}. we conclude that $II = o(1)$. In a similar fashion, we can show that $III = o(1)$, thus proving the lemma.\hfill$\blacksquare$
\section{Implementation of HAMT, NPMLE B, ASH and ASH 1}
\label{sec:npmleb_details}
\subsection{HAMT}
We use the R interface to MOSEK \citep{mosek} available in the R package \texttt{Rmosek} for solving Problem \eqref{eq:opt}. However, note that Problem \eqref{eq:opt} requires that the $KS$  parameters $\mathcal W=(\bm{w}_1^T, \ldots, \bm{w}_S^T)^T$ satisfy $Sm$ inequality constraints $B\mathcal W\succeq \bm 0_{Sm}$ and $m$ equality constraints $C\mathcal W=\bm 1_m$. Since the number of constraints is much greater than the number of parameters, solvers such as MOSEK may throw an
infeasibility certificate if numerical instabilities prohibit the solution from satisfying these constraints, particularly the equality constraints. %Here we discuss two alternatives to solving Problem \eqref{eq:opt}.
An alternative strategy is to consider the corresponding dual of Problem \eqref{eq:opt}. This approach has also been advocated by \cite{koenker2014convex,gu2017empirical,koenker2017rebayes} in the context of empirical Bayes deconvolution using nonparametric maximum likelihood estimation. The dual of the optimization Problem \eqref{eq:opt} is
\begin{equation}
	\label{eq:hamt_dual}
	\begin{split}
	\min_{(\bm \omega,\bm \rho)\in\mathbb R^{m(S+1)}} \bm 1_m^T\bm \rho +\dfrac{1}{2}(\bm v+C^T\bm\rho-B^T\bm\omega)^TD^{-1}(\bm v+C^T\bm\rho-B^T\bm\omega)\quad \mbox{subject\ to}~\bm\omega\succeq \bm 0_{Sm},
	\end{split}
\end{equation}
where $\bm v=-m^{-1}A^T\hat{\bm \varphi}^m$ and $D=m^{-1}A^TA$.	Since $D$ is positive semi-definite in our setting, the dual Problem \eqref{eq:hamt_dual} is a standard convex quadratic optimization problem
with a non-negativity constraint on the decision variables $\bm \omega$. Moreover, this dual problem always has a solution. For instance, $\bm \rho=\bm 0_m$ and $\bm \omega=\bm 0_{Sm}$ is \textit{a} feasible solution. Furthermore, if $(\bm\rho^*,\bm \omega^*)$
denotes the optimal solution to Problem \eqref{eq:hamt_dual} then the solution to the primal Problem \eqref{eq:opt} is
$\hat{\mathcal W}_m=-D^{-1}(\bm v + C^T\bm\rho^*-B^T\bm \omega^*)$.

\rred{Another practical strategy that works particularly well in our setting is to solve a relaxed version of the primal Problem \eqref{eq:opt} where the equality constraints $C\mathcal W=\bm 1_m$ in Equation \eqref{eq:opt} are relaxed to the following inequality constraints: $a\bm 1_m\preceq C\mathcal W\preceq\bm 1_m$ with $a$ set at $0.9$. The corresponding ``relaxed primal" problem is often easier to solve than Problem \eqref{eq:opt}. Furthermore, if $\hat{\mathcal W}_m=(\hat{\bm w}_1^T,\ldots,\hat{\bm w}_S^T)^T$ denotes the optimal solution to this relaxed primal problem then we only require a simple re-scaling of the columns of the $S\times m$ matrix $\mathcal G=(\hat{\bm g}_1,\ldots,\hat{\bm g}_m)$, where $\hat{\bm g}_i=\{\hat{\bm w}_j^T\bm q(\sigma_i):1\le j\le S\}$, to ensure $\sum_{j=1}^{S}\hat{\bm w}_j^T\bm q(\sigma_i)=1$ for all $i=1,\ldots,m$.}
Our implementation in R first attempts to solve the primal Problem \eqref{eq:opt}. In case the solution status from MOSEK's conic interior-point optimizer throws an infeasibility certificate, %the dual Problem \eqref{eq:hamt_dual} is solved and the corresponding primal solution is recovered. 
it attempts to solve the ``relaxed primal" problem. The dual Problem \eqref{eq:hamt_dual} is solved and the corresponding primal solution is recovered if the solution status from solving the relaxed primal is not optimal. 

\subsection{NPMLE B, ASH and ASH.1}
We first describe the construction of {NPMLE B} and then discuss the implementation of ASH and ASH 1.

For {NPMLE B}, we continue to take $g_j(\sigma_i)=\bm w_j^T\bm q(\sigma_i)$ but estimate the $\bm w_j$'s by solving Problem \eqref{eq:npmle}. Denote $\mathcal W=(\bm w_1^T,\ldots,\bm w_S^T)$, $\bm a_{ij}=\eta_{\sigma_i}(x_i-u_j)\bm q(\sigma_i)$, $\bm a_i=(\bm a_{i1}^T,\ldots,\bm a_{iS}^T)^T$ and $A=(\bm a_1,\ldots,\bm a_m)^T$. Additionally, let $Q=[\bm 1_S\otimes \bm q(\sigma_1)\ldots \bm 1_S\otimes \bm q(\sigma_m)]^T$ where $\otimes$ denotes the usual kronecker product of two matrices, and $\bm c_r$ is an $r-$dimensional vector with all entries equal to the scalar $c\in\mathbb R$. Finally, for $s=1,\ldots,S$, let $L_s$ denote a $m\times KS$ matrix whose entries are $0$ except for the $\{K(s-1)+1\}$th column which equals $[\bm q(\sigma_1),\ldots \bm q(\sigma_m)]^T$ and denote $L=[L_1^T\ldots L_S^T]^T$. With these notations, the primal optimization problem \eqref{eq:npmle} is equivalent to
\begin{eqnarray}
	\label{eq:primal_npmleb}
	\min_{\bm u,\mathcal W}-\sum_{i=1}^{m}\log u_i~\text{subject to~}A\mathcal W=\bm u,~Q\mathcal W=\bm 1_m,~L\mathcal W\succeq \bm 0_{Sm},
\end{eqnarray}
where $\bm u=(u_1,\ldots,u_m)$. As recommended by \cite{koenker2014convex,koenker2017rebayes}, it is often more efficient to solve the dual of Problem \eqref{eq:primal_npmleb} and recover the corresponding primal solution. In our context, the dual problem is given by
\begin{eqnarray}
	\label{eq:dual_npmleb}
	\min_{\bm v}-\sum_{i=1}^{m}\log v_i~\text{subject to~}A^T\bm v\preceq Q^T\bm 1_m,~\bm v\succeq \bm 0_m,
\end{eqnarray}
where $\bm v=(v_1,\ldots,v_m)$. If $\bm v^*$ denotes the optimal solution to the dual \eqref{eq:dual_npmleb} then the primal solution is $\hat{\mathcal W}=(A^TA)^{-1}A^T\bm u^*$ where $u_i^*=1/v_i^*$. We use the R interface to MOSEK \citep{mosek} available in the R package \texttt{Rmosek} for solving Problem \eqref{eq:dual_npmleb} and fix $S=50,~K=10$ for all our numerical experiments. 

For implementing the method ASH proposed in \cite{stephens2017false}, we use the function \texttt{ash} available in the R-package \texttt{ashr} and consider two versions of this method: one with $c=0$ (ASH) and the other with $c = 1$ (ASH 1). For both these versions, we set \texttt{mode = "estimate"} and estimate the Clfdr using $3,000$ posterior samples. The \texttt{ash} function requires specifying the approximate distribution of the components in the mixture used to represent $g$ via the option \texttt{mixcompdist} and the \texttt{ashr} package provides several choices such as \texttt{"uniform"},\texttt{"normal"},\texttt{"halfuniform"},\texttt{"halfnormal"} etc. While this oracle knowledge is not available to {HAMT}, we provide this information to \texttt{ash} depending on the simulation setting at hand and report the choice in the corresponding figure captions. For the real data analysis (Section \ref{sec:realdata}), we set \texttt{mixcompdist="normal"}.
\section{Additional numerical experiments}
\label{sec:more_sims}
\subsection{Experiments involving two-sided composite null hypotheses}
\label{sec:sims_twosided}
Here we assess the numerical performance of HAMT for two-sided composite null hypotheses. We evaluate the following six competing testing procedures from Section \ref{sec:sims_onesided}: AdaPTGMM, DECONV, ASH, ASH 1, NPMLE B and OR. %Additionally, we consider the testing procedure NPMLE that relies on the deconvolution estimate obtained from nonparametric maximum likelihood estimation to construct the Lfdr statistic. 
We drop GS1 and GS2 from our comparisons since these methods were developed only for testing one sided composite null hypotheses. Additionally, we do not report the performance of BH, CAMT and IHW as these $p-$value methods exhibited substantially lower power in all our experiments involving two-sided composite nulls. 

We fix $m=10^4$, $\alpha=0.1$ and evaluate the aforementioned six methods across the following three simulation settings.
\begin{figure}
	\centering
\includegraphics[width=0.8\linewidth]{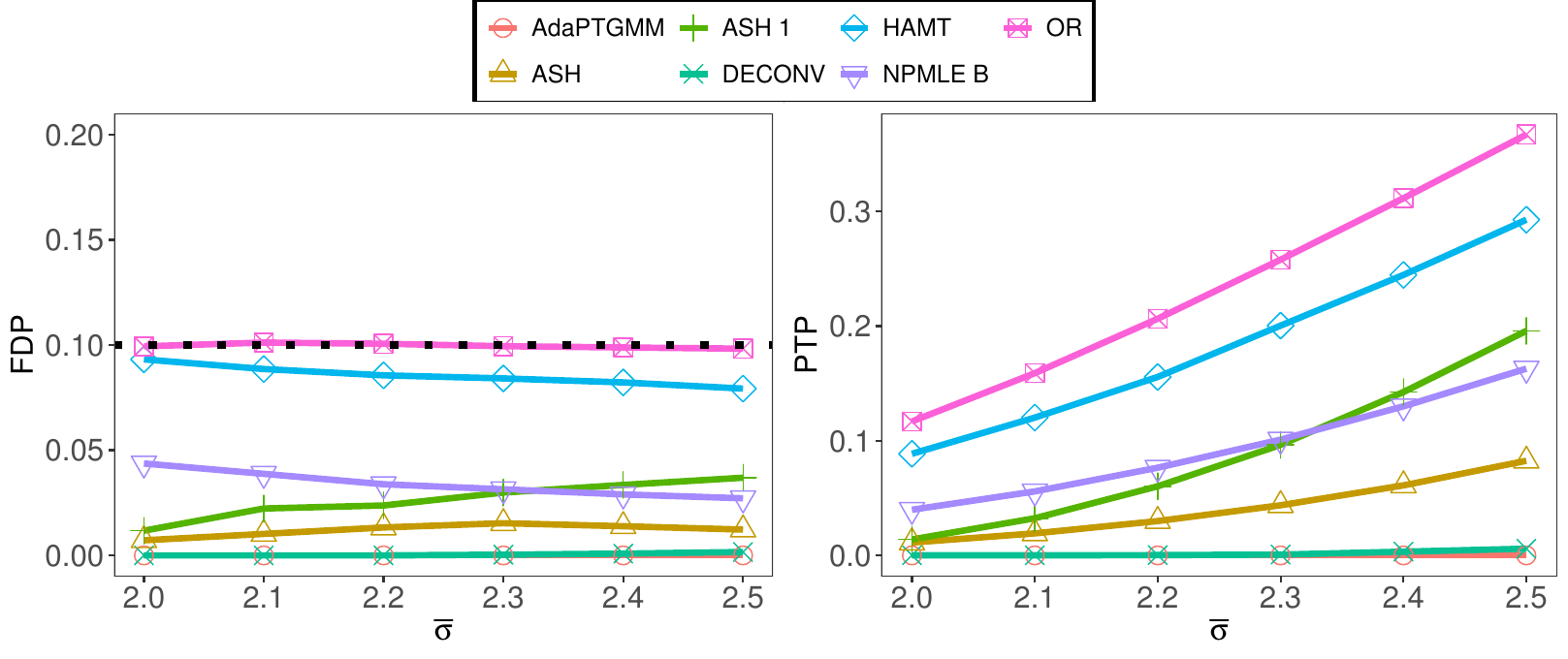}
	\caption{ Setting 1: $\sigma_i\stackrel{i.i.d.}{\sim}(1/3)\delta_{(0.5)}+(1/3)\delta_{(1)}+(1/3)\delta_{(3)}$ and conditional on $\sigma_i$, $\mu_i\sim0.9\delta_{(0)}+0.05\delta_{(\bar{\sigma}\sigma_i)}+0.05\delta_{(-\bar{\sigma}\sigma_i)}$. Here $\mathcal{A}=[-5,5]$. For ASH and ASH 1, \texttt{mixcompdist = "uniform"}.}
	\label{fig:sim_exp7_test}
	%\centering
	\includegraphics[width=0.8\linewidth]{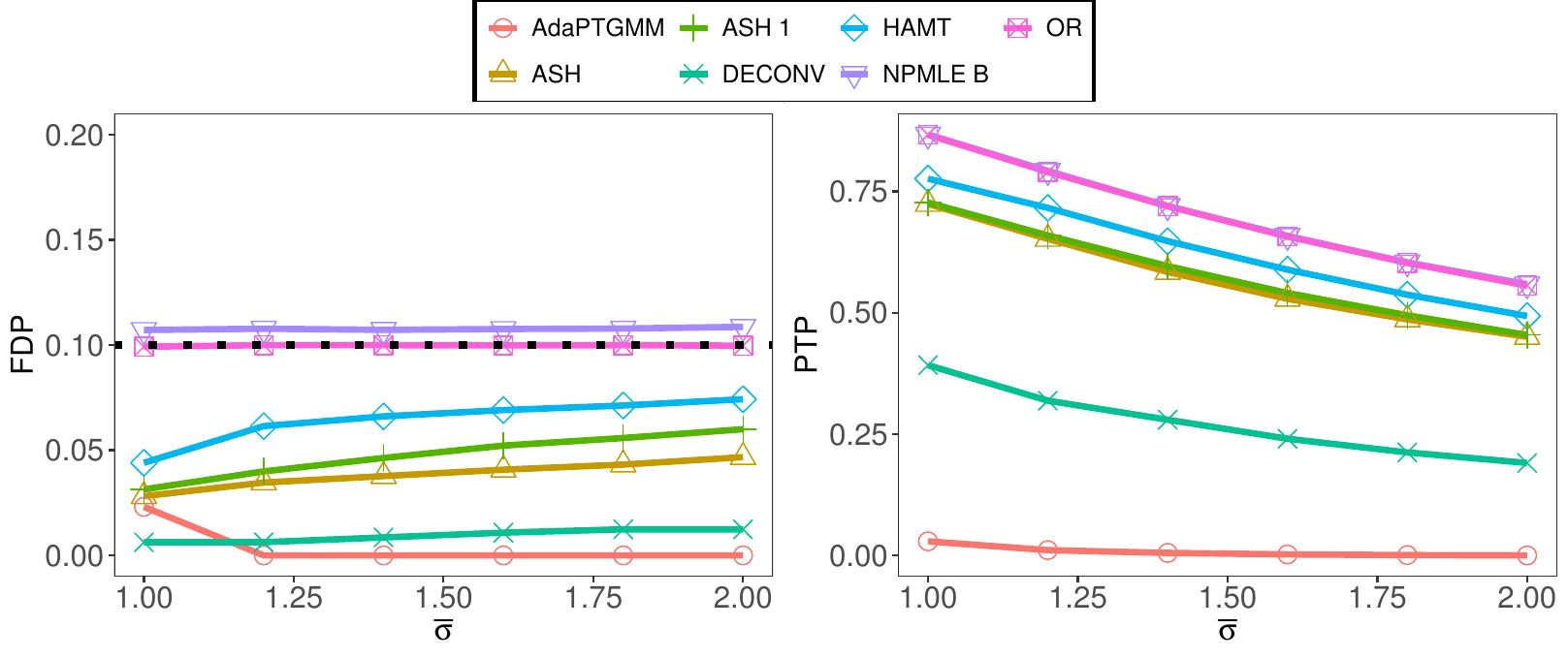}
	\caption{Setting 2: $\sigma_i\stackrel{i.i.d.}{\sim}\text{Unif}(0.5,\bar{\sigma})$. Conditional on $\sigma_i$, $\mu_i=0$ with probability $0.9$ and $\mu_i\stackrel{ind.}{\sim}N(3,\sigma_i)$ or $N(-3,\sigma_i)$ each with probability $0.05$. Here $\mathcal{A}=[-2,2]$. For ASH and ASH 1, \texttt{mixcompdist = "normal"}.}
	\label{fig:sim_exp8_test}
	%\end{figure}
	%\begin{figure}[!h]
	%	\centering
	\includegraphics[width=0.9\linewidth]{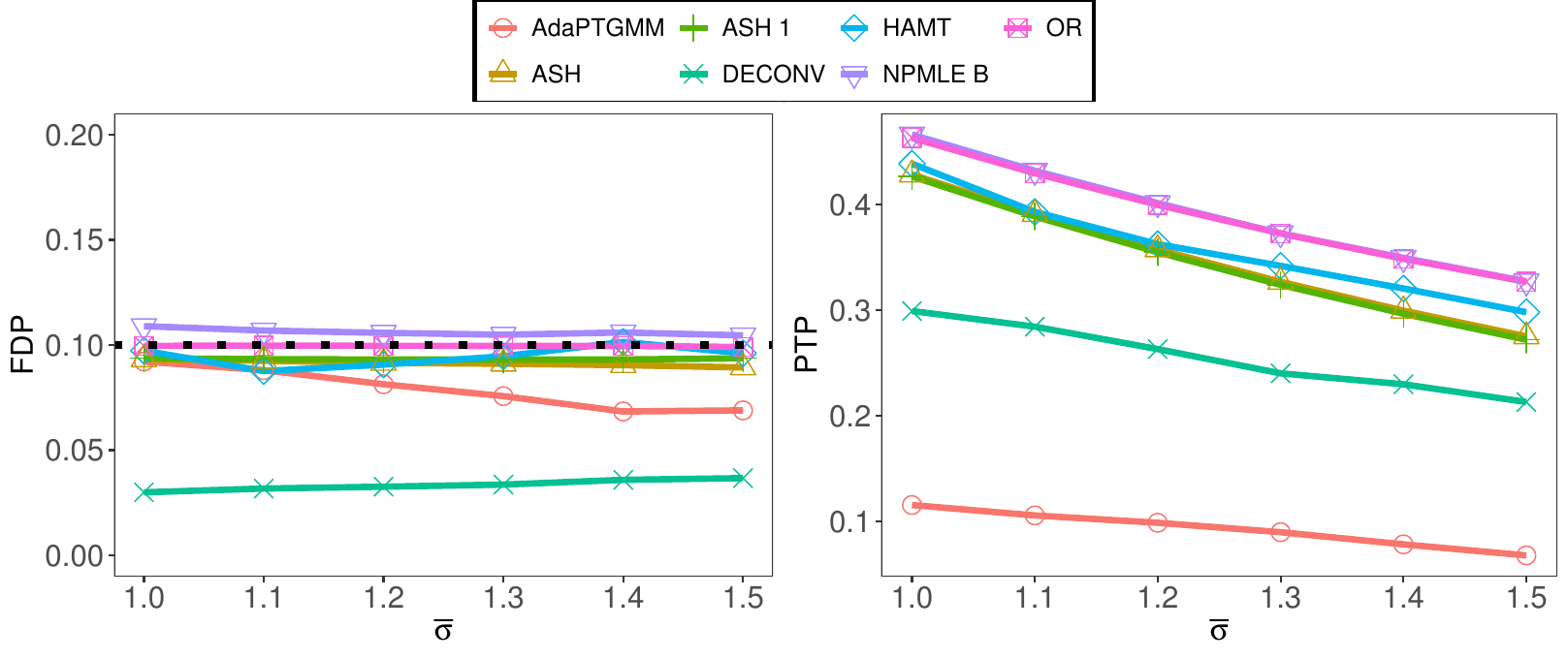}
	\caption{ Setting 3: $\sigma_i\stackrel{i.i.d.}{\sim}\text{Unif}(0.5,\bar{\sigma})$. Conditional on $\sigma_i$, $\mu_i=0$ with probability $0.8$ and $\mu_i\stackrel{ind.}{\sim}N(\sigma_i,1)$ or $N(\sigma_i,4)$ each with probability $0.1$. Here $\mathcal{A}=[-1,1]$. For ASH and ASH 1, \texttt{mixcompdist = "normal"}.}
	\label{fig:sim_set3_2s}
\end{figure}
\begin{itemize}
	\item \textbf{Setting 1 --} presented in Figure \ref{fig:sim_exp7_test}, is a modification of Setting 2 from Section \ref{sec:sims_onesided}. Here $\sigma_i=0.5,1$ or $3$ with equal probabilities. Conditional on $\sigma_i$, $\mu_i=0$ with probability $0.9$, and $\mu_i=\bar{\sigma}\sigma_i$ or $-\bar{\sigma}\sigma_i$ with probability $0.05$ each. We take $\mathcal A=[-5,5]$ for this setting and find that HAMT controls the FDR level at $\alpha$ and dominates all other methods in power.
	\item \textbf{Setting 2 --} we sample $\sigma_i$ independently from $\text{Unif}(0.5,\bar{\sigma})$ but consider a three component mixture distribution for $\mu_i$ conditional on $\sigma_i$. In particular $\mu_i=0$ with probability $0.9$ and $\mu_i\stackrel{ind.}{\sim}N(3,\sigma_i)$ or $N(-3,\sigma_i)$ each with probability $0.05$. Here we let $\mathcal A=[-2,2]$.
	\item \textbf{Setting 3 --} we continue to sample $\sigma_i$ independently from $\text{Unif}(0.5,\bar{\sigma})$. Conditional on $\sigma_i$, $\mu_i=0$ with probability $0.8$ and $\mu_i\stackrel{ind.}{\sim}N(\sigma_i,1)$ or $N(\sigma_i,4)$ each with probability $0.1$. Here we let $\mathcal A=[-1,1]$.
\end{itemize}
\red{Figures \ref{fig:sim_exp8_test} and \ref{fig:sim_set3_2s}, respectively, report the performance of the various methods across settings 2 and 3. We find that all methods, with the exception of DECONV and AdaPTGMM, are competitive with respect to their power while NPMLE B marginally fails to control the FDR at $10\%$.}
%\section{More Numerical Experiments}
\subsection{Non-Gaussian Likelihood}
\label{sec:sims_non_gaussian}
In this section we assess the performance of HAMT when $\epsilon_{i}$ in Model \eqref{eq:model1} are not necessarily standard normal random variables. Note that (i) we drop GS1, GS2 and DECONV from our comparisons as these methods rely on a Gaussian likelihood, and (ii) BH, CAMT and IHW are not reported whenever the simulation setting involves testing two-sided composite nulls.
\begin{figure}
	\centering
	\includegraphics[width=0.9\linewidth]{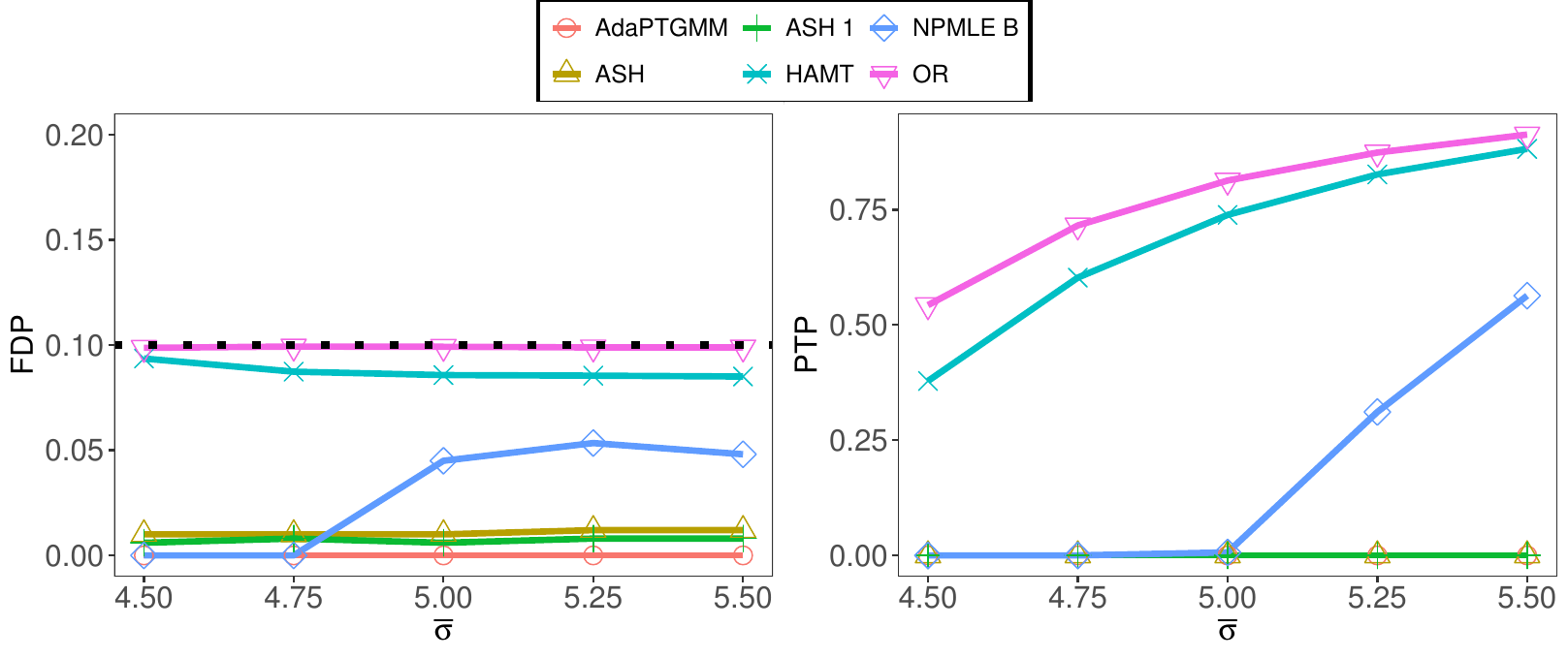}
	\caption{Setting 1: $\sigma_i\stackrel{i.i.d.}{\sim}(1/3)\delta_{(0.25)}+(1/3)\delta_{(0.75)}+(1/3)\delta_{(1.5)}$, conditional on $\sigma_i$, $\mu_i=0.9\delta_{(0)}+0.05\delta_{(\bar{\sigma}\sigma_i)}+0.05\delta_{(-\bar{\sigma}\sigma_i)}$ and, conditional on $(\mu_i,\sigma_i)$, $X_i=\mu_i+\sigma_it_5$. Here $\mathcal{A}=[-5,5]$. For ASH and ASH 1, \texttt{mixcompdist = "uniform"}}
	\label{fig:t_like}
		\centering
	\includegraphics[width=0.9\linewidth]{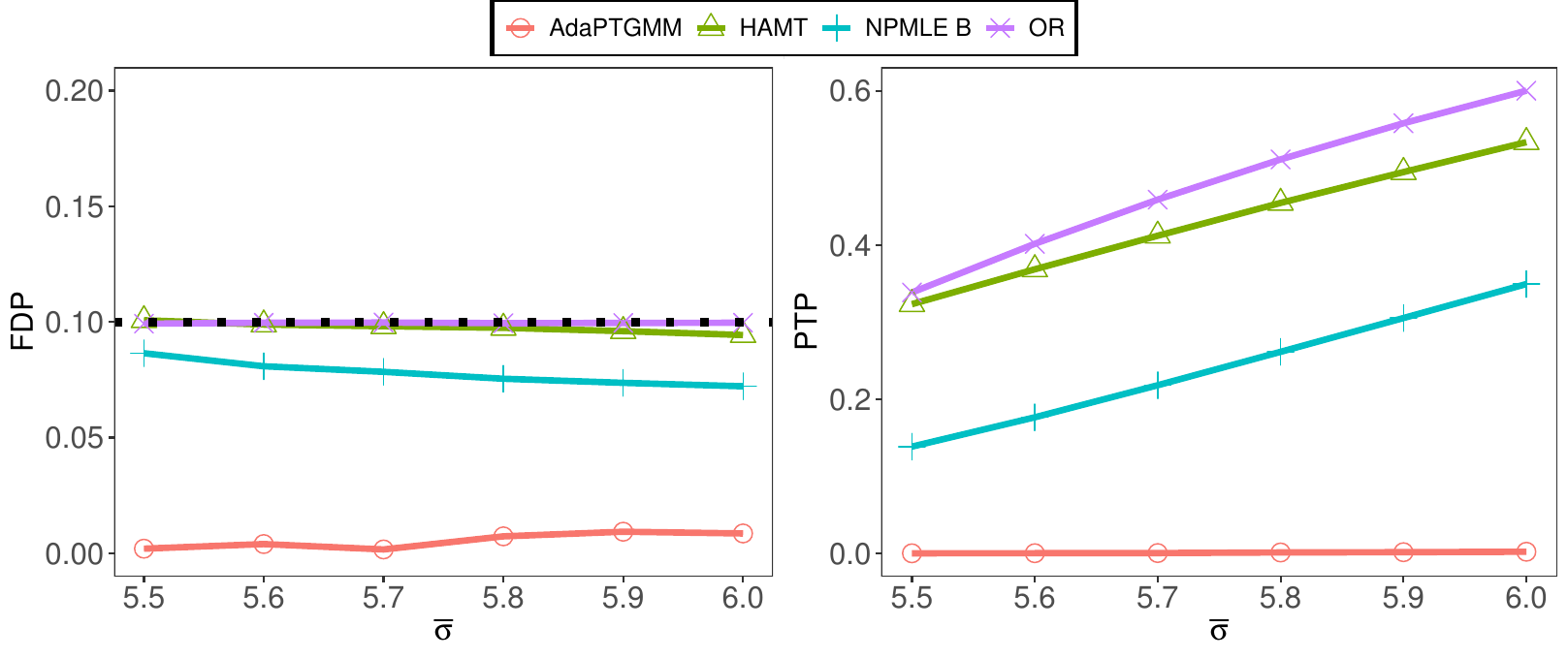}
	\caption{Setting 2: $\sigma_i\stackrel{i.i.d.}{\sim}(1/3)\delta_{(0.25)}+(1/3)\delta_{(0.75)}+(1/3)\delta_{(1.5)}$, conditional on $\sigma_i$, $\mu_i=0.9\delta_{(0)}+0.05\delta_{(\bar{\sigma}\sigma_i)}+0.05\delta_{(-\bar{\sigma}\sigma_i)}$ and, conditional on $(\mu_i,\sigma_i)$, $X_i\stackrel{ind}{\sim}Logistic(\mu_i,\sigma_i)$. Here $\mathcal{A}=[-2,2]$.}
	\label{fig:logistic_like}
		\centering
	\includegraphics[width=0.9\linewidth]{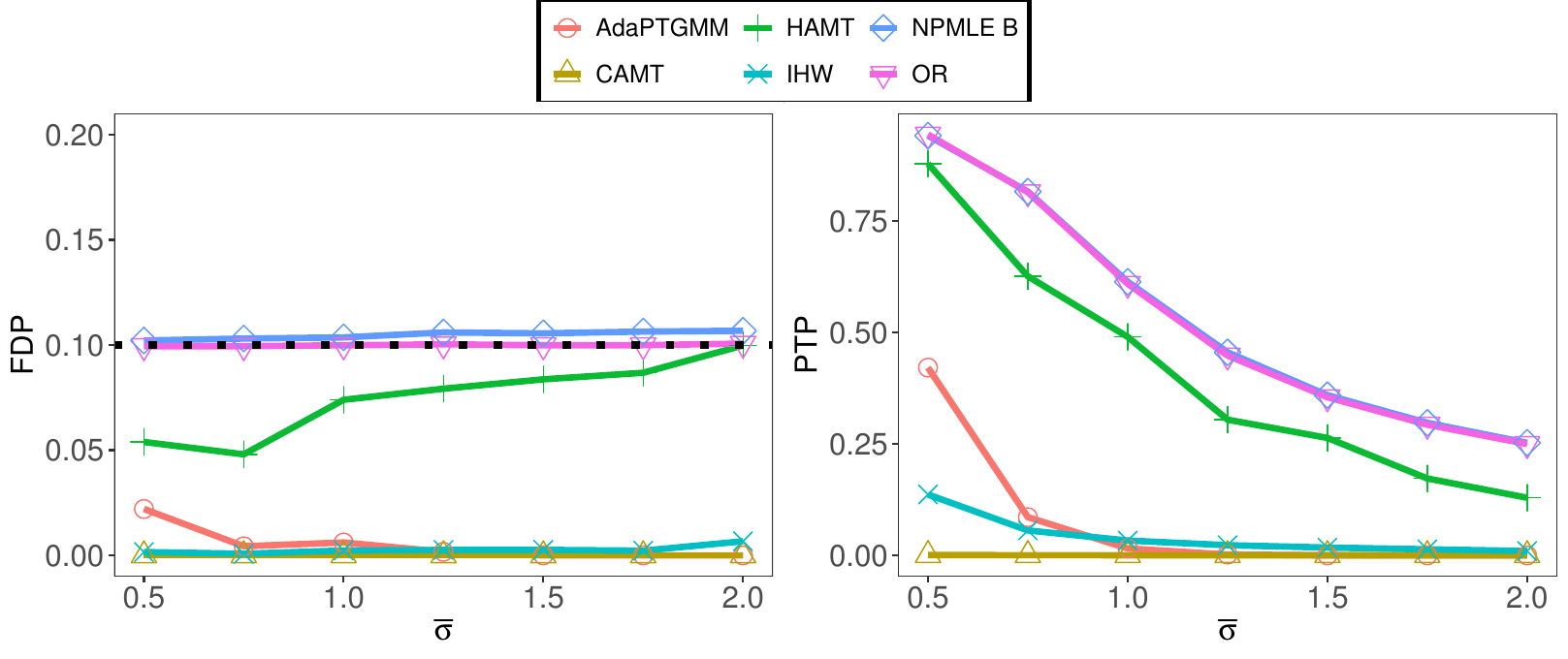}
	\caption{Setting 3: $\sigma_i\stackrel{i.i.d.}{\sim}\text{Unif}(0.3,\bar{\sigma})$, $\mu_i=0.9\delta_{(0)}+0.1N(3,1)$ and, conditional on $(\mu_i,\sigma_i)$, $X_i\stackrel{ind}{\sim}Laplace(\mu_i,\sigma_i)$. Here $\mathcal{A}=[-\infty,2]$.}
	\label{fig:laplace_like}
\end{figure}
The following three simulation settings are considered with $m=10^4$ and $\alpha=0.1$.
\begin{itemize}
	\item \textbf{Setting 1 --} $\sigma_i\stackrel{i.i.d.}{\sim}(1/3)\delta_{(0.25)}+(1/3)\delta_{(0.75)}+(1/3)\delta_{(1.5)}$, conditional on $\sigma_i$, $\mu_i=0.9\delta_{(0)}+0.05\delta_{(\bar{\sigma}\sigma_i)}+0.05\delta_{(-\bar{\sigma}\sigma_i)}$ and, conditional on $(\mu_i,\sigma_i)$, $X_i=\mu_i+\sigma_it_5$, where $t_\nu$ is a central $t-$distributed random variable with $\nu$ degrees of freedom. Here $\mathcal{A}=[-5,5]$.
	\item \textbf{Setting 2 --} same as Setting 1 except that $X_i\stackrel{ind}{\sim}Logistic(\mu_i,\sigma_i)$. Here $\mathcal{A}=[-2,2]$.
	\item \textbf{Setting 3 --} $\sigma_i\stackrel{i.i.d.}{\sim}\text{Unif}(0.3,\bar{\sigma})$, $\mu_i=0.9\delta_{(0)}+0.1N(3,1)$ and, conditional on $(\mu_i,\sigma_i)$, $X_i\stackrel{ind}{\sim}Laplace(\mu_i,\sigma_i)$. Here $\mathcal{A}=[-\infty,2]$.
\end{itemize} 
For settings 2 and 3, ASH and ASH 1 are not included in our comparisons as their corresponding R package \texttt{ash} does not provide an implementation when $\epsilon_{i}$ are Logistic or Laplace distributed. Figures \ref{fig:t_like}--\ref{fig:laplace_like} report the performance of various methods across the three simulation settings. In settings 1 and 2, HAMT dominates all methods in power. In Setting 1, NPMLE B exhibits a relatively high MC error in its FDP distribution for the first three values of $\bar\sigma$ and so for each $\bar\sigma$ we report its median FDP and PTP across the 200 MC repetitions in Figure \ref{fig:t_like}. In Setting 3, NPMLE B exhibits a marginally higher FDR than $10\%$ for larger values of $\bar{\sigma}$ but is, in general, relatively more powerful than HAMT.
\subsection{Unknown variance}
\label{sec:sims_unknownvar}
Here we investigate the performance of HAMT when the summary statistic $X_i$ are only asymptotically Normal and $\sigma_i$ are unknown, being approximated by the sample variance $S_i^2$. Specifically, conditional on $(\mu_i,\sigma_i)$ and $j=1,\ldots,n$, we let $X_{ij}=\mu_i+\sigma_ic_n\epsilon_{ij}$ where the scaling $c_n=\sqrt{n/\text{Var}(\epsilon_{ij})}$ ensures that $\text{Var}(X_{i})=\text{Var}(n^{-1}\sum_{j=1}^{n}X_{ij})=\sigma_i^2$. Denote $S_i^2=n^{-2}\sum_{j=1}^{n}(X_{ij}-X_i)^2$. In this section we assume that $X_i$ are approximate Gaussian random variables with mean $\mu_i$ and variance $S_i^2$, and consequently, assess the impact of this assumption on the quality of various testing procedures as $n$ varies. The following four settings are evaluated:
\begin{figure}
	\centering
	\includegraphics[width=0.9\linewidth]{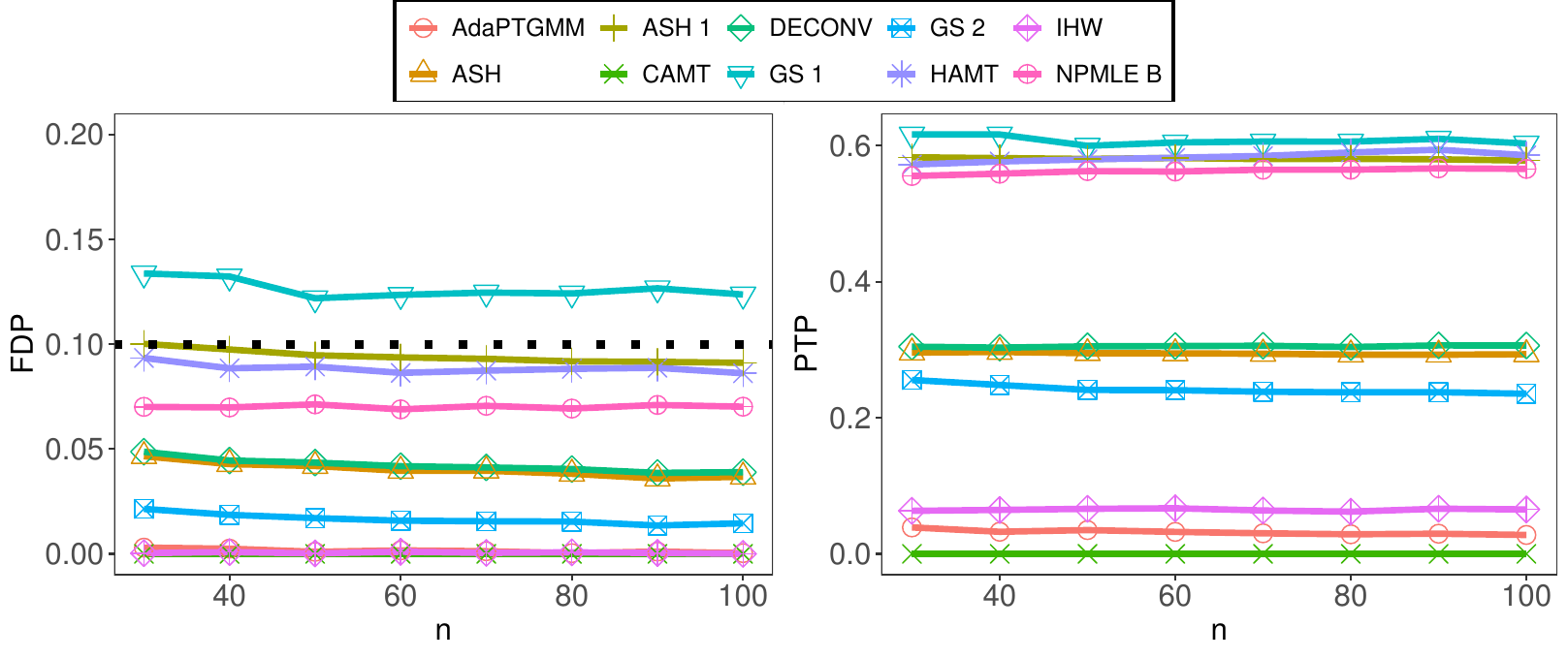}
	\caption{Setting 1: Scenario borrowed from Setting 3 of Section \ref{sec:sims_onesided} with $\sigma_i\stackrel{i.i.d}{\sim}\text{Unif}(0.5,1.5)$ and $\epsilon_{ij}\stackrel{i.i.d}{\sim}\text{Unif}(-3,3)$.}
	\label{fig:clt_unif}
		\centering
	\includegraphics[width=0.9\linewidth]{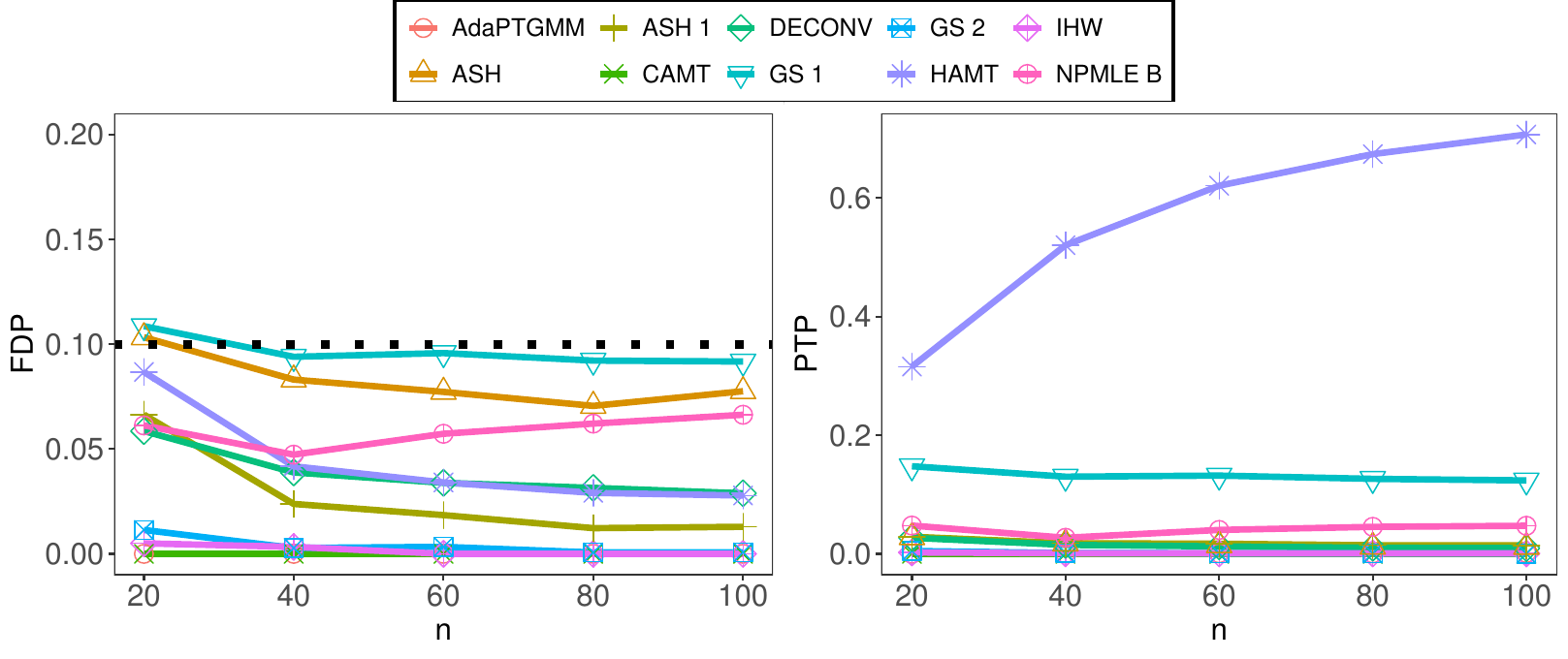}
	\caption{Setting 2: Scenario is borrowed from Setting 4 of Section \ref{sec:sims_onesided} with $\sigma_i\stackrel{i.i.d}{\sim}0.9\text{Unif}(0.5,1)+0.1\text{Unif}(1,1.5)$ and $\epsilon_{ij}$ are i.i.d central $t-$distributed random variables with $10$ degrees of freedom.}
	\label{fig:clt_t}
		\centering
	\includegraphics[width=0.9\linewidth]{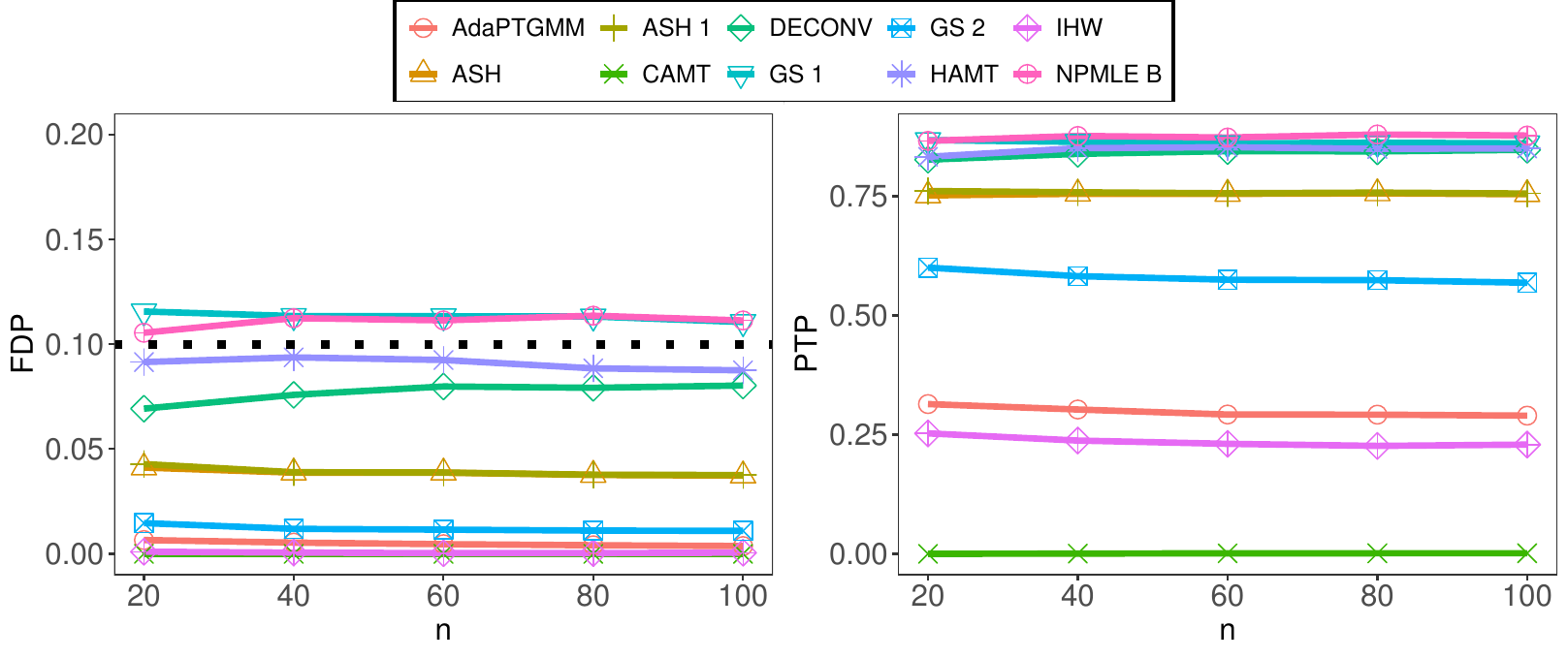}
	\caption{Setting 3: Scenario is borrowed from Setting 1 of Section \ref{sec:sims_onesided} with $\sigma_i\stackrel{i.i.d}{\sim}\text{Unif}(0.5,1)$ and $\epsilon_{ij}$ are i.i.d central $t-$distributed random variables with $10$ degrees of freedom.}
	\label{fig:clt_t_1}
\end{figure}
\begin{itemize}
	\item \textbf{Setting 1 - }this scenario is borrowed from Setting 3 of Section \ref{sec:sims_onesided} with $\sigma_i\stackrel{i.i.d}{\sim}\text{Unif}(0.5,1.5)$ and $\epsilon_{ij}\stackrel{i.i.d}{\sim}\text{Unif}(-3,3)$. In Figure \ref{fig:clt_unif}, we find that as $n$ varies, HAMT, ASH 1 and NPMLE B exhibit similar power while GS1 fails to control the FDR at $10\%$.
	\item \textbf{Setting 2 - }this scenario is borrowed from Setting 4 of Section \ref{sec:sims_onesided} with $\sigma_i\stackrel{i.i.d}{\sim}0.9\text{Unif}(0.5,1)+0.1\text{Unif}(1,1.5)$ and $\epsilon_{ij}$ are i.i.d central $t-$distributed random variables with $10$ degrees of freedom. In Figure \ref{fig:clt_t}, HAMT dominates all other methods in power as $n$ varies while controlling the FDR at $10\%$.
	\item \textbf{Setting 3 - }this scenario is borrowed from Setting 1 of Section \ref{sec:sims_onesided} with $\sigma_i\stackrel{i.i.d}{\sim}\text{Unif}(0.5,1)$ and $\epsilon_{ij}$ are i.i.d central $t-$distributed random variables with $10$ degrees of freedom. In Figure \ref{fig:clt_t_1}, GS1 and NPMLE B marginally fail to control the FDR at $10\%$. In terms of power, HAMT and DECONV exhibit similar profiles that are closely followed by ASH and ASH 1.
	\item \textbf{Setting 4 - }this scenario is borrowed from Setting 1 of Section \ref{sec:sims_non_gaussian} with $\mu_i=0.9\delta_{(0)}+0.05\delta_{(4.5\sigma_i)}+0.05\delta_{(-4.5\sigma_i)}$ and $\epsilon_{ij}$ are i.i.d Logistic random variables with location $=0$ and scale $=1$. In Figure \ref{fig:clt_logistic}, while all methods control the FDR at $10\%$, HAMT exhibits the best power followed by ASH 1 and NPMLE B.
\end{itemize}
\begin{figure}
	\centering
	\includegraphics[width=0.85\linewidth]{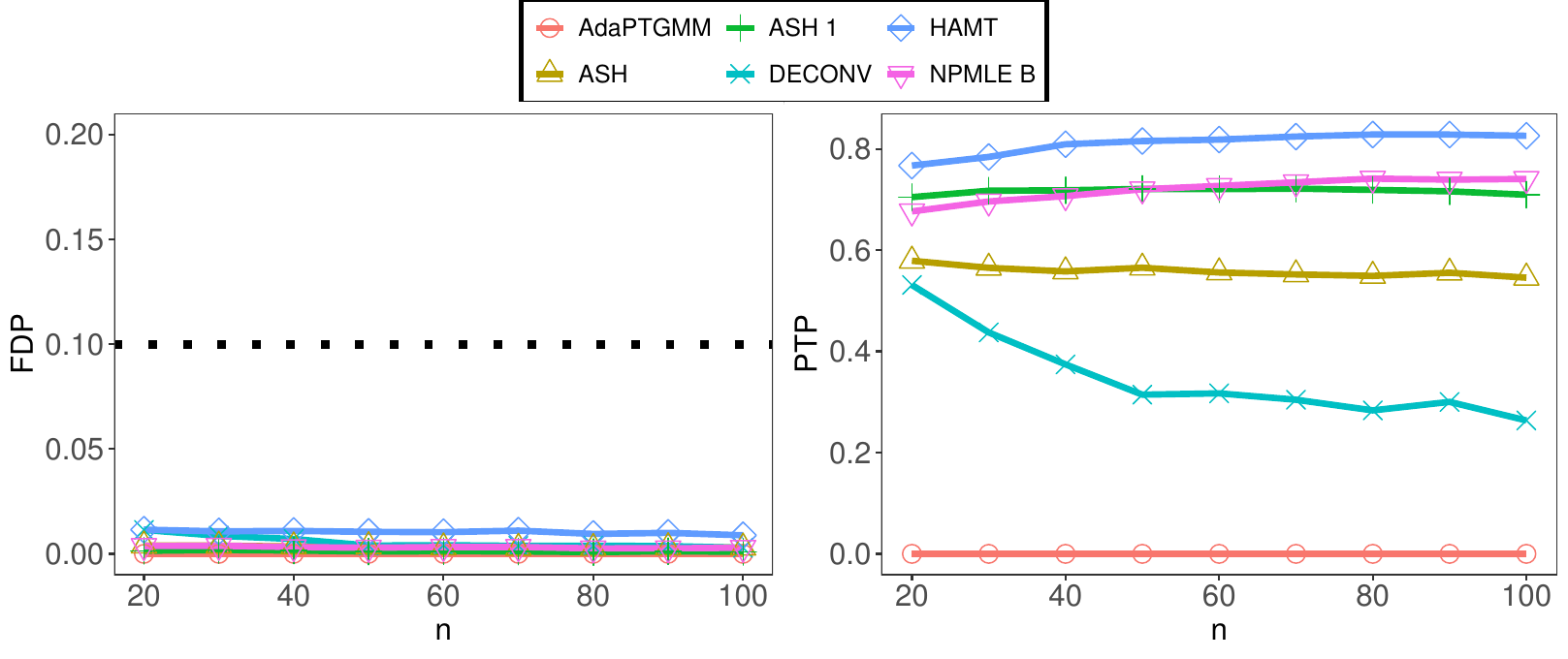}
	\caption{Setting 4: Scenario borrowed from Setting 1 of Section \ref{sec:sims_non_gaussian} with $\mu_i=0.9\delta_{(0)}+0.05\delta_{(4.5\sigma_i)}+0.05\delta_{(-4.5\sigma_i)}$ and $\epsilon_{ij}$ are i.i.d Logistic random variables with location $=0$ and scale $=1$.}
	\label{fig:clt_logistic}
\end{figure}
\section{Further discussion of the numerical performances of various methods} 
\label{sec:more_discuss}
\subsection{HAMT and NPMLE B}
We note that the multiple testing procedures underlying HAMT and NPMLE B differ only with respect to how they estimate $g_\mu(\cdot\mid\sigma)$. Specifically, NPMLE B relies on Problem \eqref{eq:npmle} to estimate $\mathcal W$ while HAMT relies on Problem \eqref{eq:opt}. However, as some of the numerical experiments in sections \ref{sec:sims_onesided} and \ref{sec:more_sims} suggest, this difference can lead to contrasting power and FDR profiles for the two methods. We elaborate on this observation through the lens of the simulation settings for figures \ref{fig:sim_exp4_test} and \ref{fig:sim_exp6_test}. We choose the simulation scenarios accompanying figures \ref{fig:sim_exp4_test} and \ref{fig:sim_exp6_test} because they represent two out of the four settings where the power of HAMT is substantially higher than that of NPMLE B at the same FDR level. The other settings being figures \ref{fig:t_like} and \ref{fig:clt_t}.
\\[1ex]
\noindent\textbf{Analysis of Figure \ref{fig:sim_exp4_test} - }the simulation setting underlying this figure is as follows: $X_i|\mu_i,\sigma_i\stackrel{ind.}{\sim}N(\mu_i,\sigma_i^2),~\sigma_i\stackrel{i.i.d}{\sim}0.9 \text{Unif}(0.5,1)+0.1 \text{Unif}(1,\bar\sigma)$ and $\mu_i=0$, if $\sigma_i\le 1$ and $2/\sigma_i$, otherwise. Here $\mathcal{A}=(-\infty,1]$. Thus, in this setting the signal strength, in terms of the magnitude of the non-null $\mu_i$, decreases as $\bar{\sigma}$ increases from 1.2 to 2. For $\bar{\sigma}=1.2$ and for a random dataset generated from the above hierarchical model, Figure \ref{fig7_u_nn1_rev2} plots the estimate of $f(\cdot\mid\sigma)$ at $\sigma=0.7$ (left), $\sigma=0.95$ (center) and $\sigma=\bar{\sigma}-0.1$ (right). For NPMLE B, these densities are plotted for four different combinations of $(S,K)$ while HAMT relies on the default choices of $S=50, K=10$. We note that while the estimates of $f(\cdot\mid 0.7)$ obtained from HAMT and the NPMLE B variants are virtually indistinguishable from the ground truth, HAMT appears to be relatively better at estimating $f(\cdot\mid0.95)$ and $f(\cdot\mid\bar{\sigma}-0.1)$. Furthermore, the four variants of NPMLE B are distinguishable only in the third panel and their respective PTP values, 0.85 $(S=50, K=10)$, 0.857 $(S=100, K=30)$, 0.842 $(S=100,K=10)$ and 0.878 $(S=50, K=30)$, are marginally less than the PTP of 0.887 returned by HAMT on this random dataset.
Figures \ref{fig7_u_nn2_rev2} and \ref{fig7_u_nn3_rev2}, respectively, represent the cases $\bar{\sigma}=1.4$ and $\bar{\sigma}=1.8$. In both these figures, HAMT is better at estimating $f(\cdot\mid\bar{\sigma}-0.1)$ than the NPMLE B variants. This is also evident in the PTP values that HAMT and the NPMLE B variants return. For instance, in Figure \ref{fig7_u_nn2_rev2} HAMT returns a PTP of 0.857 which is substantially better than 0.109 obtained from the NPMLE B variant with $S=50, K=30$. Furthermore, in Figure \ref{fig7_u_nn3_rev2} the PTP of HAMT is 0.816 which far exceeds the PTP from the NPMLE B variants. Moreover, the choices of $S$ and $K$ seem to have some impact on the power of NPMLE B when $\bar\sigma=1.4$ but relatively less so when $\bar\sigma=1.8$. 
 \begin{figure}
  	\centering
  \includegraphics[width=1\textwidth]{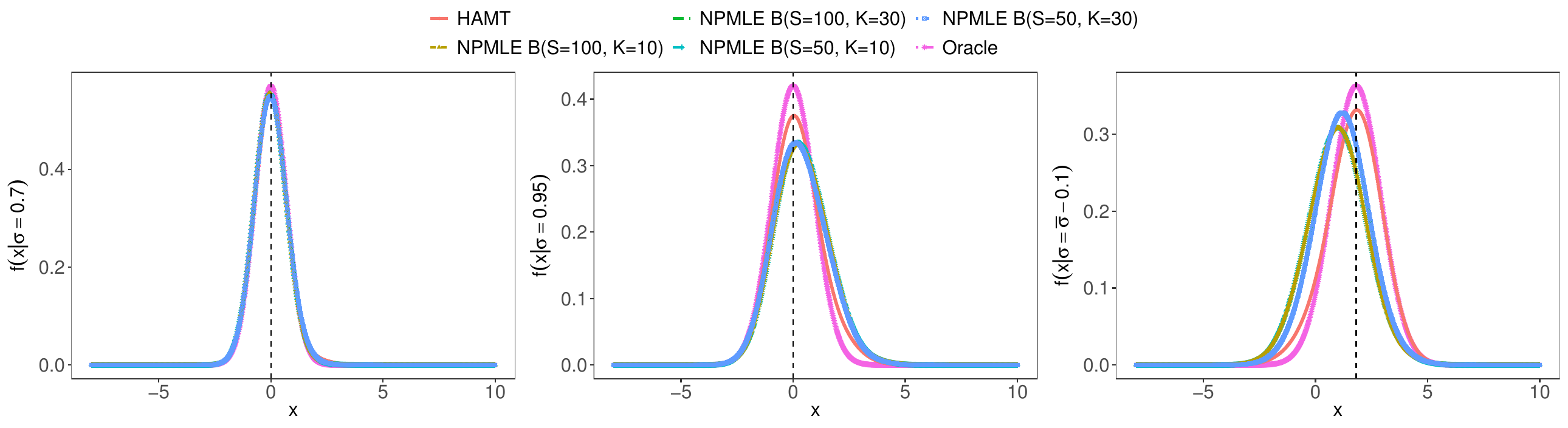}
 	\caption{The simulation setting for Figure \ref{fig:sim_exp4_test} with $\bar{\sigma}=1.2$. For a random dataset generated from the underlying hierarchical model, the three panels present the estimate of $f(\cdot\mid\sigma)$ at $\sigma=0.7$ (left), $\sigma=0.95$ (center) and $\sigma=\bar{\sigma}-0.1$ (right) obtained from HAMT and NPMLE B. The PTP values from the four variants of NPMLE B are: 0.85 $(S=50, K=10)$, 0.857 $(S=100, K=30)$, 0.842 $(S=100,K=10)$ and 0.878 $(S=50, K=30)$. The PTP from HAMT is 0.887.}
  	\label{fig7_u_nn1_rev2}
 \includegraphics[width=1\textwidth]{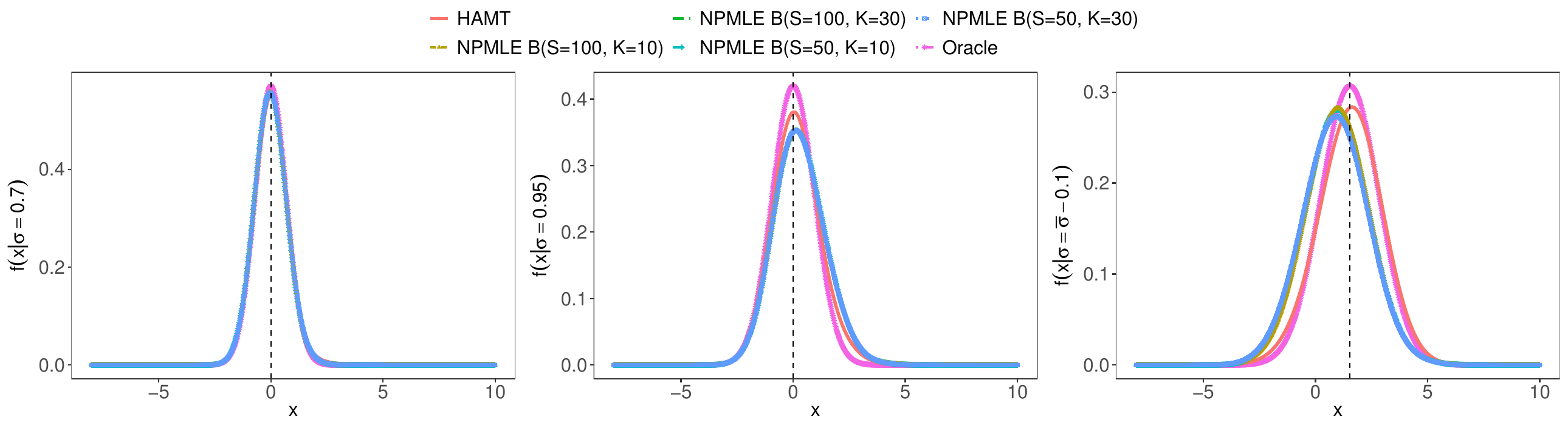}
  	\caption{Same as Figure \ref{fig7_u_nn1_rev2} but with $\bar{\sigma}=1.4$. The PTP values from the four variants of NPMLE B are: 0.089 $(S=50, K=10)$, 0.103 $(S=100, K=30)$, 0.073 $(S=100,K=10)$ and 0.109 $(S=50, K=30)$. The PTP from HAMT is 0.857.}
 	\label{fig7_u_nn2_rev2}
   \includegraphics[width=1\textwidth]{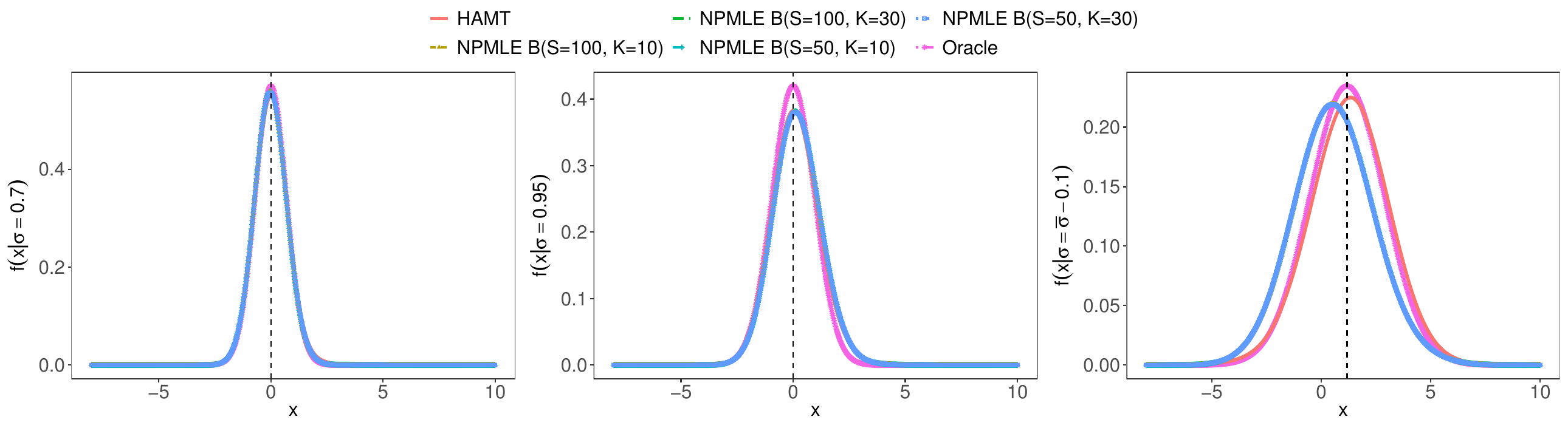}
  	\caption{Same as Figure \ref{fig7_u_nn1_rev2} but with $\bar{\sigma}=1.8$. The PTP values from the four variants of NPMLE B are: 0.018 $(S=50, K=10)$, 0.01 $(S=100, K=30)$, 0.01 $(S=100,K=10)$ and 0.017 $(S=50, K=30)$. The PTP from HAMT is 0.816.}
 	\label{fig7_u_nn3_rev2}
  \end{figure}
  \begin{figure}
   	\centering
  \includegraphics[width=0.85\textwidth]{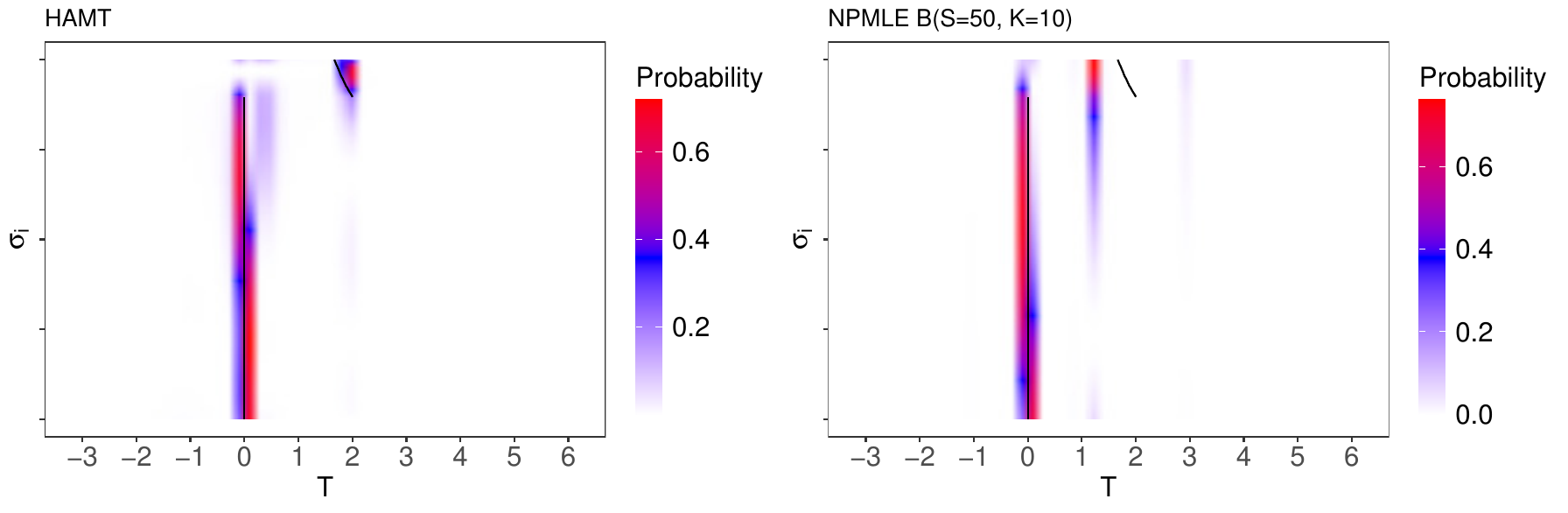}
  	\caption{The simulation setting for Figure \ref{fig:sim_exp4_test} with $\bar{\sigma}=1.2$. The left panel presents the $m\times S$ matrix $\mathcal G=(\hat{\bm g}_1,\ldots,\hat{\bm g}_m)^T$ of the estimated prior probabilities obtained from {HAMT}, where $\hat{\bm g}_i=\{\hat{\bm w}_j^T\bm q(\sigma_i):1\le j\le S\}$. The black vertical line depicts $\mu_i=0$ whenever $\sigma_i\le 1$ and the black oblique line is $\mu_i=2/{\sigma_i}$ for $\sigma_i>1$. The right panel represents the estimated prior probabilities from NPMLE B with $S=50, K=10$. }
   	\label{fig7_nn1_pp_rev2}
   \includegraphics[width=0.85\textwidth]{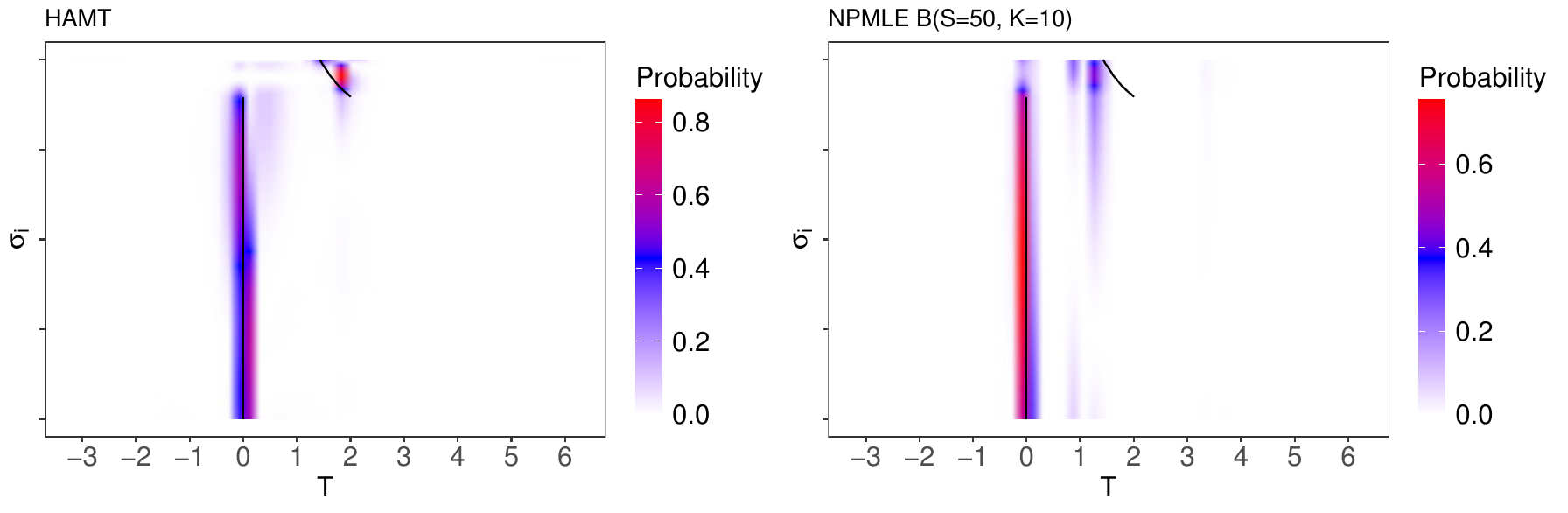}
   	\caption{Same as Figure \ref{fig7_nn1_pp_rev2} but with $\bar{\sigma}=1.4$.}
  	\label{fig7_nn2_pp_rev2}
  \includegraphics[width=0.85\textwidth]{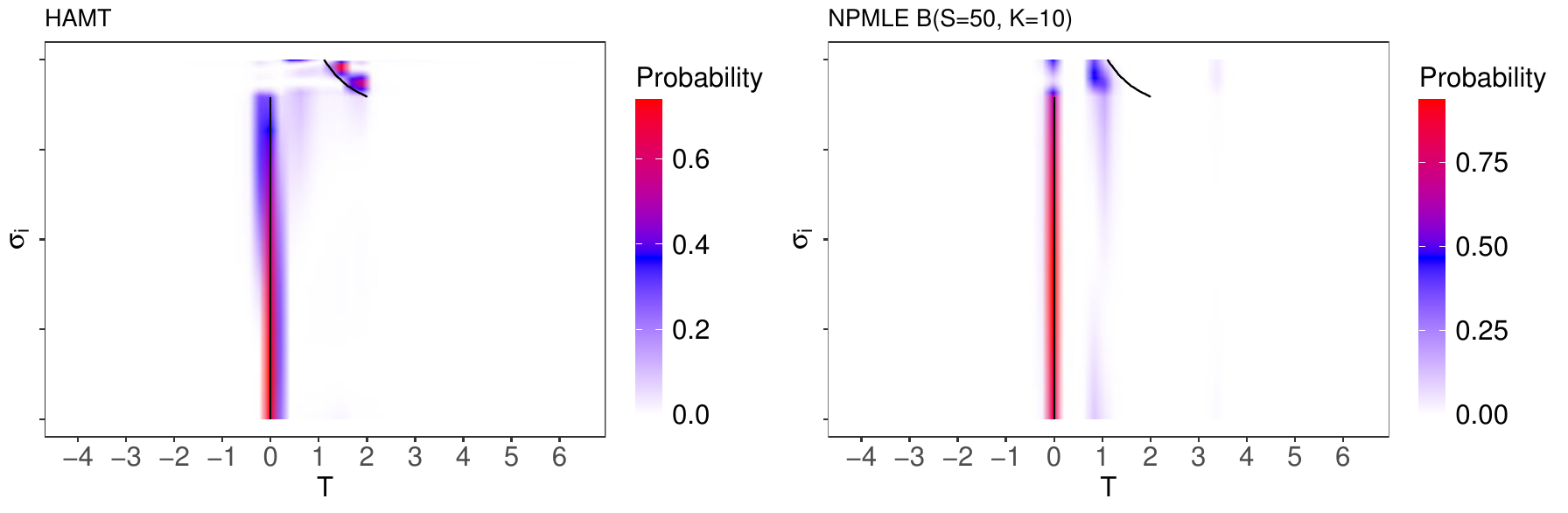}
   	\caption{Same as Figure \ref{fig7_nn1_pp_rev2} but with $\bar{\sigma}=1.8$.}
   	\label{fig7_nn3_pp_rev2}
 \end{figure}

To further understand why the marginal density estimates from HAMT and NPMLE B differ, we examine their respective estimates of $g_\mu(\cdot\mid\sigma)$. When $\bar{\sigma}=1.2$, the left panel of Figure \ref{fig7_nn1_pp_rev2} presents the $m\times S$ matrix $\mathcal G=(\hat{\bm g}_1,\ldots,\hat{\bm g}_m)^T$ of the estimated prior probabilities obtained from {HAMT}, where $\hat{\bm g}_i=\{\hat{\bm w}_j^T\bm q(\sigma_i):1\le j\le S\}$. The horizontal axis represents the support of $\mu_i$ which is given by the grid $\mathcal T$ and the vertical axis is ${\sigma}_i$ sorted in decreasing order from top to bottom. The black vertical line depicts $\mu_i=0$ whenever $\sigma_i\le 1$ and the black oblique line is $\mu_i=2/{\sigma_i}$ for $\sigma_i>1$. In the same spirit, the right panel depicts the estimated prior probabilities obtained from {NPMLE B} with $S=50, K=10$. 

We make several observations from Figure \ref{fig7_nn1_pp_rev2}. First, when $\sigma_i\le 1$ both HAMT and NPMLE B assign probability masses around zero as they should since $\mu_i=0$ whenever $\sigma_i\le 1$. However, unlike HAMT, NPMLE B also assigns some mass around $1$. Second, when $\sigma_i>1$ HAMT assigns substantially less mass around $0$ than NPMLE B. Finally, while HMAT correctly assigns probability mass around the true non-null $\mu_i$'s, NPMLE B does not and instead assigns a relatively large mass around $1$ and $3$. The aforementioned differences in the estimated prior probabilities from the two methods manifest themselves in the corresponding conditional marginal density estimates and hence in the respective rankings of the $m$ hypotheses obtained from their Clfdr statistics. Figure \ref{fig7_nn2_pp_rev2} represents the case $\bar{\sigma}=1.4$ and reveals that, unlike HAMT, NPMLE B fails to assign any prior mass around most of the non-null $\mu_i$'s, thus explaining its low power in Figure \ref{fig:sim_exp4_test}. Furthermore, in contrast to HAMT, NPMLE B assigns relatively more probability mass at $0$ even when $\sigma_i> 1$. Finally, Figure \ref{fig7_nn3_pp_rev2} represents the case $\bar{\sigma}=1.8$ where NPMLE B continues to assign zero probability mass around most of the non-null $\mu_i$'s. 
\\[1ex]
\noindent\textbf{Analysis of Figure \ref{fig:sim_exp6_test} - }the simulation setting underlying this figure provides another opportunity to analyze the differences between HAMT and NPMLE B. Here $X_i|\mu_i,\sigma_i\stackrel{ind.}{\sim}N(\mu_i,\sigma_i^2),~\sigma_i\stackrel{i.i.d.}{\sim}\text{Unif}(0.25,\bar\sigma)$, conditional on $\sigma_i$, $\mu_i=3\sigma_i$ and $\mathcal{A}=(-\infty,4]$. Thus, as $\bar\sigma$ increases from 1.5 to 2, the signal strength increases. With $\bar{\sigma}=1.5$ and for a random dataset generated from the above hierarchical model, Figure \ref{fig8_u_nn1_rev2} plots the estimates of $f(\cdot\mid\sigma)$ obtained from HAMT and the four variants of NPMLE B at $\sigma=0.5$ (left), $\sigma=1$ (center) and $\sigma=\bar{\sigma}-0.1$ (right).  Figures \ref{fig8_u_nn2_rev2} and \ref{fig8_u_nn3_rev2}, respectively, represent the cases $\bar{\sigma}=1.7$ and $\bar{\sigma}=1.9$. Overall, the marginal density estimates at $\sigma=1$ and $\bar{\sigma}-0.1$ are substantially closer to the ground truth in case of HAMT. {Also, the choices of $S$ and $K$ do not seem to generate any visible differences in the marginal density estimates from NPMLE B and this observation is also supported by the relatively similar PTP values that these variants return.}
\begin{figure}
   	\centering
  \includegraphics[width=1\textwidth]{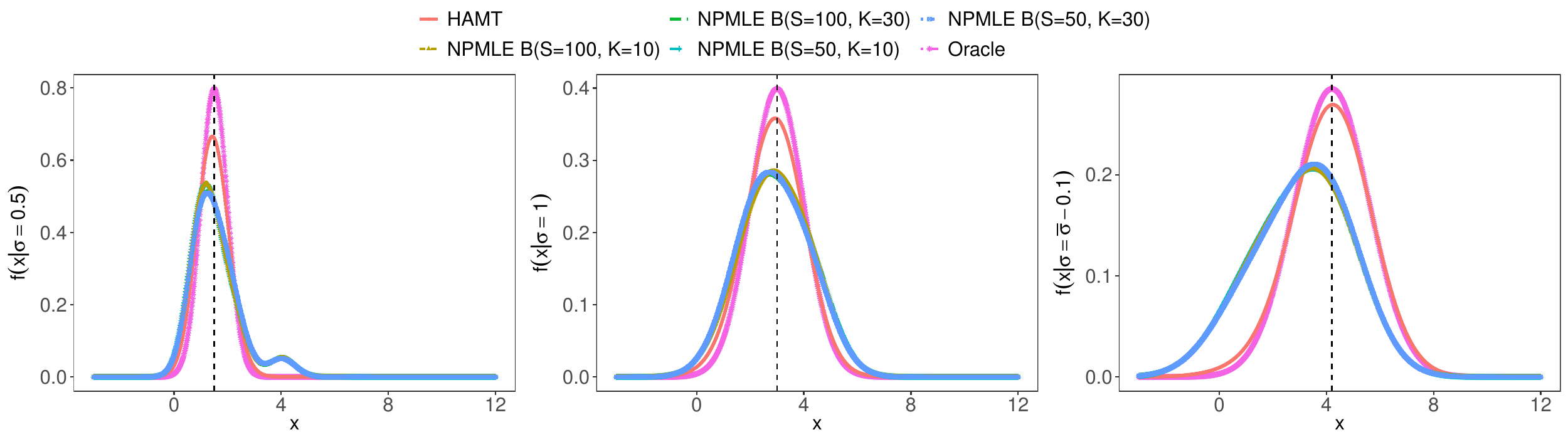}
  	\caption{The simulation setting for Figure \ref{fig:sim_exp6_test} with $\bar{\sigma}=1.5$. For a random dataset generated from the underlying hierarchical model, the three panels present the estimate of $f(\cdot\mid\sigma)$ at $\sigma=0.5$ (left), $\sigma=1$ (center) and $\sigma=\bar{\sigma}-0.1$ (right) obtained from HAMT and NPMLE B. The PTP values from the four variants of NPMLE B are: 0.557 $(S=50, K=10)$, 0.551 $(S=100, K=30)$, 0.559 $(S=100,K=10)$ and 0.538 $(S=50, K=30)$. The PTP from HAMT is 0.873.}
   	\label{fig8_u_nn1_rev2}
   \includegraphics[width=1\textwidth]{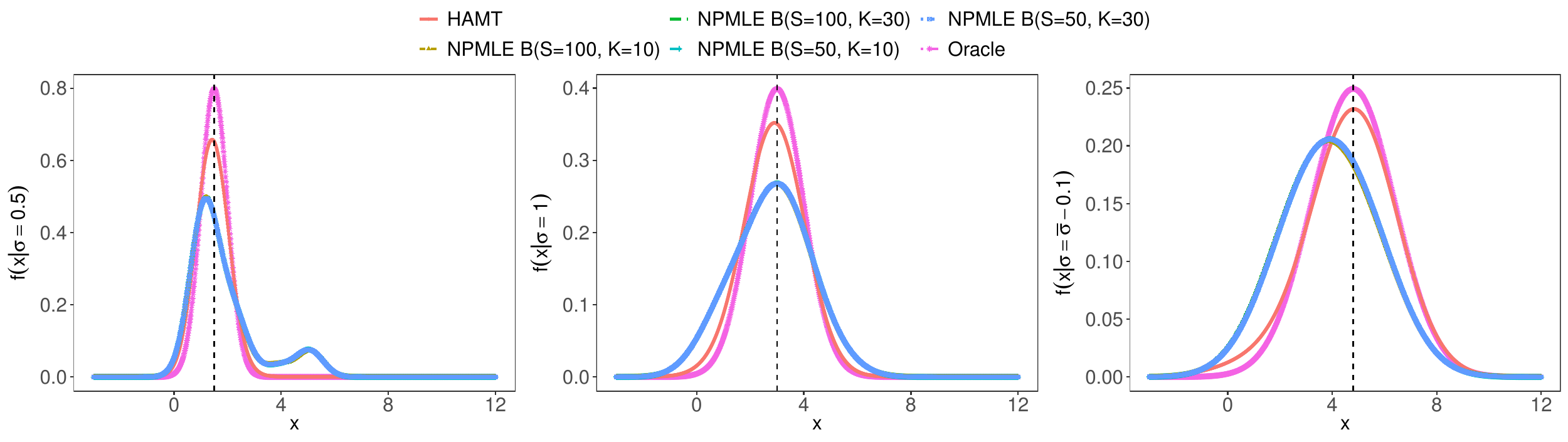}
   	\caption{Same as Figure \ref{fig8_u_nn1_rev2} but with $\bar{\sigma}=1.7$. The PTP values from the four variants of NPMLE B are all approximately 0.024. The PTP from HAMT is 0.797.}
  	\label{fig8_u_nn2_rev2}
    \includegraphics[width=1\textwidth]{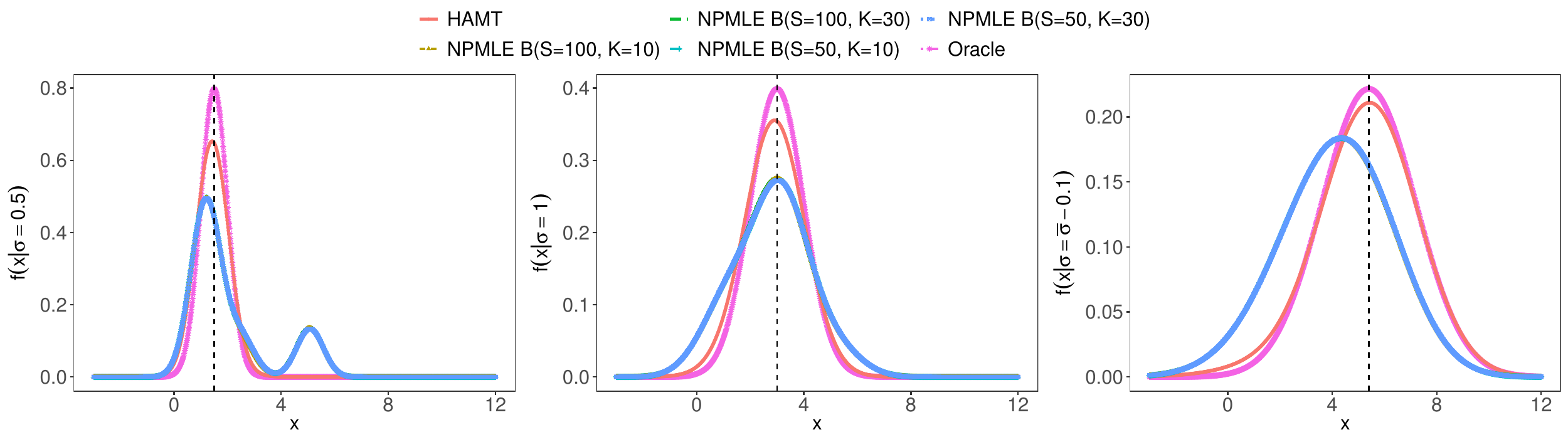}
   	\caption{Same as Figure \ref{fig8_u_nn1_rev2} but with $\bar{\sigma}=1.9$. The PTP values from the four variants of NPMLE B are: 0.468 $(S=50, K=10)$, 0.488 $(S=100, K=30)$, 0.489 $(S=100,K=10)$ and 0.470 $(S=50, K=30)$. The PTP from HAMT is 0.882.}
  	\label{fig8_u_nn3_rev2}
   \end{figure}
      \begin{figure}
 	\centering
 \includegraphics[width=0.85\textwidth]{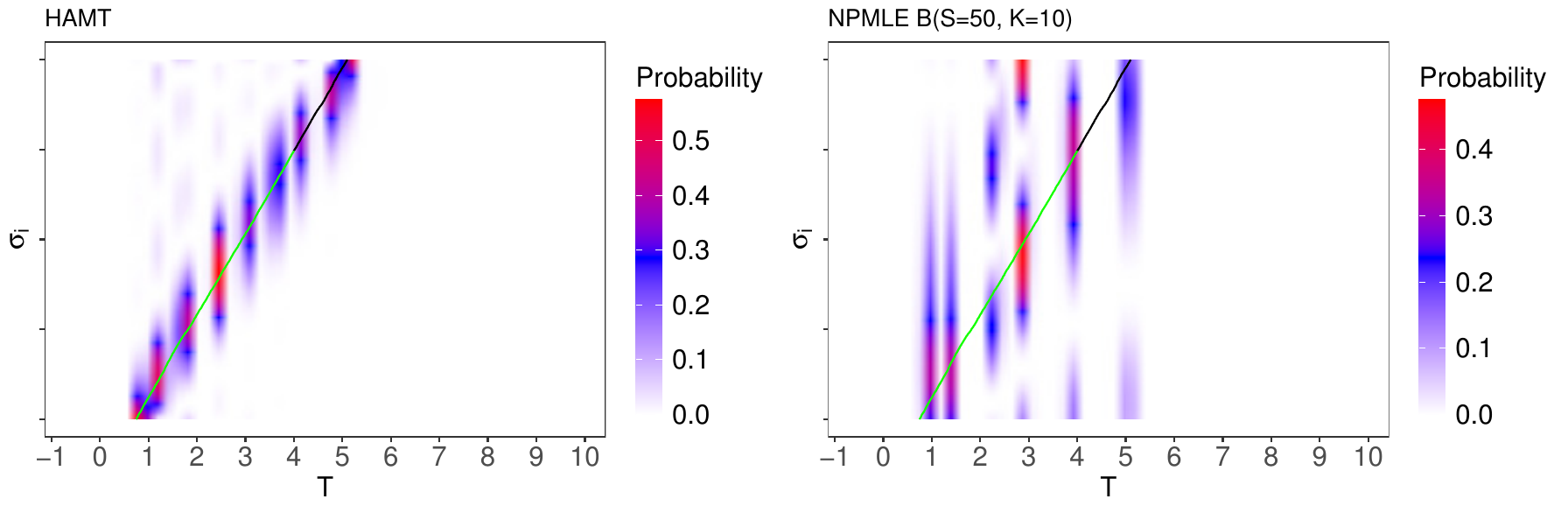}
 	\caption{The simulation setting for Figure \ref{fig:sim_exp6_test} with $\bar{\sigma}=1.7$. The left panel presents the $m\times S$ matrix $\mathcal G=(\hat{\bm g}_1,\ldots,\hat{\bm g}_m)^T$ of the estimated prior probabilities obtained from {HAMT}, where $\hat{\bm g}_i=\{\hat{\bm w}_j^T\bm q(\sigma_i):1\le j\le S\}$. The green and black  oblique lines depict, respectively, the null and non-null $\mu_i$'s. The right panel represents the estimated prior probabilities from NPMLE B with $S=50, K=10$.}
\label{fig8_nn2_pp_rev2}
 	\centering
\includegraphics[width=0.85\textwidth]{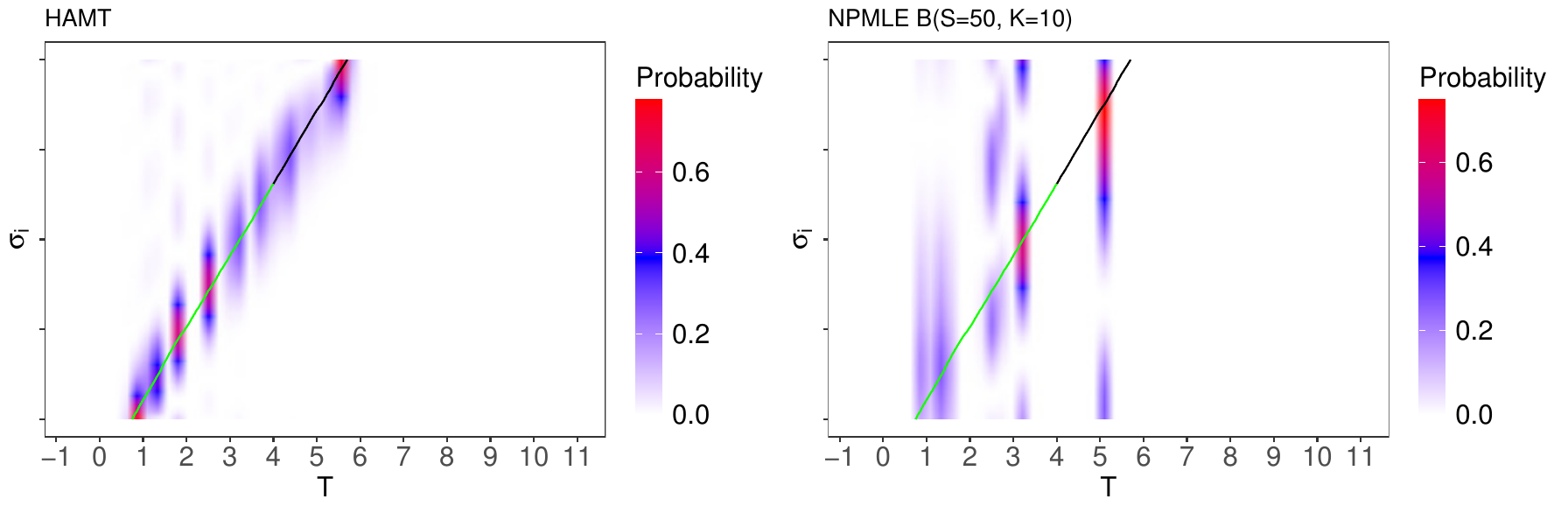}
 	\caption{Same as Figure \ref{fig8_nn2_pp_rev2} but with $\bar{\sigma}=1.9$.}
 \label{fig8_nn3_pp_rev2}
 \end{figure}

When $\bar{\sigma}=1.7$, Figure \ref{fig8_nn2_pp_rev2} presents the estimated prior probabilities obtained from {HAMT} (left panel) and NPMLE B (right panel) with $S=50, K=10$. The green and black  oblique lines depict, respectively, the null and non-null $\mu_i$'s. The disparity in the prior probability estimates from HAMT and NPMLE B in this setting is stark. The probabilities from HAMT, in particular, align closely with the true prior distribution that assigns all probability mass at $3\sigma_i$. For instance, when $\sigma_i=1$, HAMT assigns virtually no probability mass at 2 and below, and 5 and above. NPMLE B, in contrast, assigns mass around $1$ and $5$. Furthermore, NPMLE B does not assign any probability mass to the non-null $\mu_i$'s in $(4,5]$ while HAMT does. These differences manifest themselves in the corresponding marginal density estimates of $f(\cdot\mid\sigma)$ as seen in Figure \ref{fig8_u_nn2_rev2}. Figure \ref{fig8_nn3_pp_rev2} presents the same information at $\bar{\sigma}=1.9$. For the non-null $\mu_i$'s, NPMLE B assigns probability mass only around 5, thus missing a substantial fraction of the non-null $\mu_i$'s. HAMT, in contrast, is relatively better in estimating $g_\mu(\cdot\mid\sigma)$ in this setting, which ultimately results in its improved power and FDR profiles compared to NPMLE B. 
\subsection{Insights on when each method is expected to perform well.}
\label{sec:insights}
In the simulation studies of sections \ref{sec:sims_onesided} and \ref{sec:more_sims}, there are six procedures, besides HAMT, that rely on different approaches for estimating $g_\mu(\cdot|\sigma)$. These include two procedures from \cite{gu2018oracle} (GS1 and GS2), a procedure based on the deconvolution method of \cite{efron2016empirical} (DECONV), two procedures from \cite{stephens2017false} (ASH and ASH.1) and NPMLE B. We discuss when each of these methods can be expected to outperform HAMT in power at the same FDR level. 
\\[1ex]
\noindent\textbf{GS1, GS2 and DECONV - }GS1 and GS2 are based on the standardized statistic $Z_i=(X_i-\mu_0)/\sigma_i$ and rely on the deconvolution estimate obtained from nonparametric maximum likelihood estimation to construct the Lfdr statistic. DECONV, on the other hand, ignores the dependence between $\mu_i$ and $\sigma_i$. Recall that the discussions in sections \ref{sec:standardization} and \ref{sec:dependence} reveal that these three procedures may exhibit substantial power loss when $\mu_i$ and $\sigma_i$ are correlated. However when $\mu_i$ is indeed independent of $\sigma_i$, GS1, GS2 and DECONV can be more powerful than HAMT. An example of this scenario is presented in Figure \ref{fig:sim_exp1_test} where GS1 and DECONV are competitive to HAMT in power.
\\[1ex]
\noindent\textbf{NPMLE B - } this is the method that is  closest to HAMT in its construction. While the simulation experiments in sections \ref{sec:sims_onesided} and \ref{sec:more_sims} reveal several settings where HAMT is more powerful than NPMLE B, there are also scenarios where NPMLE B is competitive. These include the cases where (i) the distribution of $\mu_i$ depends on $\sigma_i$, such as Figure \ref{fig:sim_exp8_test}, and (ii) the distribution of $X_i$ given $\mu_i,\sigma_i$ is non Gaussian, such as Figure \ref{fig:laplace_like}. A particular setting where NPMLE B can indeed outperform HAMT is when $\mu_i$ and $\sigma_i$ are independent. Figures \ref{fig:ash_vs_hamt_1} -- \ref{fig:ash_vs_hamt_3} represent three such scenarios. Intuitively, when $g_\mu(\cdot\mid\sigma_i)\coloneqq g_\mu(\cdot)$ is independent of $\sigma_i$, directly maximizing the marginal log-likelihood $\sum_{i=1}^{m}\log f(x_i\mid\sigma_i)$ (Problem \eqref{eq:npmle}) is a better approach for learning $g_\mu(\cdot)$ than the density matching approach of HAMT (Problem \eqref{eq:opt}) because the $\sigma_i$'s do not encode any extra information regarding the $\mu_i$'s that the bivariate kernel density estimator $\hat{\varphi}_{(-i)}^m(x_i,\sigma_i)$ (Equation \eqref{eq:kde_hart_jacknife}) of $f(x_i\mid\sigma_i)$ can utilize. However, as discussed in Section \ref{sec:deconvolution}, much research is needed to fully comprehend the conditions under which NPMLE B can guarantee asymptotic FDR control and how the hyper-parameters $S$ and $K$ impact its power.
\\[1ex]
\noindent\textbf{ASH and ASH.1 - }an important assumption underlying these two procedures is that the distribution of the non-null $\mu_i$ is unimodal. As \cite{stephens2017false} comment, this unimodal assumption (UA) is naturally satisfied in many practical settings. However, in the simulation studies presented in sections \ref{sec:sims_onesided} and \ref{sec:more_sims}, the distribution of the non-null $\mu_i$ do not satisfy the UA. This is the main reason why ASH and its variant ASH.1 are typically less powerful than HAMT in our numerical experiments. In this section, we borrow three simulation settings from \cite{stephens2017false} and demonstrate that when the UA holds, ASH can be more powerful than HAMT. 
\begin{figure}
		\centering
		\includegraphics[width=0.8\textwidth]{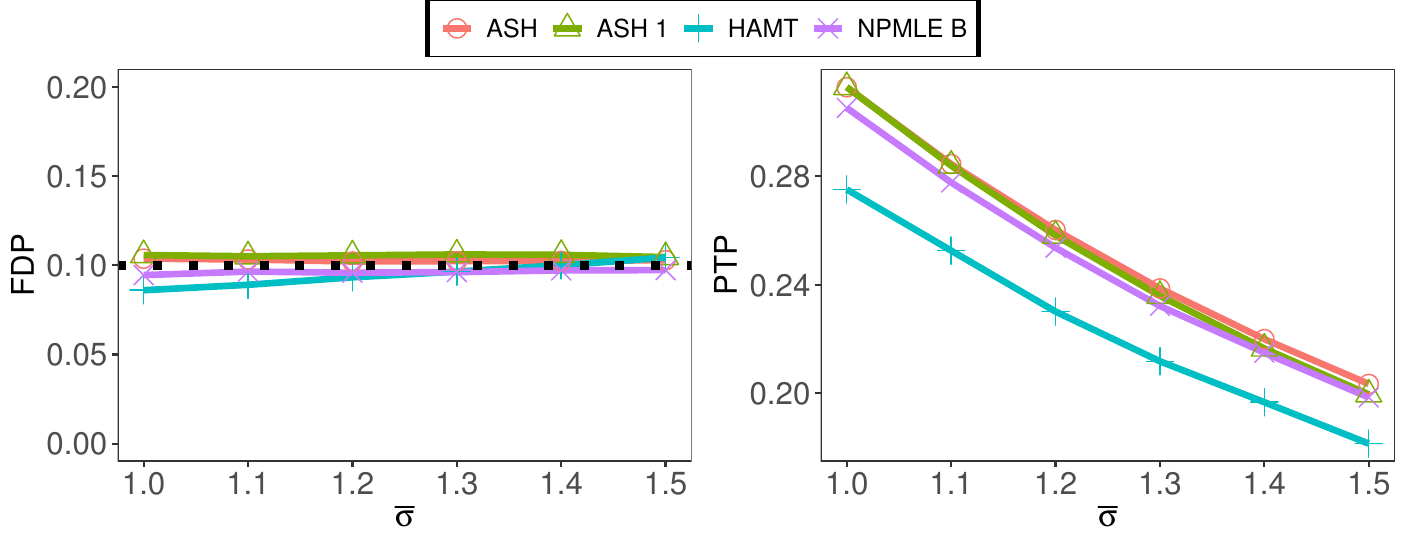}
		\caption{Setting 1: $\sigma_i\stackrel{i.i.d}{\sim}\text{Unif}(0.25,\bar{\sigma})$, $\mu_i\stackrel{i.i.d}{\sim}0.9\delta_{(0)}+0.1F_a$,
			$X_i|\mu_i,\sigma_i\stackrel{ind.}{\sim}N(\mu_i,\sigma_i^2)$. $F_a=0.4N(0.0.25^2)+0.2N(0,0.5^2)+0.2N(0,1)+0.4(0,2^2)$ and $\mathcal A=(-\infty, 1/2]$.}
		\label{fig:ash_vs_hamt_1}
	\end{figure}
	\begin{figure}
		\centering
		\includegraphics[width=0.8\textwidth]{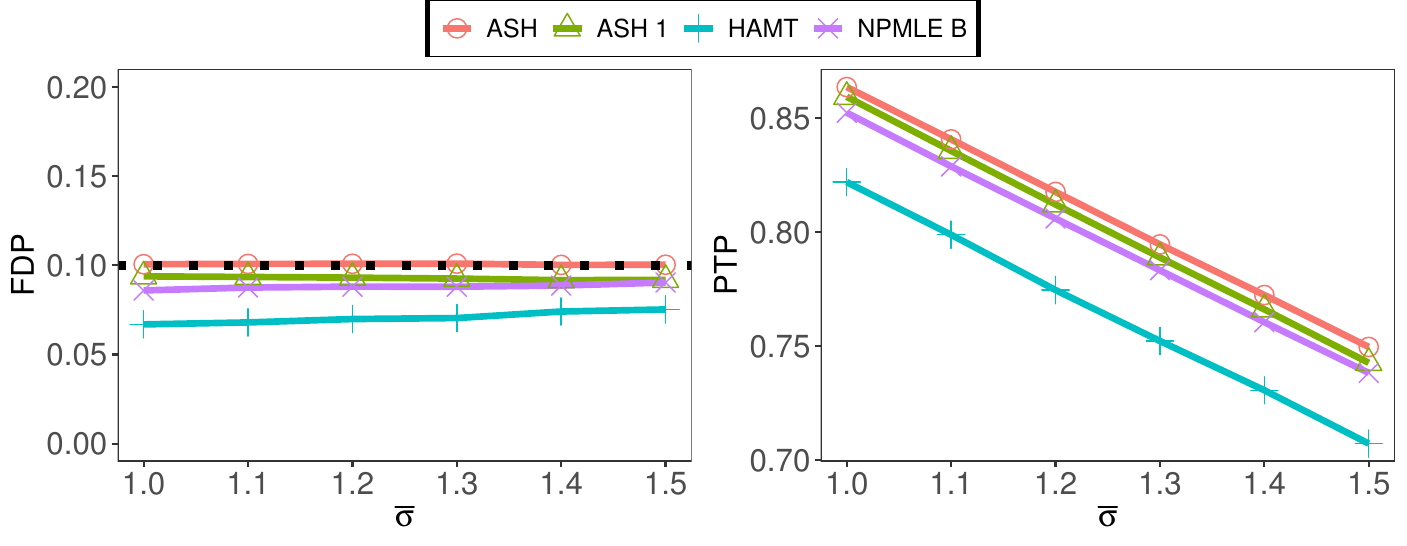}
		\caption{Setting 2: Same as Figure \ref{fig:ash_vs_hamt_1} but $F_a=N(0,4^2)$ and $\mathcal A=(-\infty, 1]$.}
		\label{fig:ash_vs_hamt_2}
	\end{figure}
	\begin{figure}
		\centering
		\includegraphics[width=0.8\textwidth]{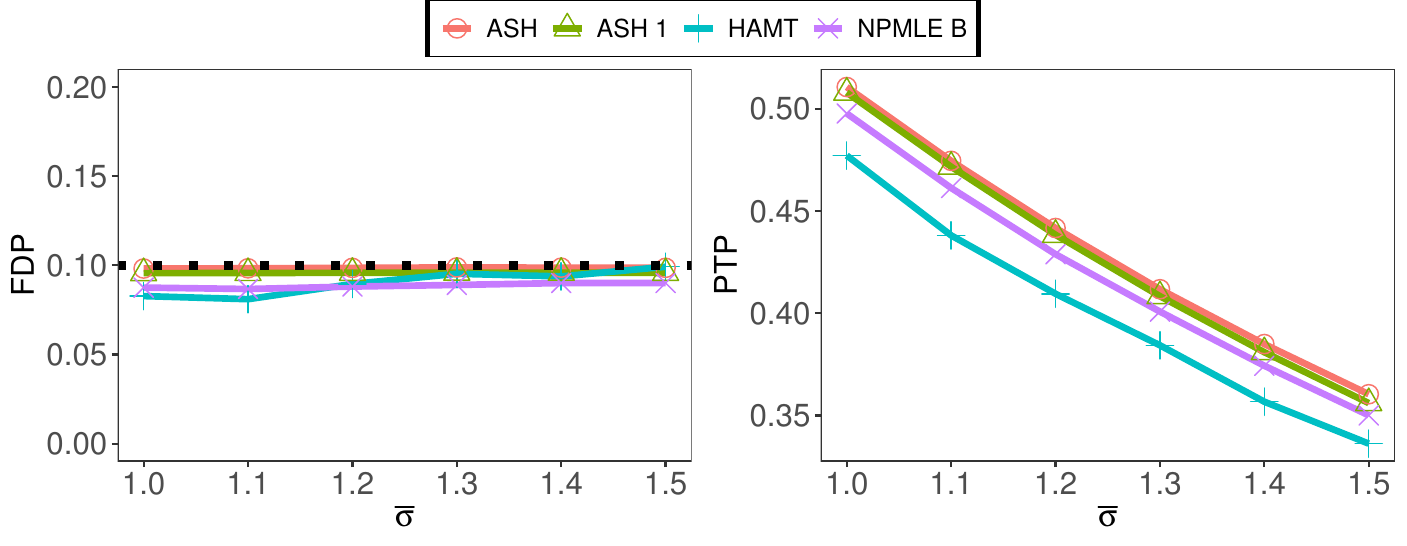}
		\caption{Setting 3: Same as Figure \ref{fig:ash_vs_hamt_1} but $F_a=(2/3)N(0,1)+(1/3)N(0,2^2)$.}
		\label{fig:ash_vs_hamt_3}
	\end{figure}

We let $\sigma_i\stackrel{i.i.d}{\sim}\text{Unif}(0.25,\bar{\sigma})$, $\mu_i\stackrel{i.i.d}{\sim}0.9\delta_{(0)}+0.1F_a$,
	$X_i|\mu_i,\sigma_i\stackrel{ind.}{\sim}N(\mu_i,\sigma_i^2)$, set $\alpha=0.1$ and vary $\bar\sigma\in\{1,1.1,1.2,1.3,1.4,1.5\}$. We are interested in testing $m=10^4$ hypotheses of the form $H_{0i}:\mu_i\in\mathcal A~vs~H_{1i}:\mu_i\notin\mathcal A$ where $\mathcal A=(-\infty, \mu_0]$. The distributions $F_a$ are borrowed from Table S.3 of \cite{stephens2017false} as follows:
	\begin{itemize}
		\item \textbf{Setting 1 - }in this scenario  $F_a=0.4N(0.0.25^2)+0.2N(0,0.5^2)+0.2N(0,1)+0.4(0,2^2)$ is a `Spiky' distribution. Here $\mu_0=1/2$.
		\item \textbf{Setting 2 - }here $F_a=N(0,4^2)$ is the `big-Normal' and $\mu_0=1$.
		\item \textbf{Setting 3 - }here $F_a=(2/3)N(0,1)+(1/3)N(0,2^2)$ is the `near-Normal' distribution and $\mu_0=1/2$.
	\end{itemize}
Figures \ref{fig:ash_vs_hamt_1} -- \ref{fig:ash_vs_hamt_3} report the average FDP and PTP of HAMT, ASH, ASH.1 and NPMLE B across 200 Monte-Carlo repetitions of the data generating process. Unsurprisingly, ASH and ASH.1 are the top two methods in terms of their power, further suggesting that in applications where the UA assumption is known to hold, ASH and its variant are indeed powerful alternatives to HAMT. Under these three settings, the distribution of $\mu_i$ is independent of $\sigma_i$ and we find that NPMLE B is also more powerful than HAMT.
\section{Real data analysis}
\label{sec:realdata}
In this section we analyze a dataset from \cite{banerjee2019large} that hold daily player-level gaming information over 60 days from a mobile app game. For monetization of these games, managers are often interested in identifying a group of players who are most engaged with the game so that personalized promotional offers can be pushed to their devices. While there are several ways of measuring game engagement, such as engagement via purchases or through social media activity, here we use the daily duration of play as a measure of how engaged each player is with the game. However, a positive daily duration of play does not necessarily mean that the player is highly engaged. Rather, from a game manager's perspective, sustained playing activities translate to high levels of engagement, either through purchases or social media activities. Thus, in this analysis we focus on players who have logged-in to the game for at least 5 days in the 60 day period and the goal is to select those players whose mean daily duration of play exceeds $30$ minutes. 

Formally, let $Y_{ij}>0$ denote the duration of play in minutes for player $i$ on day $j$ where $j=1,\ldots,n_i$ and $i=1,\ldots,m$. Here $n_i\in[5,60]$ denotes the number days that player $i$ has logged-in to the game and there are $m=10,336$ such players in our data. Following \cite{banerjee2019large}, we work with the log duration of play $X_{ij}=\log Y_{ij}$ and denote $X_i=n_i^{-1}\sum_{j=1}^{n_i}\log Y_{ij}$. We assume that $X_i\mid (n_i,\mu_i,\sigma_i)\stackrel{ind.}{\sim}N(\mu_i,\sigma_i^2)$, and test $H_{0,i}:\mu_i\le \log(30)~vs~H_{1,i}:\mu_i>\log(30)$. Since $\sigma_i$ are unknown in this example, we calculate the sample standard deviation $S_i$ and consider the $m$ pairs $(X_i,{\sigma}_i)$ for the testing problem, where we set ${\sigma}_i= S_i/\sqrt{n_i}$ with some abuse of notation. %The FDR level is set at $\alpha=0.1$ and we include GS 1, GS2 and DECONV along with HAMT. S_i/\sqrt{n_i}
\begin{figure}
	\centering
	\includegraphics[width=0.65\linewidth]{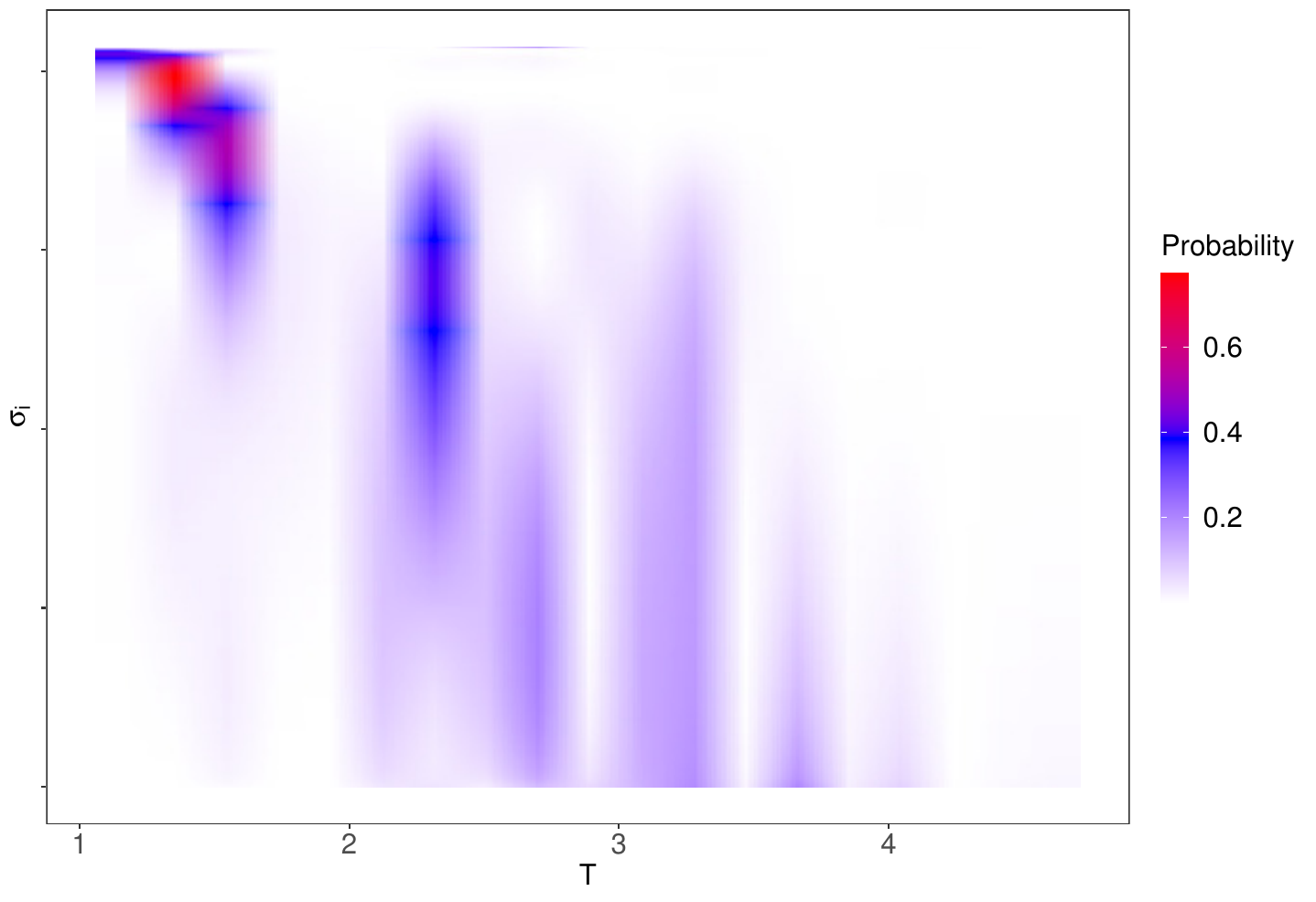}
	\caption{The heatmap representing the $m\times S$ matrix $\mathcal G=(\hat{\bm g}_1,\ldots,\hat{\bm g}_m)^T$ where $\hat{\bm g}_i=\{\hat{\bm w}_j^T\bm q(\sigma_i):1\le j\le S\}$. The horizontal axis represents the support of $\mu_i$ which is give by the grid $\mathcal T$, truncated to $[1,4.6]$, and the vertical axis is the standard error ${\sigma}_i$.}
	\label{fig:realdata_gameheatmap}
\end{figure}

We first discuss the estimate of prior probabilities arising from the deconvolution estimator that HAMT relies on. The heatmap in Figure \ref{fig:realdata_gameheatmap} presents the $m\times S$ matrix $\mathcal G=(\hat{\bm g}_1,\ldots,\hat{\bm g}_m)^T$ of the estimated prior probabilities where $\hat{\bm g}_i=\{\hat{\bm w}_j^T\bm q(\sigma_i):1\le j\le S\}$. The horizontal axis represents the support of $\mu_i$ which is give by the grid $\mathcal T$, truncated to $[1,4.6]$ for ease of presentation, and the vertical axis is ${\sigma}_i$ arranged in an increasing order from bottom to top. It is interesting to note that when ${\sigma}_i$ are small, most of the prior mass is concentrated in $[2,4]$. As ${\sigma}_i$ increases, the deconvolution estimator adjusts and assigns more mass in the interval $[1,3]$. This is further elucidated in Figure \ref{fig:realdata_gamepriorprobs} where we plot $\hat{\bm g}_i$ %the estimated prior density $\hat{g}_{ij}= \hat{g}_\mu(\mu=u_j|{\sigma}_i=\sigma)$ 
for $\sigma_i=\sigma\in\{0.1,0.5,1\}$, %$u_j\in S$
and notice a change in the spread of the estimated prior density as $\sigma$ increases from left to right. Deconvolution estimators that ignore the dependence between $\mu_i$ and ${\sigma}_i$ are incapable of demonstrating such patterns in the estimated prior density.
\begin{figure}
	\centering
	\includegraphics[width=1\linewidth]{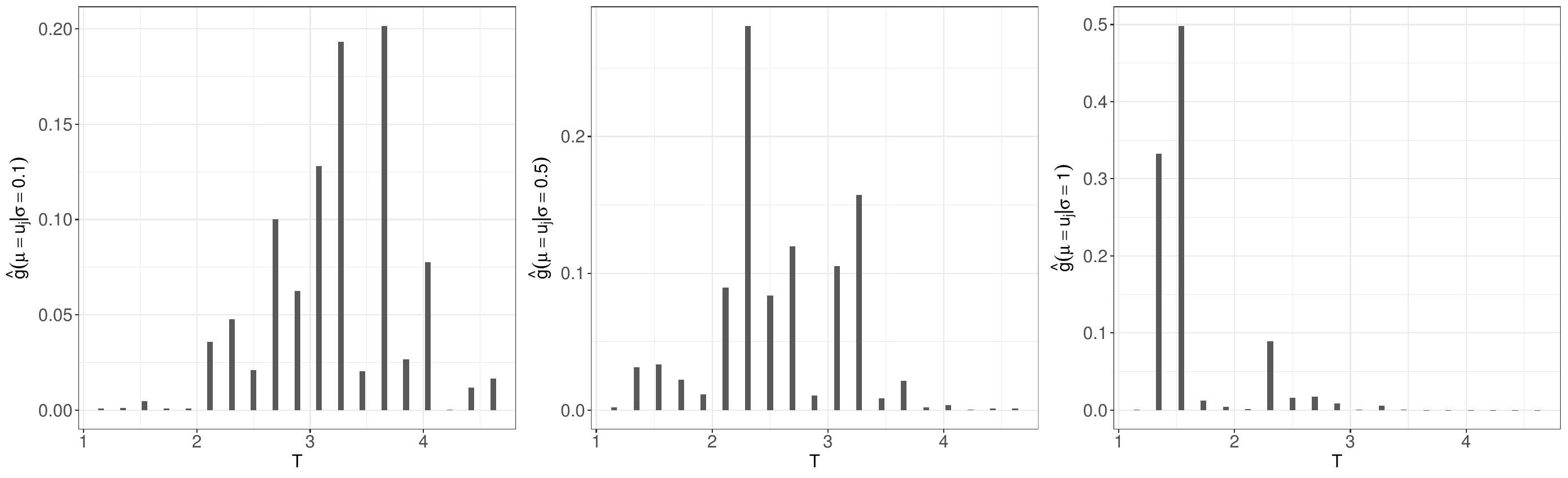}
	\caption{Plot of the estimated prior masses $\hat{\bm g}_i$ for $\sigma_i=\sigma\in\{0.1,0.5,1\}$. %density $\hat{g}_{ij}= \hat{g}_\mu(\mu=S|{\sigma}_i=\sigma)$ where $\sigma\in\{0.1,0.5,1\}$. 
			The horizontal axis is truncated to $[1,4.6]$ as the estimated probability mass is negligible outside this interval.}
	\label{fig:realdata_gamepriorprobs}
\end{figure}

For the multiple testing problem described earlier, HAMT relies on the deconvolution estimates $\hat{\bm g}_{i}$ to estimate the oracle Clfdr statistic $T_i^{\sf OR}$. Table \ref{tab:realdata_game} reports the percentage of hypotheses rejected by each method for different choices of the FDR level $\alpha$ and we find that HAMT rejects more hypotheses than GS 1 and GS 2 while ASH 1 rejects the most. 
\begin{table}[!t]
	\centering
	\caption{Percentage of hypotheses rejected by each method.}
	\begin{tabular}{cccccccc}
			\hline
			$\alpha$ & GS 1  & GS 2  & DECONV & NPMLE B & ASH & ASH 1 &HAMT \\
			\hline
			0.05 & 2.38\%  & 1.34\% &3.40\% & 3.98\%&3.51\% & 4.54\%&3.37\% \\
			\hline
			0.1   & 3.30\%  & 1.75\% &4.76\% & 5.49\%& 4.88\%& 6.57\%& 4.44\% \\
			\hline
			0.15 & 4.04\%  & 2.11\% & 6.07\% &6.86\% &6.18\% &8.52\% & 5.38\% \\
			\hline
		\end{tabular}%
	\label{tab:realdata_game}%
\end{table}%
\begin{figure}
	\centering
	\includegraphics[width=0.75\linewidth]{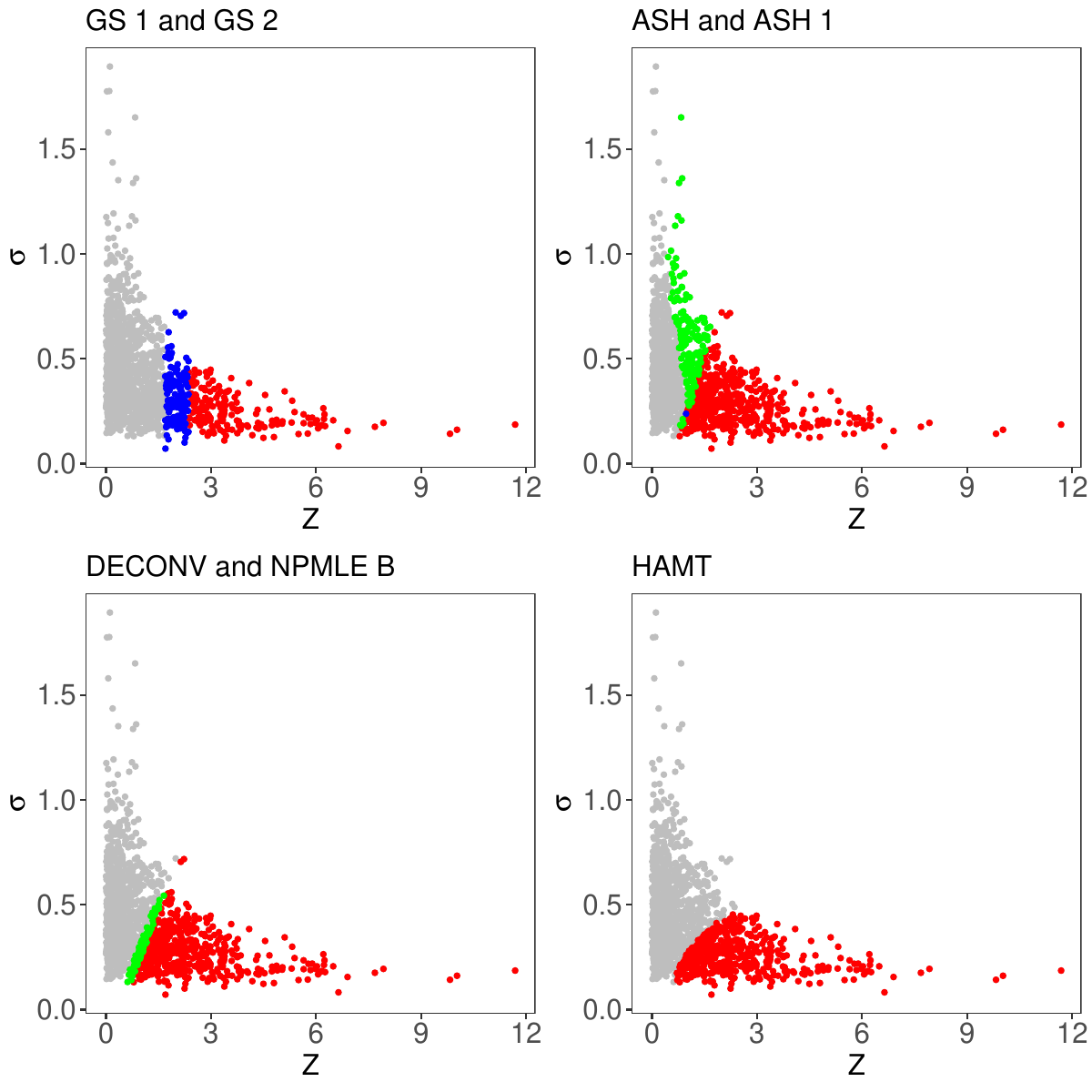}
	\caption{Scatter plot of $(Z_i,\sigma_i),~i=1,\ldots,m$ where $Z_i=(X_i-\mu_0)/\sigma_i$. The red dots in the bottom right panel indicate the hypotheses rejected by HAMT at $\alpha=0.1$. In the remaining three panels, we compare the rejections of GS 1 versus GS 2 (top left), ASH versus ASH 1 (top right) and DECONV versus NPMLE B (bottom left). In these three panels, the red dots indicate hypotheses rejected by both method 1, such as GS 1, and method 2, such as GS 2. Similarly, the blue dots represent hypotheses rejected by method 1 but not by method 2 and the green dots depict hypotheses rejected by method 2 but not by method 1. The horizontal axis is truncated below $0$ as all rejections are made when $Z_i>0$.}
	\label{fig:realdata_gamezvals}
\end{figure}
The red dots in the bottom right panel of Figure \ref{fig:realdata_gamezvals} indicate the hypotheses rejected by HAMT at $\alpha=0.1$. In the remaining three panels of Figure \ref{fig:realdata_gamezvals}, we compare the rejections of GS 1 versus GS 2 (top left), ASH versus ASH 1 (top right) and DECONV versus NPMLE B (bottom left). In these three panels, the red dots indicate hypotheses rejected by both method 1, such as GS 1, and method 2, such as GS 2. Similarly, the blue dots represent hypotheses rejected by method 1 but not by method 2 and the green dots depict hypotheses rejected by method 2 but not by method 1. We note that the rejection regions of GS 1 and GS2 depend only on $Z_i=(X_i-\mu_0)/\sigma_i$ and ASH 1 rejects relatively more hypotheses than ASH, particularly when $\sigma_i$ is large and $Z_i$ is small. In contrast, the rejection regions of HAMT and NPMLE B depend on both $Z_i$ and $\sigma_i$. Moreover, in comparison to the other methods, HAMT rejects more hypotheses when $\sigma_i$ is small and does not reject any hypotheses when $\sigma_i$ is bigger than $0.5$. The rejection regions of DECONV, ASH and ASH 1 give the impression that they depend on both $Z_i$ and $\sigma_i$, however as seen in our simulation experiments, these methods may fail to control the FDR at the desired level in case $\mu_i$ and $\sigma_i$ are correlated as their deconvolution estimator is not designed to capture this dependence.
\clearpage
%\newpage
\bibliographystyle{chicago}
\bibliography{refs}

\begin{thebibliography}{}

\bibitem[\protect\citeauthoryear{Banerjee, Fu, James, Mukherjee, and
  Sun}{Banerjee et~al.}{2023}]{banerjee2020nonparametric}
Banerjee, T., L.~J. Fu, G.~M. James, G.~Mukherjee, and W.~Sun (2023).
\newblock Nonparametric empirical bayes estimation on heterogeneous data.
\newblock {\em arXiv preprint arXiv:2002.12586\/}.

\bibitem[\protect\citeauthoryear{Banerjee, Mukherjee, Dutta, and
  Ghosh}{Banerjee et~al.}{2019}]{banerjee2019large}
Banerjee, T., G.~Mukherjee, S.~Dutta, and P.~Ghosh (2019).
\newblock A large-scale constrained joint modeling approach for predicting user
  activity, engagement, and churn with application to freemium mobile games.
\newblock {\em Journal of the American Statistical Association\/}.

\bibitem[\protect\citeauthoryear{Barber and Cand{\`e}s}{Barber and
  Cand{\`e}s}{2015}]{barber2015controlling}
Barber, R.~F. and E.~J. Cand{\`e}s (2015).
\newblock Controlling the false discovery rate via knockoffs.
\newblock {\em The Annals of Statistics\/}~{\em 43\/}(5), 2055--2085.

\bibitem[\protect\citeauthoryear{Basu, Cai, Das, and Sun}{Basu
  et~al.}{2018}]{basu2018weighted}
Basu, P., T.~T. Cai, K.~Das, and W.~Sun (2018).
\newblock Weighted false discovery rate control in large-scale multiple
  testing.
\newblock {\em Journal of the American Statistical Association\/}~{\em
  113\/}(523), 1172--1183.

\bibitem[\protect\citeauthoryear{Bates, Cand{\`e}s, Lei, Romano, and
  Sesia}{Bates et~al.}{2021}]{Batetal21}
Bates, S., E.~Cand{\`e}s, L.~Lei, Y.~Romano, and M.~Sesia (2021).
\newblock Testing for outliers with conformal p-values.
\newblock {\em arXiv:2104.08279\/}, Preprint.

\bibitem[\protect\citeauthoryear{Benjamini and Hochberg}{Benjamini and
  Hochberg}{1995}]{benjamini1995controlling}
Benjamini, Y. and Y.~Hochberg (1995).
\newblock Controlling the false discovery rate: a practical and powerful
  approach to multiple testing.
\newblock {\em Journal of the Royal statistical society: series B
  (Methodological)\/}~{\em 57\/}(1), 289--300.

\bibitem[\protect\citeauthoryear{Benjamini, Krieger, and Yekutieli}{Benjamini
  et~al.}{2006}]{benjamini2006adaptive}
Benjamini, Y., A.~M. Krieger, and D.~Yekutieli (2006).
\newblock Adaptive linear step-up procedures that control the false discovery
  rate.
\newblock {\em Biometrika\/}~{\em 93\/}(3), 491--507.

\bibitem[\protect\citeauthoryear{Boca and Leek}{Boca and
  Leek}{2018}]{boca2018direct}
Boca, S.~M. and J.~T. Leek (2018).
\newblock A direct approach to estimating false discovery rates conditional on
  covariates.
\newblock {\em PeerJ\/}~{\em 6}, e6035.

\bibitem[\protect\citeauthoryear{Cai and Sun}{Cai and
  Sun}{2009}]{cai2009simultaneous}
Cai, T.~T. and W.~Sun (2009).
\newblock Simultaneous testing of grouped hypotheses: Finding needles in
  multiple haystacks.
\newblock {\em Journal of the American Statistical Association\/}~{\em
  104\/}(488), 1467--1481.

\bibitem[\protect\citeauthoryear{Cao, Chen, and Zhang}{Cao
  et~al.}{2022}]{cao2022optimal}
Cao, H., J.~Chen, and X.~Zhang (2022).
\newblock Optimal false discovery rate control for large scale multiple testing
  with auxiliary information.
\newblock {\em The Annals of Statistics\/}~{\em 50\/}(2), 807--857.

\bibitem[\protect\citeauthoryear{Chao and Fithian}{Chao and
  Fithian}{2021}]{chao2021adapt}
Chao, P. and W.~Fithian (2021).
\newblock Adapt-gmm: Powerful and robust covariate-assisted multiple testing.
\newblock {\em arXiv preprint arXiv:2106.15812\/}.

\bibitem[\protect\citeauthoryear{Chen}{Chen}{2024}]{chen2022empirical}
Chen, J. (2024).
\newblock Empirical bayes when estimation precision predicts parameters.
\newblock {\em arXiv preprint arXiv:2212.14444\/}.

\bibitem[\protect\citeauthoryear{Dicker and Zhao}{Dicker and
  Zhao}{2016}]{dicker2016high}
Dicker, L.~H. and S.~D. Zhao (2016, 02).
\newblock {High-dimensional classification via nonparametric empirical Bayes
  and maximum likelihood inference}.
\newblock {\em Biometrika\/}~{\em 103\/}(1), 21--34.

\bibitem[\protect\citeauthoryear{Efron}{Efron}{2004}]{efron2004large}
Efron, B. (2004).
\newblock Large-scale simultaneous hypothesis testing: the choice of a null
  hypothesis.
\newblock {\em Journal of the American Statistical Association\/}~{\em
  99\/}(465), 96--104.

\bibitem[\protect\citeauthoryear{Efron}{Efron}{2008}]{efron2008microarrays}
Efron, B. (2008).
\newblock Microarrays, empirical bayes and the two-groups model.
\newblock {\em Statistical science\/}~{\em 23\/}(1), 1--22.

\bibitem[\protect\citeauthoryear{Efron}{Efron}{2012}]{efron2012large}
Efron, B. (2012).
\newblock {\em Large-scale inference: empirical Bayes methods for estimation,
  testing, and prediction}, Volume~1.
\newblock Cambridge University Press.

\bibitem[\protect\citeauthoryear{Efron}{Efron}{2014}]{efron2014two}
Efron, B. (2014).
\newblock Two modeling strategies for empirical bayes estimation.
\newblock {\em Statistical science: a review journal of the Institute of
  Mathematical Statistics\/}~{\em 29\/}(2), 285.

\bibitem[\protect\citeauthoryear{Efron}{Efron}{2016}]{efron2016empirical}
Efron, B. (2016).
\newblock Empirical bayes deconvolution estimates.
\newblock {\em Biometrika\/}~{\em 103\/}(1), 1--20.

\bibitem[\protect\citeauthoryear{Efron and Tibshirani}{Efron and
  Tibshirani}{2007}]{efron2007testing}
Efron, B. and R.~Tibshirani (2007).
\newblock On testing the significance of sets of genes.
\newblock {\em The annals of applied statistics\/}~{\em 1\/}(1), 107--129.

\bibitem[\protect\citeauthoryear{Fu, Gang, James, and Sun}{Fu
  et~al.}{2022}]{fu2022heteroscedasticity}
Fu, L., B.~Gang, G.~M. James, and W.~Sun (2022).
\newblock Heteroscedasticity-adjusted ranking and thresholding for large-scale
  multiple testing.
\newblock {\em Journal of the American Statistical Association\/}~{\em
  117\/}(538), 1028--1040.

\bibitem[\protect\citeauthoryear{Genovese and Wasserman}{Genovese and
  Wasserman}{2002}]{genovese2002operating}
Genovese, C. and L.~Wasserman (2002).
\newblock Operating characteristics and extensions of the false discovery rate
  procedure.
\newblock {\em Journal of the Royal Statistical Society: Series B (Statistical
  Methodology)\/}~{\em 64\/}(3), 499--517.

\bibitem[\protect\citeauthoryear{G'Sell, Wager, Chouldechova, and
  Tibshirani}{G'Sell et~al.}{2016}]{g2016sequential}
G'Sell, M.~G., S.~Wager, A.~Chouldechova, and R.~Tibshirani (2016).
\newblock Sequential selection procedures and false discovery rate control.
\newblock {\em Journal of the royal statistical society: series B (statistical
  methodology)\/}~{\em 78\/}(2), 423--444.

\bibitem[\protect\citeauthoryear{Gu and Koenker}{Gu and
  Koenker}{2017}]{gu2017empirical}
Gu, J. and R.~Koenker (2017).
\newblock Empirical bayesball remixed: Empirical bayes methods for longitudinal
  data.
\newblock {\em Journal of Applied Econometrics\/}~{\em 32\/}(3), 575--599.

\bibitem[\protect\citeauthoryear{Gu and Shen}{Gu and Shen}{2018}]{gu2018oracle}
Gu, J. and S.~Shen (2018).
\newblock Oracle and adaptive false discovery rate controlling methods for
  one-sided testing: theory and application in treatment effect evaluation.
\newblock {\em The Econometrics Journal\/}~{\em 21\/}(1), 11--35.

\bibitem[\protect\citeauthoryear{Guan and Tibshirani}{Guan and
  Tibshirani}{2022}]{guan2022prediction}
Guan, L. and R.~Tibshirani (2022).
\newblock Prediction and outlier detection in classification problems.
\newblock {\em Journal of the Royal Statistical Society. Series B, Statistical
  Methodology\/}~{\em 84\/}(2), 524.

\bibitem[\protect\citeauthoryear{Hu, Zhao, and Zhou}{Hu
  et~al.}{2010}]{hu2010false}
Hu, J.~X., H.~Zhao, and H.~H. Zhou (2010).
\newblock False discovery rate control with groups.
\newblock {\em Journal of the American Statistical Association\/}~{\em
  105\/}(491), 1215--1227.

\bibitem[\protect\citeauthoryear{Ignatiadis and Huber}{Ignatiadis and
  Huber}{2021}]{ignatiadis2021covariate}
Ignatiadis, N. and W.~Huber (2021).
\newblock Covariate powered cross-weighted multiple testing.
\newblock {\em Journal of the Royal Statistical Society: Series B (Statistical
  Methodology)\/}~{\em 83\/}(4), 720--751.

\bibitem[\protect\citeauthoryear{Jin and Cai}{Jin and
  Cai}{2007}]{jin2007estimating}
Jin, J. and T.~T. Cai (2007).
\newblock Estimating the null and the proportion of nonnull effects in
  large-scale multiple comparisons.
\newblock {\em Journal of the American Statistical Association\/}~{\em
  102\/}(478), 495--506.

\bibitem[\protect\citeauthoryear{Kiefer and Wolfowitz}{Kiefer and
  Wolfowitz}{1956}]{kiefer1956consistency}
Kiefer, J. and J.~Wolfowitz (1956).
\newblock Consistency of the maximum likelihood estimator in the presence of
  infinitely many incidental parameters.
\newblock {\em The Annals of Mathematical Statistics\/}, 887--906.

\bibitem[\protect\citeauthoryear{Koenker and Gu}{Koenker and
  Gu}{2017}]{koenker2017rebayes}
Koenker, R. and J.~Gu (2017).
\newblock Rebayes: An r package for empirical bayes mixture methods.
\newblock {\em Journal of Statistical Software\/}~{\em 82\/}(1), 1--26.

\bibitem[\protect\citeauthoryear{Koenker and Mizera}{Koenker and
  Mizera}{2014}]{koenker2014convex}
Koenker, R. and I.~Mizera (2014).
\newblock Convex optimization, shape constraints, compound decisions, and
  empirical bayes rules.
\newblock {\em Journal of the American Statistical Association\/}~{\em
  109\/}(506), 674--685.

\bibitem[\protect\citeauthoryear{Laird}{Laird}{1978}]{laird1978nonparametric}
Laird, N. (1978).
\newblock Nonparametric maximum likelihood estimation of a mixing distribution.
\newblock {\em Journal of the American Statistical Association\/}~{\em
  73\/}(364), 805--811.

\bibitem[\protect\citeauthoryear{Lei and Fithian}{Lei and
  Fithian}{2016}]{lei2016power}
Lei, L. and W.~Fithian (2016).
\newblock Power of ordered hypothesis testing.
\newblock In {\em International conference on machine learning}, pp.\
  2924--2932. PMLR.

\bibitem[\protect\citeauthoryear{Lei and Fithian}{Lei and
  Fithian}{2018}]{lei2018adapt}
Lei, L. and W.~Fithian (2018).
\newblock Adapt: an interactive procedure for multiple testing with side
  information.
\newblock {\em Journal of the Royal Statistical Society: Series B (Statistical
  Methodology)\/}~{\em 80\/}(4), 649--679.

\bibitem[\protect\citeauthoryear{Leung and Sun}{Leung and
  Sun}{2021}]{leung2021zap}
Leung, D. and W.~Sun (2021).
\newblock Zap: $ z $-value adaptive procedures for false discovery rate control
  with side information.
\newblock {\em arXiv preprint arXiv:2108.12623\/}.

\bibitem[\protect\citeauthoryear{Li and Barber}{Li and
  Barber}{2017}]{li2017accumulation}
Li, A. and R.~F. Barber (2017).
\newblock Accumulation tests for fdr control in ordered hypothesis testing.
\newblock {\em Journal of the American Statistical Association\/}~{\em
  112\/}(518), 837--849.

\bibitem[\protect\citeauthoryear{Li and Barber}{Li and
  Barber}{2019}]{li2019multiple}
Li, A. and R.~F. Barber (2019).
\newblock Multiple testing with the structure-adaptive benjamini--hochberg
  algorithm.
\newblock {\em Journal of the Royal Statistical Society: Series B (Statistical
  Methodology)\/}~{\em 81\/}(1), 45--74.

\bibitem[\protect\citeauthoryear{Liu, Sarkar, and Zhao}{Liu
  et~al.}{2016}]{liu2016new}
Liu, Y., S.~K. Sarkar, and Z.~Zhao (2016).
\newblock A new approach to multiple testing of grouped hypotheses.
\newblock {\em Journal of Statistical Planning and Inference\/}~{\em 179},
  1--14.

\bibitem[\protect\citeauthoryear{Love, Huber, and Anders}{Love
  et~al.}{2014}]{love2014moderated}
Love, M.~I., W.~Huber, and S.~Anders (2014).
\newblock Moderated estimation of fold change and dispersion for rna-seq data
  with deseq2.
\newblock {\em Genome biology\/}~{\em 15\/}(12), 1--21.

\bibitem[\protect\citeauthoryear{Lu and Stephens}{Lu and
  Stephens}{2019}]{lu2019empirical}
Lu, M. and M.~Stephens (2019).
\newblock Empirical bayes estimation of normal means, accounting for
  uncertainty in estimated standard errors.
\newblock {\em arXiv preprint arXiv:1901.10679\/}.

\bibitem[\protect\citeauthoryear{MOSEK}{MOSEK}{2019}]{mosek}
MOSEK, A. (2019).
\newblock {\em MOSEK Rmosek package 10.0.34}.

\bibitem[\protect\citeauthoryear{Pop-Eleches and Urquiola}{Pop-Eleches and
  Urquiola}{2013}]{pop2013going}
Pop-Eleches, C. and M.~Urquiola (2013).
\newblock Going to a better school: Effects and behavioral responses.
\newblock {\em American Economic Review\/}~{\em 103\/}(4), 1289--1324.

\bibitem[\protect\citeauthoryear{Scott, Kelly, Smith, Zhou, and Kass}{Scott
  et~al.}{2015}]{scott2015false}
Scott, J.~G., R.~C. Kelly, M.~A. Smith, P.~Zhou, and R.~E. Kass (2015).
\newblock False discovery rate regression: an application to neural synchrony
  detection in primary visual cortex.
\newblock {\em Journal of the American Statistical Association\/}~{\em
  110\/}(510), 459--471.

\bibitem[\protect\citeauthoryear{Silverman}{Silverman}{1986}]{silverman1986density}
Silverman, B.~W. (1986).
\newblock {\em Density estimation for statistics and data analysis}, Volume~6.
\newblock CRC press.

\bibitem[\protect\citeauthoryear{Stephens}{Stephens}{2017}]{stephens2017false}
Stephens, M. (2017).
\newblock False discovery rates: a new deal.
\newblock {\em Biostatistics\/}~{\em 18\/}(2), 275--294.

\bibitem[\protect\citeauthoryear{Sun and Cai}{Sun and
  Cai}{2007}]{sun2007oracle}
Sun, W. and T.~T. Cai (2007).
\newblock Oracle and adaptive compound decision rules for false discovery rate
  control.
\newblock {\em Journal of the American Statistical Association\/}~{\em
  102\/}(479), 901--912.

\bibitem[\protect\citeauthoryear{Sun and McLain}{Sun and
  McLain}{2012}]{sun2012multiple}
Sun, W. and A.~C. McLain (2012).
\newblock Multiple testing of composite null hypotheses in heteroscedastic
  models.
\newblock {\em Journal of the American Statistical Association\/}~{\em
  107\/}(498), 673--687.

\bibitem[\protect\citeauthoryear{Tansey, Wang, Blei, and Rabadan}{Tansey
  et~al.}{2018}]{tansey2018black}
Tansey, W., Y.~Wang, D.~Blei, and R.~Rabadan (2018).
\newblock Black box fdr.
\newblock In {\em International conference on machine learning}, pp.\
  4867--4876. PMLR.

\bibitem[\protect\citeauthoryear{Uffelmann, Huang, Munung, De~Vries, Okada,
  Martin, Martin, Lappalainen, and Posthuma}{Uffelmann
  et~al.}{2021}]{uffelmann2021genome}
Uffelmann, E., Q.~Q. Huang, N.~S. Munung, J.~De~Vries, Y.~Okada, A.~R. Martin,
  H.~C. Martin, T.~Lappalainen, and D.~Posthuma (2021).
\newblock Genome-wide association studies.
\newblock {\em Nature Reviews Methods Primers\/}~{\em 1\/}(1), 1--21.

\bibitem[\protect\citeauthoryear{Wand and Jones}{Wand and
  Jones}{1994}]{wand1994kernel}
Wand, M.~P. and M.~C. Jones (1994).
\newblock {\em Kernel smoothing}, Volume~60.
\newblock CRC press.

\bibitem[\protect\citeauthoryear{Wasserman}{Wasserman}{2006}]{wasserman2006all}
Wasserman, L. (2006).
\newblock {\em All of nonparametric statistics}.
\newblock Springer Science \& Business Media.

\bibitem[\protect\citeauthoryear{Weinstein, Ma, Brown, and Zhang}{Weinstein
  et~al.}{2018}]{weinstein2018group}
Weinstein, A., Z.~Ma, L.~D. Brown, and C.-H. Zhang (2018).
\newblock Group-linear empirical bayes estimates for a heteroscedastic normal
  mean.
\newblock {\em Journal of the American Statistical Association\/}, 1--13.

\bibitem[\protect\citeauthoryear{Xie, Kou, and Brown}{Xie
  et~al.}{2012}]{xie2012sure}
Xie, X., S.~Kou, and L.~D. Brown (2012).
\newblock Sure estimates for a heteroscedastic hierarchical model.
\newblock {\em Journal of the American Statistical Association\/}~{\em
  107\/}(500), 1465--1479.

\bibitem[\protect\citeauthoryear{Zhang, Xia, and Zou}{Zhang
  et~al.}{2019}]{zhang2019fast}
Zhang, M.~J., F.~Xia, and J.~Zou (2019).
\newblock Fast and covariate-adaptive method amplifies detection power in
  large-scale multiple hypothesis testing.
\newblock {\em Nature communications\/}~{\em 10\/}(1), 1--11.

\bibitem[\protect\citeauthoryear{Zhang and Chen}{Zhang and
  Chen}{2022}]{zhang2022covariate}
Zhang, X. and J.~Chen (2022).
\newblock Covariate adaptive false discovery rate control with applications to
  omics-wide multiple testing.
\newblock {\em Journal of the American Statistical Association\/}~{\em
  117\/}(537), 411--427.

\end{thebibliography}
\end{document}